\documentclass[reprint, amsmath, amssymb, showpacs, floatfix, longbibliography, aps, twocolumn, superscriptaddress, nofootinbib]{revtex4-1}
\usepackage{CJKutf8}
\usepackage{amsthm}
\usepackage{verbatim}
\usepackage{amsmath}
\usepackage{graphicx}
\usepackage{dcolumn}
\usepackage{bm}
\usepackage{amsfonts,stmaryrd}
\usepackage{physics}
\usepackage{dsfont}
\usepackage{graphicx}
\usepackage{xcolor}
\usepackage{graphicx}
\usepackage{cancel}
\usepackage{natbib}
\usepackage{color}
\usepackage{tikz}
\usepackage{mathrsfs}
\usepackage{relsize}
\usepackage{subfigure} 
\usepackage[linktocpage]{hyperref}
\usepackage[all]{xy}
\hypersetup{colorlinks=true,citecolor=blue,linkcolor=blue, urlcolor=blue, breaklinks=true}

\usepackage{yhmath}
\usepackage{blkarray}

\usepackage{multirow}
\usepackage{environ}
\NewEnviron{eqs}{%
\begin{equation}\begin{split}
    \BODY
\end{split}\end{equation}}

\newcommand{\ZZ}{{\mathbb Z}}
\newcommand{\RR}{{\mathbb R}}

\newcommand{\bv}{\boldsymbol v}
\newcommand{\be}{\boldsymbol e}
\newcommand{\bface}{\boldsymbol f}

\newcommand{\ba}{\boldsymbol a}
\newcommand{\bb}{\boldsymbol b}
\newcommand{\bc}{\boldsymbol c}

\newcommand{\lr}[1]{ \langle {#1} \rangle}

\newcommand{\mS}{{\mathcal{S}}}

\newcommand{\mA}{{\mathcal{A}}}

\newtheorem{theorem}{Theorem}[section]

\newtheorem{remark}{Remark}[section]
\newtheorem{definition}[theorem]{Definition}
\newtheorem{conjecture}[theorem]{Conjecture}

\usetikzlibrary{decorations.markings}

\graphicspath{{./Figures/}}

\newcommand{\change}{\color{black}}

\begin{document}

\title{Generalized Statistics on Lattices}

\author{Ryohei Kobayashi}
\thanks{These authors contributed equally to this work.}
\affiliation{School of Natural Sciences, Institute for Advanced Study, Princeton, NJ 08540, USA}

\author{Yuyang Li}
\thanks{These authors contributed equally to this work.}
\affiliation{International Center for Quantum Materials, School of Physics, Peking University, Beijing 100871, China}

\author{Hanyu Xue}
\thanks{These authors contributed equally to this work.}
\affiliation{International Center for Quantum Materials, School of Physics, Peking University, Beijing 100871, China}
\affiliation{Yuanpei College, Peking University, Beijing 100871, China}

\author{Po-Shen Hsin}
\email[E-mail: ]{po-shen.hsin@kcl.ac.uk}
\affiliation{Department of Mathematics, King’s College London, Strand, London WC2R 2LS, UK}

\author{Yu-An Chen}
\email[E-mail: ]{yuanchen@pku.edu.cn}
\affiliation{International Center for Quantum Materials, School of Physics, Peking University, Beijing 100871, China}

\date{\today}

\begin{abstract}

The statistics of particles and extended excitations, such as loops and membranes, are fundamental to modern condensed matter physics, high-energy physics, and quantum information science, yet a comprehensive lattice‑level framework for computing them remains elusive.
In this work, we develop a universal microscopic method to determine the generalized statistics of Abelian excitations on lattices of arbitrary dimension, and demonstrate it by deriving the statistics of particles, loops, and membranes in up to three spatial dimensions.
Our approach constructs a sequence of local unitary operators whose many‑body Berry phase encodes the desired statistical invariant. The required sequence is generated automatically from the Smith normal form of locality constraints and therefore needs no extra physical input.
We prove that the resulting invariants are quantized, provide an algorithm that computes them efficiently, and show how they unify familiar braiding and fusion data of particles while also uncovering new self- and mutual-statistics of loop and membrane excitations.
We further demonstrate that each statistical invariant corresponds to an ’t Hooft anomaly of a generalized symmetry; we show that a non‑trivial invariant both (i) obstructs gauging that symmetry and (ii) forbids any short‑range‑entangled (symmetry‑preserving) ground state.
This establishes a precise connection between microscopic lattice anomalies and many-body dynamics, providing a generalization of the Lieb–Schultz–Mattis theorem that constrains a wide class of quantum lattice systems.

\end{abstract}

\maketitle
\tableofcontents

\section{Introduction}

\begin{figure*}[t]
    \centering
    \subfigure[(1+1)D]{\raisebox{1.5cm}{\includegraphics[scale=0.7]{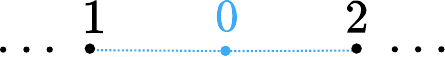}}\label{fig: complex (a)}}
    \hspace{0.05\linewidth}
    \subfigure[(2+1)D]{\includegraphics[scale=0.7]{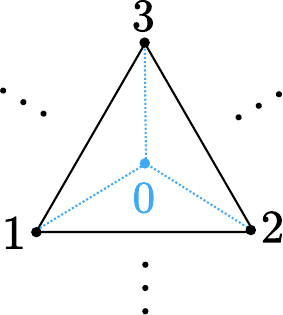}\label{fig: complex (b)}}
    \hspace{0.1\linewidth}
    \subfigure[(3+1)D]{\includegraphics[scale=0.7]{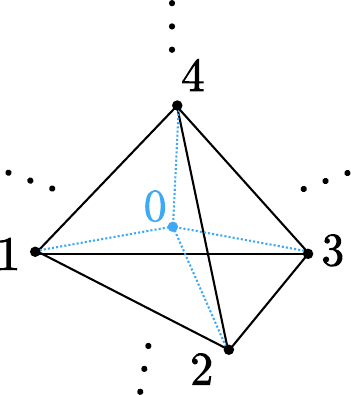}\label{fig: complex (c)}}
    \caption{In each dimension, we consider different simplicial complexes. 
    In (1+1)D, we use a segment subdivided by a vertex; in (2+1)D, a triangle with a central vertex; and in (3+1)D, a tetrahedron subdivided by a central vertex.
    These complexes are embedded into a larger spatial manifold (omitted by $\cdots$), with edge lengths chosen to be much larger than the system's correlation length.
    For computational convenience, we often compactify the manifold, wrapping the segment into a circle $S^1$, the triangle into a 2-sphere $S^2$, and the tetrahedron into a 3-sphere $S^3$.
    {\change
    We assume that for any simplex $\Delta$, there exists a finite-depth unitary $U(g)_\Delta$ that creates invertible $g$-excitations on its boundary $\partial\Delta$. For example, $U(g)_{01}$ is a string operator that produces a $g$ particle at vertex $1$ and a $g^{-1}$ particle at vertex $0$. Similarly, on the 2-simplex $\Delta_{012}$, the membrane operator $U(h)_{012}$ generates a $h$-loop excitation along the boundary edges of $\Delta_{012}$. Here, elements $g$ and $h$ belong to the fusion groups of particles and loops, respectively.
    }
    }
    \label{fig: complex}
\end{figure*}

{\change
Statistics of excitations is fundamental to quantum physics and underlie many phenomena in condensed matter and high-energy physics. Nontrivial statistics forbid their condensation.
For instance, bosons can condense and lead to superfluidity or superconductivity \cite{Bose1924Plancks, Einstein2005Quantentheorie, Landau1941Theory, Onsager1949Statistical, Ginzburg1950superconductivity, Bardeen1957Superconductivity}, while fermions cannot condense in isolation \cite{Gu2014supercohomology, Gaiotto:2015zta, Bhardwaj:2016clt, Barkeshli:2021ypb}.
In two spatial dimensions, the particle statistics broadens to anyons, enabling fractional quantum Hall physics \cite{Wilczek1982Quantum, Laughlin1983Anomalous, Stormer1999NobelLecture, Moore1991Nonabelions} and fault-tolerant quantum computation \cite{Freedman2002AModular, freedman2003topological, Kitaev2003Fault, Nayak2008NonAbelian}. Braided fusion categories provide a complete mathematical description of particle statistics in two spatial dimensions \cite{Kitaev:2005hzj, Levin2005String, rowell2006quantum, Kitaev2012Models, Kong2014Anyon}.

Extended excitations, such as loops and membranes naturally arising in three and higher dimensions, are still under active development. Their statistics govern confinement in gauge theories, constrain possible phase transitions, and control the logical operators of higher-dimensional quantum codes \cite{Wang2014braiding, Jiang2014Generalized, Jian2014Layer, Wang2015Field, Wang2016Bulk, Ye2016Topological, Pavel2017Braiding, Chan2018Braiding, Wang2019Topological, Zhang2021Compatible, Zhang2022Topological, Barkeshli2023Codimension, Barkeshli2024higher}. Yet, unlike the particle case, a general and practical framework for defining and computing these statistics on a lattice has remained elusive.
Recent studies show that loop statistics in (3+1)D $\mathbb{Z}_{2}$ gauge theories with fermionic particles, appearing in certain superconducting phases, realize discrete gravitational anomalies~\cite{Thorngren:2014pza, Kobayashi2019gapped, Johnson-Freyd:2020twl,  Hsin:2021qiy, CH21, FHH21, Tata2022anomalies}. This anomaly is believed to prevent any lattice realization of the gauge theory in three spatial dimensions. Remarkably, the fermionic loop statistics also give rise to a nontrivial quantum cellular automaton, a locality-preserving unitary that plays a central role in the classification of unitary operators on lattice quantum systems \cite{Haah:2018jdf, CH21, Chen2023Highercup, fidkowski2024qca}.
}

The statistics of topological excitations also provide 't Hooft anomaly of the corresponding symmetry, which constrains the dynamics of the quantum systems and distinguishes different phases of matter. For instance, the 1-form symmetry in (2+1)D is generated by Abelian anyons \cite{Hsin:2018vcg, Wang2020InAbelian}, and the statistics determines whether the 1-form symmetry can be gauged.
{\change 
In lattice models with a tensor-product Hilbert space, higher-form symmetries are typically emergent. The associated symmetry generators become genuinely topological only once the Gauss law is energetically enforced at low energies. At the same time, families of symmetry operators supported on closed submanifolds can often be identified as exact microscopic symmetries, commuting with the full lattice Hamiltonian. Any quantum system with an anomalous higher-form symmetry cannot be a trivially gapped phase, as demonstrated in various lattice models~\cite{Barkeshli2015GeneralizedKitaev, Chen:2017fvr, Chen:2018nog, Chen2020, Yang:2024zir}.\footnote{\change More precisely, an anomalous higher-form symmetry $S$ cannot preserve any short-range entangled state; that is, there is no state $|\Psi\rangle$, obtained from a product state by a finite-depth quantum circuit, that satisfies $S|\Psi\rangle \propto |\Psi\rangle$.}}
The 't Hooft anomaly matching condition also allows us to distinguish different phases of matter by the statistics of topological excitations, as used extensively in Refs.~\cite{Levin2012Braiding, Ellison2022Pauli, Ellison2023paulitopological, Shirley2022ThreeQCA, Chen2024Chiral, Ruba2024Homological, Liang2024Extracting, Liang2024Operator}.

In this work, we develop {\change a general framework} on lattices to compute all possible statistics for excitations of arbitrary dimensions, including particles and extended excitations such as loop and membrane excitations. Statistics are extracted from multi-step operations on lattices that transform excitations back to themselves, such as moving particle excitations using string operators, and the statistics arise as the Berry phase of the operation.
At each step, microscopic phase ambiguities may arise, modifying the state locally; the operations are designed so that these ambiguities cancel, yielding well-defined, universal statistics.
{\change We refer to such constructions as \emph{statistical processes}.
Examples of statistical processes have been discussed in Refs.~\cite{Levin2003Fermions, Kawagoe2020Microscopic} for particle excitations and in Ref.~\cite{FHH21} for loop excitations, but these studies do not treat the processes systematically nor extend them to excitations in arbitrary dimensions.
}

We develop a systematic method based on the Smith normal form to construct processes (sequences of unitary operators) with clear physical interpretations. Using this method, we unify and improve known statistical processes for both particle and loop excitations and also uncover new self and mutual statistics of membrane excitations.
{\change The generalized statistics in this framework encompass not only conventional braiding processes but also statistics arising from distinct fusion pathways, such as the fusion of particles in one dimension, loops in two dimensions, and membranes in three dimensions.}

Moreover, the statistics of excitations often imply a nontrivial low-energy spectrum, as particles with nontrivial statistics cannot condense. We find that these microscopic definitions of statistics are directly related to 't Hooft anomalies in lattice models, and we demonstrate that these statistics prevent the realization of a short-range entangled state. This connection provides insight into the dynamical consequences of 't Hooft anomalies in microscopic lattice models.
We focus on invertible excitations, whose fusion follows group multiplication rules. The generalization to non-invertible excitations will be explored in future work.

\begin{figure*}
    \centering
    \subfigure[Particle excitations]{\includegraphics[scale=0.53]{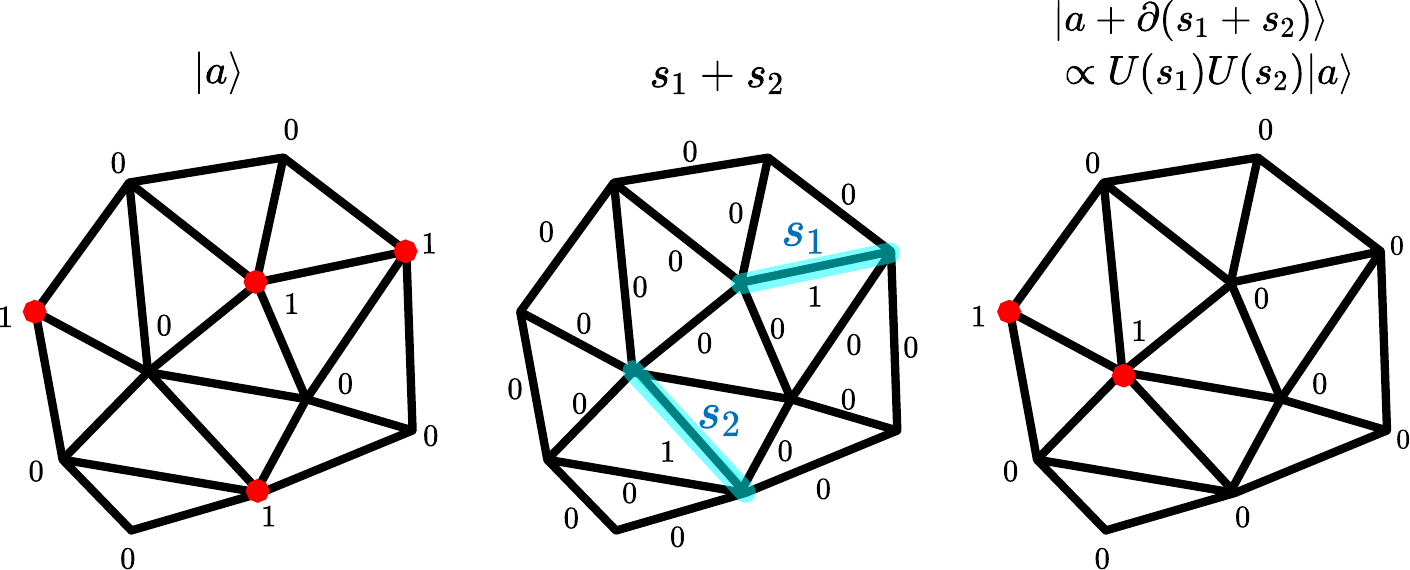}}\\
    \raisebox{0ex}{\subfigure[Loop excitations]{\includegraphics[scale=0.53]{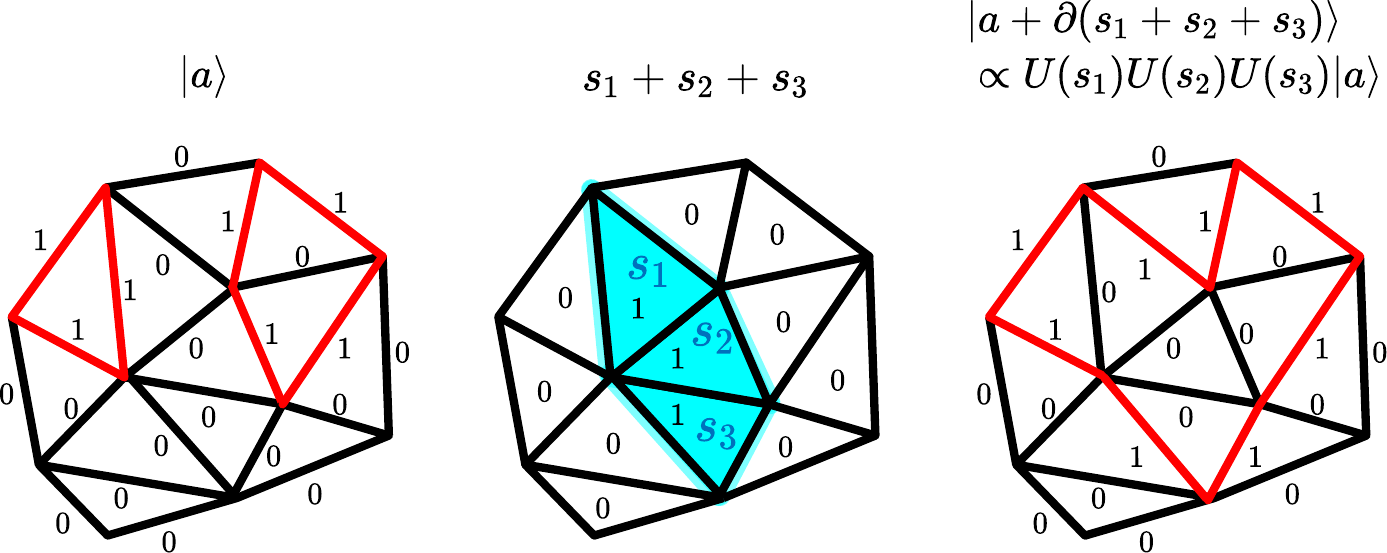}}}
    \caption{Particles and loop excitations.
    (a) For simplicity, we consider particles with the fusion group $G = \mathbb{Z}_2$. The configuration state $\ket{a}$ is labeled by a $\mathbb{Z}_2$ field at each vertex, indicating the presence or absence of particle excitations. Edges $s_1$ and $s_2$ have coefficients $1 \in \mathbb{Z}_2$. The unitary operator $\prod_i U(s_i)$ is the string operator supported on these edges, creating particles at the endpoints.
    (b) For loop excitations with the fusion group $G = \mathbb{Z}_2$, the configuration state $\ket{a}$ is labeled by a $\mathbb{Z}_2$ field on each edge, with edges labeled by $1 \in \mathbb{Z}_2$ forming closed loops. Faces $s_1$, $s_2$, $s_3$ have coefficients $1 \in \mathbb{Z}_2$.
    The unitary operator $\prod_i U(s_i)$ is the membrane operator supported on these faces, creating loop excitations along the boundaries.
    Applying $\prod_i U(s_i)$ to $\ket{a}$ in both cases yields a configuration state proportional to $\ket{a + \partial \left( \sum_i s_i \right)}$.
    These examples can be generalized to any Abelian group $G$, with additional considerations for the branching structure.
    }
    \label{fig: example of a and s}
\end{figure*}

{\change
\subsection*{Summary of results}

This work introduces a unified framework for defining the statistics of $p$-dimensional invertible excitations, generated by unitary operators supported on $(p+1)$-dimensional submanifolds, in lattices of arbitrary spatial dimension $d$.
Invertibility guarantees that the corresponding unitary circuit has finite depth\footnote{\change In the literature, this is sometimes referred to as a \emph{shallow-depth circuit}, meaning that the circuit depth is small compared to the system size under consideration.}; for example, Abelian anyons arise at the endpoints of finite-depth string operators.
Causality imposes locality constraints on these operators: for example, the commutator of two unitaries supported on regions $A$ and $B$ is localized within a neighborhood of their intersection $A \cap B$.
These locality constraints determine the possible structure of generalized statistics on lattices.

\subsubsection{Review of detecting particle statistics on lattices}\label{sec:Review_of_particle}

To illustrate our approach, we begin with the well-known case of particle statistics.
From quantum mechanics, we know that the statistics of particles is captured by the phase acquired when two identical particles are exchanged.
The challenge is to define such an exchange precisely on a lattice without being obscured by microscopic details.
For concreteness, here we focus on particles obeying $\mathbb{Z}_2$ fusion, where two identical particles can fuse into the trivial (vacuum) particle.

Consider two identical particles located at sites 1 and 2 on the lattice, as shown in Fig.~\ref{fig: complex (b)}.  
We perform the following steps:  
\begin{enumerate}
    \item Move the particle from site 1 to site 0,
    \item Move the particle from site 2 to site 1,
    \item Move the particle from site 0 to site 2.
\end{enumerate}
At the end of this sequence, the particles at sites 1 and 2 have been exchanged.  
If $U_{ij}$ denotes the string operator that moves a particle from site $i$ to site $j$, this process is represented by  
\begin{equation}
    U_{02}U_{21}U_{10}~.
\label{eq: wrong exchange}
\end{equation}
While this operator exchanges the two particles, it can also introduce unwanted phases into the quantum state, unlike in a classical exchange process.
For example, moving the particle from site 1 to site 0 may produce an additional phase in the state, potentially altering the outcome of the entire process.
Therefore, we seek a process that captures the particle statistics while remaining insensitive to such phase ambiguities, which generically arise in lattice systems.  

We label each site in the particle-number basis, with $0$ denoting no particle and $1$ denoting a particle; a total configuration state $|a\rangle$ on a finite lattice is thus specified by a $0$-chain, a set of $\mathbb{Z}_2$ values assigned to the vertices that indicate particle occupancy at each site.
Let $U(s)$ be the unitary string operator that moves a particle along a string $s$, which can also be interpreted as creating two particles at the endpoints $\partial s$.
Explicit examples of $|a\rangle$ and $U(s)$ are shown in Fig.~\ref{fig: example of a and s}.
In general,  
\begin{equation}
    U(s)|a\rangle \propto |a+\partial s\rangle~,
\end{equation}
where $|a\rangle$ is the initial state and $|a+\partial s\rangle$ is the state with two additional particles at $\partial s$.  
The phase difference between $U(s)|a\rangle$ and $|a+\partial s\rangle$ is gauge-dependent: it depends on the chosen overall phases of $|a\rangle$ and $|a+\partial s\rangle$, as well as on the specific convention used to define $U(s)$.  
We can construct an exchange process that is immune to such phase ambiguities.  
Consider the lattice in Fig.~\ref{fig: complex (b)}, where the two particles at sites 1 and 2 are exchanged via the operators~\cite{Levin2003Fermions}  
\begin{equation}
    U_{02}U_{03}^{-1}U_{01}U_{02}^{-1}U_{03}U_{01}^{-1}~.
\label{eq:T-junction in the introduction}
\end{equation}
When acting on the initial state $\left|\hbox{ \raisebox{-1ex}{\includegraphics[width=.6cm]{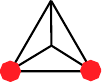}}} \right\rangle$, where particles occupy vertices 1 and 2, this T-junction process can be visualized as follows:
\begin{widetext}
\begin{equation}
    \vcenter{\hbox{\includegraphics[width=0.94\linewidth]{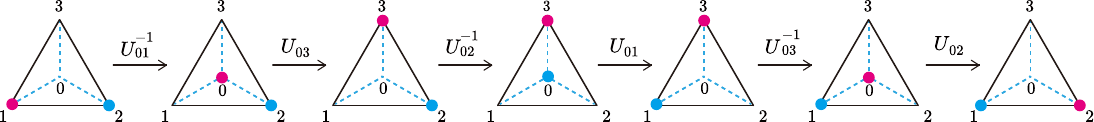}}}~~.
\end{equation}
\end{widetext}
In contrast to Eq.~\eqref{eq: wrong exchange}, each string operator here appears together with its inverse, ensuring that the microscopic phase contributions cancel exactly.  
This construction thus yields the particle statistics without contamination from local, lattice-dependent phase factors.

In this work, we generalize this construction to detect the statistics of extended excitations.  
We formulate lattice processes that map between excitation configurations in a nontrivial manner, analogous to particle exchange, while ensuring that microscopic phase ambiguities cancel throughout the process.
Additional examples of such statistical processes involving particles, loops, and membranes in various spatial dimensions are presented in Sec.~\ref{sec:Examples_Statistics}.
}

{\change
\subsubsection{Universal microscopic description for statistics}

We now present the general setting for our construction.
This work develops a universal framework for defining generalized statistics on lattices across different dimensions. We consider systems with invertible excitations, whose loci can be moved by finite-depth quantum circuits. An excitation may correspond to a violation of certain Hamiltonian terms, a defect that locally modifies the Hamiltonian, or, more generally, a symmetry defect.
Hence, this framework applies to both \emph{gapped} and \emph{gapless} systems.
An invertible excitation possesses an inverse, forming a fusion group $G$. For simplicity, we first summarize the results for Abelian groups $G$, with the understanding that the framework extends to non-Abelian fusion groups as well. These excitations may also be spatially extended, spanning $p$ dimensions—for example, $p=0$ corresponds to quasiparticles, $p=1$ to loop excitations such as magnetic flux loops, and higher $p$ to more extended objects.

Our framework begins by specifying the possible configurations of excitations, denoted by $\mathcal{A}$, on a simplicial complex $X$ embedded in space. We consider the Hilbert space basis labeled by excitations $a \in \mathcal{A}$, where each $p$-dimensional simplex is associated with an element of the group $G$. Each configuration is represented by the \textbf{configuration state} $|a\rangle$.
This state is not uniquely defined, as it can be redefined through the action of unitary operators supported at the locations of the excitations. To address this ambiguity, we fix a particular choice of states ${\ket{a}}$ and construct invariants that remain independent of these definitions.

Excitations are generated by local unitary operators $U(s)$, supported on a $(p+1)$-dimensional simplex, where $s$ represents the $(p+1)$-simplex with a coefficient in $G$. We assume these excitations are deconfined, meaning that the operator $U(s)$ creates excitations only localized around its boundary $\partial s$.\footnote{\change In condensed matter physics, a particle is called \emph{deconfined} if it can be separated from its antiparticle at arbitrarily large distances without incurring an energy cost that grows with distance. In contrast, a \emph{confined} particle has an energy cost proportional to the separation between excitations, since the corresponding string operator $U(s)$ violates Hamiltonian terms all along its length, not just near the endpoints.}
When a sum of $s$, such as $s_1 + \cdots + s_n$, is closed (having no boundary), the product operator $U(s_n) \cdots U(s_1)$ preserves excitations and generates a generalized symmetry of the system.
Since the excitations are assumed to be invertible, this generalized symmetry forms a group. In general, the symmetries can form a higher group \cite{Kapustin:2013uxa,Benini:2018reh}, and they are symmetries supported on subsystems of the lattice~\cite{Barkeshli2024higher}.

We assume that the unitaries $U(s)$ generating the symmetry can be realized by a finite-depth quantum circuit—for example, they do not involve lattice isometries such as translations or rotations. This is always the case when $U$ generates an internal symmetry, and we will restrict our attention to such cases from now on. This assumption will play an important role in our later formulation of invariants for generalized statistics.

The unitaries move the configuration of the excitations.
In general, a unitary operator can be used to connect two states $\ket{a}$ and $\ket{a'}$ through the relation $U(s)\ket{a} \propto \ket{a'}$, where $a + \partial s = a'$ and $+$ represents the fusion of Abelian excitations.
Once the states $\ket{a}$ and the unitary operators $U(s)$ are chosen, the phase factor $\theta(s, a)$ can be specified as:
\begin{align}
    U(s)\ket{a} =e^{i\theta(s,a)} \ket{a+\partial s}~,
\end{align}
with $\theta(s, a)$ depending on both the unitary operator $U(s)$ and the initial state $\ket{a}$.
From the relation $U^{-1}U=1$, it follows that
\begin{align}
    U(s)^{-1}\ket{a} =e^{-i\theta(s,a-\partial s)} \ket{a-\partial s}.
\end{align}
These phase factors encode essential information about the system.  
As an illustration, the overall phase in the T-junction process~\eqref{eq:T-junction in the introduction} acting on the initial state $\left|\hbox{ \raisebox{-1ex}{\includegraphics[width=.6cm]{T_junction_state_12.pdf}}} \right\rangle$ is given by
\begin{equation}
    e^{i\Theta} = ~ \left\langle \hbox{ \raisebox{-1ex}{\includegraphics[width=.6cm]{T_junction_state_12.pdf}}} \right|
    U_{02}U_{03}^{-1}U_{01}U_{02}^{-1}U_{03}U_{01}^{-1}
    \left| \hbox{ \raisebox{-1ex}{\includegraphics[width=.6cm]{T_junction_state_12.pdf}}} \right\rangle,
\end{equation}
with
\begin{eqs}
    \Theta=&~\theta \left(e_{02},\hbox{ \raisebox{-1ex}{\includegraphics[width=.6cm]{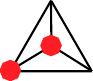}}} \right)
    -\theta \left(e_{03},\hbox{ \raisebox{-1ex}{\includegraphics[width=.6cm]{T_junction_state_01.pdf}}} \right) 
    + \theta \left(e_{01},\hbox{ \raisebox{-1ex}{\includegraphics[width=.6cm]{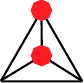}}} \right) \\ 
    &
    - \theta \left(e_{02},\hbox{ \raisebox{-1ex}{\includegraphics[width=.6cm]{T_junction_state_03.pdf}}} \right) 
    + \theta \left(e_{03},\hbox{ \raisebox{-1ex}{\includegraphics[width=.6cm]{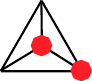}}} \right) 
    - \theta \left(e_{01},\hbox{ \raisebox{-1ex}{\includegraphics[width=.6cm]{T_junction_state_02.pdf}}} \right)~,
    \nonumber
\end{eqs}
where the simplices $s$ correspond to edges $e_{ij}$ in Fig.~\ref{fig: complex (b)}.
A single phase factor $\theta(s,a)$ does not by itself have a direct physical meaning, as it depends on the specific choices of $\{\ket{a}\}$ and $\{U(s)\}$. Changes in these choices can be viewed as ``gauge transformations,'' under which the ``Berry connection'' $\theta(s,a)$ varies. However, certain linear combinations of these phases, such as $\Theta$, can be constructed to be independent of the choices of $\{\ket{a}\}$ and $\{U(s)\}$. These gauge-invariant quantities are physically meaningful and will be referred to as \textbf{generalized statistics}. In this sense, generalized statistics encode the intrinsic properties of the excitations, independent of arbitrary conventions.

We then consider a sequence of unitaries starting with the specific configuration of excitations back to itself,
\begin{eqs}
     &\bra{a_{0}}U(s_{n-1}) \dots U(s_j)\dots U(s_0)\ket{a_0}  \\
     &= \exp\left(i\sum_{j=0}^{n-1}\theta(s_j,a_j)\right)~,
\end{eqs}
with $a_{j+1}:=a_j + \partial s_j$.
The above sequence can be regarded as the ``closed path'' in the Hilbert space, along which we measure the Berry phase associated with the family of states.
For a practical purpose, it is convenient to treat the above phase as the formal sum of the objects $\theta(s,a)$:
\begin{align}
    E=\bigoplus_{s, a}\mathbb{Z} \theta(s,a)~.
\label{eq: E definition}
\end{align}
The phase defined above may or may not remain invariant under possible deformations of the states ${\ket{a}}$ and the unitaries ${U(s)}$, such as the redefinition by phases or local perturbations modifying the unitaries $U\rightarrow UU'$ at each step for unitaries $U'$ close to the identity.
The sequence needs to be carefully designed to be invariant under such deformations. We will see that the necessary and sufficient conditions for the invariance can be formulated as the constraints on the $\ZZ$ coefficients in Eq.~\eqref{eq: E definition}. We define the subgroup $E_{\mathrm{inv}} \subset E$, which contains the elements that qualify as invariants associated with the excitations.

Some elements of $E_{\mathrm{inv}}$ correspond to trivial invariants that reduce to the identity operator. A trivial sequence of unitaries arises from the locality properties of the unitary operators. For instance, if the supports of two operators $U(s_{1})$ and $U(s_{2})$ do not overlap, i.e., $s_{1} \cap s_{2} = \varnothing$, their commutator is trivial
\begin{eqs}
    [U(s_{2}), U(s_{1})] = 1~,
    \label{eq: Us1 Us2 commutator}
\end{eqs}
where $[A, B] := A^{-1} B^{-1} A B$. This property extends to higher commutators involving multiple operators, such as:
\begin{eqs}
    &[U(s_n),[\cdots,[U(s_2),U(s_1)]]] = 1~,
    \label{eq: Us1 Us2 ... Usn commutator}
\end{eqs}
if $s_1 \cap s_2 \cap \dots \cap s_n = \varnothing$.
This property follows from each unitary being a finite-depth circuit, which ensures that the commutator $[U(s_{2}), U(s_{1})]$ has support only near $s_{1}\cap s_{2}$, since the circuit depth is much smaller than the length scale of the complexes under consideration.
These higher commutators acting on any initial configuration state $\ket{a_0}$, form the subgroup $E_{\mathrm{id}} \subset E_{\mathrm{inv}}$.
The {\bf genuine invariants} of the excitations are characterized by the quotient group
\begin{equation}
    T := E_{\mathrm{inv}} / E_{\mathrm{id}},
\end{equation}
where the trivial phases arising from locality have been factored out. We show in general that, for a finite symmetry group $G$, the invariants $T$ form a finite Abelian group.
This implies that the invariants $T$ must be quantized into the discrete values. This group of invariants can be explicitly computed using a computer, given the possible configurations of the excitations and the unitaries, as well as the group $G$ describing the fusion of excitations.

\begin{remark}
    The genuine invariants may appear to depend on the choice of underlying cellulation, and indeed they can vary for small complexes. 
    However, based on our numerical verification, we conjecture that the invariants $T$ become universal once the $d$-dimensional complex is a cellulation of a $d$-dimensional manifold that is sufficiently large to embed $\partial \Delta_{d+1}$, the boundary of a $(d+1)$-simplex, as in the triangulations of spheres illustrated in Fig.~\ref{fig: complex}. 
    In other words, once the underlying complex is refined enough to support the relevant statistical processes (such as particle exchange), any triangulation or cellulation yields the same genuine invariants.  
    We further conjecture that these universal invariants $T$ coincide with the cohomology of the Eilenberg–MacLane space, as summarized in Table~\ref{tab: particle loop membrane}.
\end{remark}

\subsubsection{Statistics as anomalies and their dynamical consequences}

The invariant $\Theta \in T$ provides a microscopic definition of the statistics of excitations. 
At the same time, $\Theta$ can be interpreted as a microscopic definition of an 't~Hooft anomaly of the global symmetry, which represents an obstruction to gauging the symmetry. This can be understood as follows.  

To gauge a symmetry, one introduces gauge fields and promotes the global symmetry to a local gauge symmetry generated by Gauss law operators $G_\Delta$ supported on local regions $\Delta$. The physical Hilbert space is then restricted to states satisfying $G_\Delta = 1$. If this gauge-invariant Hilbert space is empty, the gauging procedure is obstructed—namely, an 't~Hooft anomaly is present.  
In Sec.~\ref{sec:Anomaly}, we further show that the generalized statistics can be expressed as a product of Gauss law operators,
\begin{equation}
    e^{i \Theta} = \prod G_\Delta.
\end{equation}
A nontrivial phase $\Theta$ obtained in this way signals an obstruction to enforcing the constraint $G_\Delta = 1$ everywhere. In this sense, a nontrivial $\Theta$ characterizes an 't~Hooft anomaly.

The presence of 't~Hooft anomalies constrains the low-energy spectrum of the theory. In particular, they forbid the existence of a unique gapped ground state. We find that the invariants of microscopic lattice systems $\Theta$ directly lead to such a dynamical consequence: when the invariant $\Theta$ with the symmetry defects is nontrivial, the state cannot be a short-range entangled (SRE) state preserving the symmetry. We show this statement in full generality, assuming the tensor network representation of the state with excitations. This is reminiscent of the Lieb-Schultz-Mattis theorem \cite{LIEB1961407, oshikawalsm, Hastings_2004, Kobayashi2019lsm, PhysRevB.101.224437, Cheng:2022sgb, Seifnashri2024lsm}, which constrains the low-energy spectrum of the state based on a given action of the internal and crystalline symmetries.

A large class of 't Hooft anomalies can be described through group cohomology by employing the group cohomology SPT phase in the bulk via bulk-boundary correspondence \cite{Dijkgraaf:1989pz, Chen:2011pg, Else2014Classifying}.
The mathematical results presented in Table~\ref{tab: particle loop membrane} show the group cohomology of higher-form symmetry, i.e., the cohomology of the Eilenberg–MacLane space, for finite Abelian groups \cite{WangCheng}.
Our generalized statistics provides a microscopic perspective of 't Hooft anomalies on the lattice, and in all examples we have evaluated on a computer, its classification matches the results in Table~\ref{tab: particle loop membrane}.
Although the computational power limits us from verifying arbitrarily large groups $G$, the agreement in the small cases suggests the correspondence. This consistency leads us to conjecture that our generalized statistics are classified by the group cohomology of higher groups. In the following section, we explicitly demonstrate examples of generalized statistics for small groups $G$, illustrating their correspondence to the group cohomology of higher-form symmetries.

The paper is organized as follows.  
Sec.~\ref{sec:Examples_Statistics} describes a variety of examples of generalized statistics for particles, loops, and membranes in up to three spatial dimensions, including exactly solvable models whose domain-wall excitations demonstrate the nontrivial statistics of each process.
Sec.~\ref{sec:Framework} defines generalized statistics from axioms for invertible excitations and shows that they take quantized values.  
Sec.~\ref{sec:Computation} presents an algorithm that computes these statistics by constructing microscopic processes as sequences of lattice operators, suitable for direct computer implementation.  
Using this framework, Sec.~\ref{sec:Anomaly} gives a microscopic definition of ’t~Hooft anomalies as obstructions to gauging symmetries, and Sec.~\ref{sec:SRE} proves that nontrivial statistics forbid short-range entangled states, revealing their dynamical consequences.  
We conclude in Sec.~\ref{sec:discussions} with possible future directions.  
Background concepts and extended discussions are provided in the appendices.

}

\begin{widetext}


\begin{table*}[thb]
\centering
\renewcommand{\arraystretch}{1.5}
\begin{tabular}{|c|c|c|c|}
\hline
 & $G$-particles with $G= \prod_i \mathbb{Z}_{N_i}$ & $G$-loops with $G= \prod_i \mathbb{Z}_{N_i}$ & $G$-membranes with $G= \prod_i \mathbb{Z}_{N_i}$\\
\hline
(1+1)D & 
$\begin{aligned}
\vphantom{H^{H^H}}& H^3(B G, U(1)) \\
=& \textstyle\prod_i \mathbb{Z}_{N_i} \textstyle\prod_{i<j} \mathbb{Z}_{(N_i, N_j)} \\
&\textstyle\prod_{i<j<k} \mathbb{Z}_{(N_i, N_j, N_k)} 
\end{aligned}$
& &\\ 
\hline
(2+1)D & 
$\begin{aligned}
&H^4(B^2 G, U(1)) \\
=& \textstyle\prod_{i} \mathbb{Z}_{(N_i, 2) \times N_i} \textstyle\prod_{i<j} \mathbb{Z}_{(N_i, N_j)}
\end{aligned}$
& 
$\begin{aligned}
\vphantom{H^{H^H}}&H^4(BG, U(1)) \\
=& \textstyle\prod_{i<j} \mathbb{Z}_{(N_i,N_j)}^2 \textstyle\prod_{i<j<k} \mathbb{Z}_{(N_i,N_j,N_k)}^2 \\
&\textstyle\prod_{i<j<k<l} \mathbb{Z}_{(N_i,N_j,N_k,N_l)}
\end{aligned}$
&\\
\hline
(3+1)D & 
$\begin{aligned}
&H^5(B^3G, U(1)) \\
=& \textstyle\prod_{i} \mathbb{Z}_{(N_i, 2)}
\end{aligned}$
& 
$\begin{aligned}
&H^5(B^2G, U(1)) \\
=& \textstyle\prod_{i} \mathbb{Z}_{(N_i, 2)} \textstyle\prod_{i<j} \mathbb{Z}_{(N_i,N_j)}
\end{aligned}$
& 
$\begin{aligned}
\vphantom{H^{H^H}}&H^5(BG, U(1)) \\
=& \textstyle\prod_{i} \mathbb{Z}_{N_i} \textstyle\prod_{i<j} \mathbb{Z}^2_{(N_i, N_j)} \\
&\textstyle\prod_{i<j<k} \mathbb{Z}^4_{(N_i, N_j, N_k)}  \\
&\textstyle\prod_{i<j<k<l} \mathbb{Z}^3_{(N_i,N_j,N_k,N_l)} \\
&\textstyle\prod_{i<j<k<l<m} \mathbb{Z}_{(N_i,N_j,N_k,N_l,N_m)}
\end{aligned}$ \\
\hline
\end{tabular}
\caption{\change The cohomology of the Eilenberg–MacLane space $B^n G := K(G, n)$ for the finite Abelian group $G = \prod_i \mathbb{Z}_{N_i}$~\cite{WangCheng}. The notation $(N_i, N_j, \cdots)$ denotes the greatest common divisor among the integers. This cohomology classifies the anomaly as an obstruction to gauging the higher-form $G$ symmetry, which corresponds to symmetry-protected topological (SPT) phases in one higher dimension.
We conjecture that these data precisely match the generalized statistics of particle, loop, and membrane excitations as defined by Eq.~\eqref{eq: quotient group} in Sec.~\ref{sec:Framework of invariants in generic dimensions}. Specifically, $p$-dimensional excitations in $(d+1)$-dimensional spacetime have generalized statistics characterized by $H^{d+2}(B^{d-p}G, U(1))$.
In Sec.~\ref{sec:Computational algorithm SNF}, we verify this conjecture for small groups $G$ (specifically $G=\mathbb{Z}_N$ with $N \leq 12$ or $\mathbb{Z}_N \times \mathbb{Z}_N$ with $N \leq 5$), with explicit unitary operator sequences for the generalized statistics given in Sec.~\ref{sec:Examples_Statistics}.
}
\label{tab: particle loop membrane}
\end{table*}

{\change
\section{Examples of generalized statistics}\label{sec:Examples_Statistics}

To illustrate our construction, we compute explicit statistical invariants for particle, loop, and membrane excitations in dimensions $d\le3$, using representative Abelian fusion groups.  A systematic derivation valid for arbitrary excitations in any dimension will be presented in Sec.~\ref{sec:Computation}.

}

\subsection{Particle excitations}

We consider particle excitations in $(1+1)$, $(2+1)$, and $(3+1)$ spacetime dimensions.

\subsubsection{Particles in (1+1)D}

Let the fusion groups of the particles be a finite group $G$, where each particle is labeled by an element $g \in G$. While $G$ can be non-Abelian in general, we first focus on the Abelian case for demonstration.
We define the hopping operator $U(g)_{ij}$ that creates a particle $g$ at vertex $j$ and a particle $g^{-1}$ at vertex $i$. The statistics of $g$-particles are defined by: 
\begin{equation}
    Z_3(g) := [ U(g)_{02},U(g)_{01}^{|g|}]~,
    \label{eq: (1+1)D particle fusion statistics}
\end{equation}
where the vertices $0$, $1$, and $2$ are positioned along a segment, as shown in Fig.~\ref{fig: complex (a)}, {\change and we define the group commutator by
\begin{equation}
    [A,B]:= A^{-1} B^{-1}A B.
\end{equation}
}
Here, $|g|$ denotes the order of the element $g \in G$.
This index corresponds to the degree-3 cohomology class $H^3(BG, U(1))$, {\change as shown in Table~\ref{tab: particle loop membrane}}.
{\change The statistics $Z_3(g)$ can be interpreted as the partition function of the corresponding cohomology class on a lens space~\cite{Wang2015Topological, Tantivasadakarn2017Dimensional}:}
\begin{equation}
    Z_3(g) = \prod_{n=1}^{|g|} F(g, g^n, g)~,
\end{equation}
where the $F$-symbol $F(g_1, g_2, g_3)$ is {\change a sequence of unitary operators} defined microscopically in Ref.~\cite{Kawagoe2020Microscopic, Seifnashri2024lsm}.
{\change
Notably, when particles are treated as symmetry defects, the expression in Eq.~\eqref{eq: (1+1)D particle fusion statistics} coincides with the topological invariants computed from the boundaries of (2+1)D symmetry-protected topological (SPT) phases~\cite{CarolynZhangSPTEntangler}, which capture the Else–Nayak index~\cite{Else2014Classifying}. The nontrivial statistics described by Eq.~\eqref{eq: (1+1)D particle fusion statistics} signal an obstruction to satisfying $U(g)^{|g|}_{ij} = 1$, thereby reflecting an anomalous symmetry action.
}
We also emphasize that the statistics in Eq.~\eqref{eq: (1+1)D particle fusion statistics} apply to non-Abelian groups. For instance, when $G = S_3$ is the symmetric group,  which contains the Abelian subgroups $\mathbb{Z}_2$ and $\mathbb{Z}_3$, we can substitute their generators into Eq.~\eqref{eq: (1+1)D particle fusion statistics} to derive the $\mathbb{Z}_2$ and $\mathbb{Z}_3$ invariants, separately. These results are consistent with the cohomology class $H^3(S_3, U(1)) = \mathbb{Z}_6 = \mathbb{Z}_2 \times \mathbb{Z}_3$.

{\change
Now, we illustrate the simplest example of nontrivial fusion statistics. Consider the anomalous $\mathbb{Z}_2$ symmetry on a one-dimensional qubit chain~\cite{Chen2011Twodimensional, Else2014Classifying, Ji_2020, Chatterjee2023Symmetryshadow}:
\begin{equation}
    S:=\prod_i X_i \prod_i CZ_{i,i+1}~,
\label{eq: 1d anomalous Z2 symmetry}
\end{equation}
where $X_i$ and $Z_i$ are Pauli matrices on site $i$, and $CZ_{i,i+1}$ is the controlled-$Z$ gate between adjacent qubits.
This anomalous symmetry action originates at the boundary of the (2+1)D Levin–Gu $\mathbb{Z}_2$ SPT phase~\cite{Levin2012Braiding}.
The symmetry-preserving operators are generated by
\begin{eqs}
    W_{i+\frac{1}{2}}:= Z_i Z_{i+1}, \quad
    U_{i-\frac{1}{2} \rightarrow i+\frac{1}{2}} :=  X_i ~CZ_{i-1, i} ~CZ_{i, i+1}.
\label{eq: 1d domain wall and hopping}
\end{eqs}
We interpret the $W_{i+\frac{1}{2}}$ as the domain wall excitation on the edge $\langle i,i+1\rangle$, and $U_{i-\frac{1}{2} \rightarrow i+\frac{1}{2}}$ is the hopping operator of the domain wall excitation from $i-\frac{1}{2}$ to $i+\frac{1}{2}$. We omit the group label since $\ZZ_2$ has only one nontrivial element. Applying Eq.~\eqref{eq: (1+1)D particle fusion statistics}, the fusion statistic of this domain wall is
\begin{eqs}
    [ U_{i+\frac{1}{2} \rightarrow i+\frac{3}{2}},~ U_{i-\frac{1}{2} \rightarrow i+\frac{1}{2}}^2] 
    = [  X_{i+1} ~CZ_{i, i+1}~CZ_{i+1, i+2},~Z_{i-1}Z_{i+1}]
    = -1. \nonumber
\end{eqs}
The value $-1$ confirms that the $\ZZ_2$ symmetry in Eq.~\eqref{eq: 1d anomalous Z2 symmetry} is anomalous.
}

\subsubsection{Particles in (2+1)D and (3+1)D}

Invertible particles in (2+1) and (3+1)-dimensional spacetime must have an Abelian fusion group $G$.
As reviewed in Sec.~\ref{sec:Review_of_particle}, the statistics of a particle labeled by a group element $g \in G$ can be detected using the T-junction process~\cite{Levin2003Fermions, Kawagoe2020Microscopic, FHH21}:  
\begin{eqs}
    e^{i\Theta(g)}:= ~& U(g)_{02} U(g)_{03}^{-1} U(g)_{01} U(g)_{02}^{-1} U(g)_{03} U(g)_{01}^{-1}~,
\label{eq: T-junction process}
\end{eqs}
where the vertices $0$, $1$, $2$, and $3$ are shown in Fig.~\ref{fig: complex (b)} and \ref{fig: complex (c)}.
When $G = \mathbb{Z}_N$, the possible values of $\Theta$ are:
\begin{equation}
    e^{i\Theta} = \begin{cases}
        \exp\left(2\pi i \frac{k}{\mathrm{gcd}(2, N) \times N}\right) & \text{in (2+1)D}~, \\
        \pm 1 & \text{in (3+1)D}~,
    \end{cases}
\label{eq: phase of T-junction for N and D}
\end{equation}
where $0 \leq k < \mathrm{gcd}(2, N) \times N$ is an integer.
{\change
These values are characterized by the degree-4 cohomology class $H^4(B^2 G, U(1))$ in (2+1)D and the degree-5 cohomology class $H^5(B^3 G, U(1))$ in (3+1)D, associated with the $1$-form and $2$-form $G$ symmetry, respectively (see Table~\ref{tab: particle loop membrane}).
}
In (3+1)D and higher dimensions, the $\pm$ sign corresponds to boson or fermion statistics, which are related to the second Stiefel-Whitney class $w_2$ \cite{Bhardwaj:2016clt, Chen:2017fvr, Chen:2018nog, Ellison2019Disentangling, Chen2020, CET2021, Chen2023Equivalence}.

\subsection{Loop excitations}

Loop excitations can be interpreted as defect lines or flux loops, and we consider loop excitations in (2+1) and (3+1) spacetime dimensions.

\subsubsection{Loops in (2+1)D}
In this dimension, the "fusion" of loops also defines statistics, analogous to the $F$-symbol for particles. The simplest examples of nontrivial statistics arise when the fusion group is $G = \mathbb{Z}_N \times \mathbb{Z}_N$ (with generators labeled by $a$ and $b$), and there are two invariants:

{\change
\begin{equation}
    \vcenter{\hbox{\includegraphics[scale=1.0]{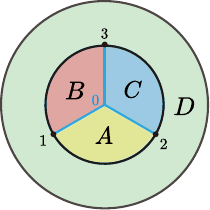}}}
    \quad \quad
    \begin{aligned}
        Z_4^I(a, b) &:= (U(a)_{B+C})^{-N} 
        \Big(U(a)_{B+C} \left[U(a)_{B}, [U(a)_{A},U(b)_{A+B+C+D} ]  \right] \Big)^N, \\
        Z_4^{II}(a, b) &:= (U(b)_{B+C})^{-N} 
        \Big(U(b)_{B+C} \left[U(b)_{B}, [U(b)_{A},U(a)_{A+B+C+D} ]  \right]  \Big)^N~,
    \label{eq: ABCD loop fusion process}
    \end{aligned}
\end{equation}
where $U(i)_R$ denotes the membrane operator that creates loop excitations labeled by $i$ on the boundaries of region $R$ (with $i = a, b$ as generators of the fusion group $G = \mathbb{Z}_N \times \mathbb{Z}_N$).  
The notation $U(i)_{I+J+\cdots}$ refers to the membrane operator acting on the union $I \cup J \cup \cdots$, defined as the product $U(i)_I U(i)_J \cdots$.\footnote{For $N=2$, the statistics can be simplified as $Z_4^I(a, b) = [U(a)_{C}, [U(a)_{B}, [U(a)_{A}, U(b)_{A+B+C+D}]]]$.}
The formulas for $N=2,3,4,5$ are obtained from our algorithm by computer calculation (Sec.~\ref{sec:Computational algorithm SNF}), and we expect them to extend to arbitrary $N$.

Similar to particle fusion, we can show that the nontrivial statistics $Z_4(a)$ and $Z_4(b)$ are obstructions for
\begin{equation}
    U(a)_{f}^N = U(b)_{f}^N = [U(a)_{f}, U(b)_{f} ] = 1~,\quad \forall f,
\end{equation}
indicating an anomalous symmetry. We also highlight that the statistics in Eq.~\eqref{eq: ABCD loop fusion process} correspond to the degree-4 cohomology class $H^4(BG, U(1))$ of the global symmetry $G$, {\change as shown in Table~\ref{tab: particle loop membrane}}.

For a concrete example of nontrivial loop fusion statistics, consider an anomalous $\ZZ_2 \times \ZZ_2$ symmetry on a two-dimensional square lattice.
Each vertex $v$ hosts two qubits $a,b$ with Pauli operators $X^a_v,Z^a_v$ and $X^b_v,Z^b_v$. We label their $Z$-eigenvalues by $a_v,b_v \in \{0,1\}$ via
\begin{equation}
    Z^a_v = (-1)^{a_v},\quad Z^b_v = (-1)^{b_v}.
\end{equation}
The anomalous $\ZZ_2 \times \ZZ_2$ symmetry is generated by
\begin{eqs}
    S^a :=& \prod_v X^a_v \prod_{f=\Box_{1234}} \left( \vcenter{\hbox{\includegraphics[scale=.4]{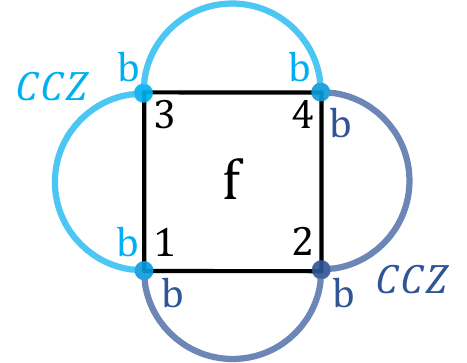}}} \right)
          = \prod_v X^a_v \prod_{f=\Box_{1234}} (-1)^{b_1 b_2 b_4 + b_1 b_3 b_4}, \\
    S^b :=& \prod_v X^b_v,
\label{eq: 2d anomalous Z2 x Z2 symmetry}
\end{eqs}
where $CCZ$ denotes the controlled-controlled-$Z$ gate.
This anomalous (2+1)D system can be derived from the boundary of a (3+1)D SPT phase with cocycle\footnote{{A different boundary theory (``anomalous projective semion states”) of this SPT phase was studied in Ref.~\cite{Wang2017Twistedgauge}.}}  
\begin{equation}
    \frac{1}{2} A_1 \cup B_1 \cup B_1 \cup B_1 \in H^4(B\ZZ_2 \times \ZZ_2, U(1)).
\end{equation}
In Appendix~\ref{sec: 2d anomalous Z2xZ2}, we show that, within the symmetric subspace, the domain walls of this anomalous $\ZZ_2 \times \ZZ_2$ symmetry become loop excitations with nontrivial fusion rules.
}

\begin{figure*}[hbt]
    \centering
    \vspace{-3em}
    \includegraphics[width=0.85\textwidth]{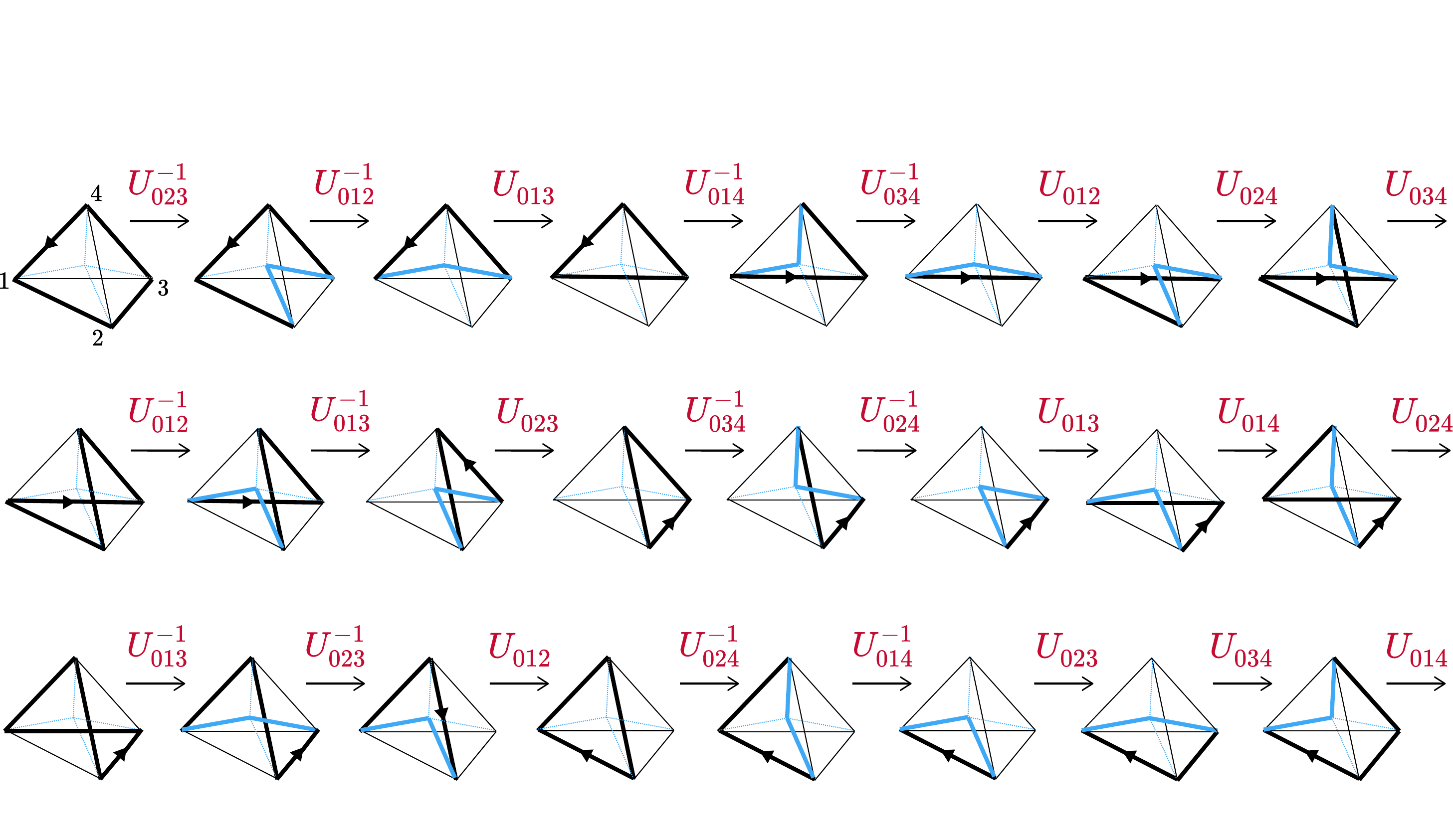}
    \vspace{-2em}
    \caption{The 24-step process for detecting the statistics of loops with $G=\mathbb{Z}_2$ fusion in (3+1)D. For $\mathbb{Z}_2$ loops, different orientations correspond to the same configuration state, indicating that the initial and final configurations are reversed and illustrating the loop-flipping process. This unitary sequence yields the same invariant as the 36-step unitary process proposed in Ref.~\cite{FHH21}. We prove that this 24-step process is optimal, as no shorter sequence can achieve the same invariant.}
    \label{fig: 24 step process}
    \vspace{1em}
    \centering
    \hspace{5ex}
    \includegraphics[width=0.72\textwidth]{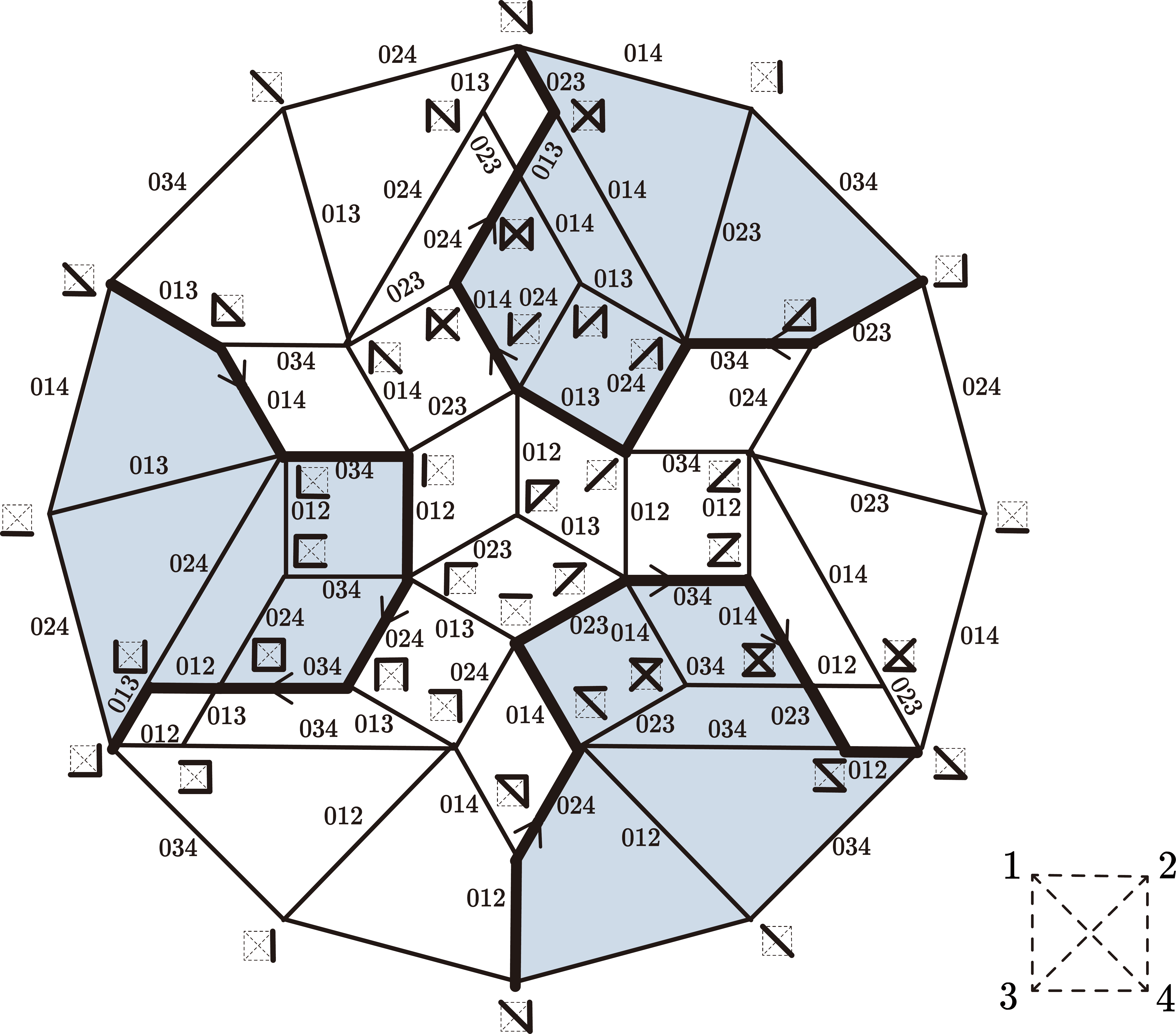}
    \vspace{-0.5em}
    \caption{The space of single-loop configuration states. Each vertex represents a single-loop configuration, where the loop is drawn only on edges between vertices $1$, $2$, $3$, and $4$ in Fig.~\ref{fig: complex (c)} since the edges adjacent to vertex $0$ can be inferred from these. Each edge of this space is labeled by $\lr{0ij}$, and we can apply the operator $U_{0ij}$ or $U^{-1}_{0ij}$ to create or annihilate the loop on edge $ij$. This configuration space forms an $\mathbb{RP}^2$ structure \cite{FHH21}, where antipodal vertices represent the same configuration state. The 24-step process is illustrated by the black directed line, exhibiting the $C_3$ rotational symmetry and corresponding to the nontrivial element of $\pi_1(\mathbb{RP}^2) = \ZZ_2$.
    }
    \label{fig: 24 step process configuration space}
\end{figure*}

\subsubsection{Loops in (3+1)D}

In (3+1)D, the loop statistics is determined by the loop-flipping process for $G = \mathbb{Z}_2$. We introduce a novel 24-step process to detect this statistics: 
\begin{eqs}
    \mu_{24} := ~& U_{014} U_{034} U_{023} U_{014}^{-1} U_{024}^{-1} U_{012} U_{023}^{-1} U_{013}^{-1} \\
    \times & U_{024} U_{014} U_{013} U_{024}^{-1} U_{034}^{-1} U_{023} U_{013}^{-1} U_{012}^{-1} \\
    \times & U_{034} U_{024} U_{012} U_{034}^{-1} U_{014}^{-1} U_{013} U_{012}^{-1} U_{023}^{-1}~,
    \label{eq: 24-step process}
\end{eqs}
which is shown in Fig.~\ref{fig: 24 step process} explicitly. 
Each line in Eq.~\eqref{eq: 24-step process} is obtained from the previous one by applying the substitutions $1 \to 2$, $2 \to 3$, and $3 \to 1$.\footnote{{In the case of loop excitations, the membrane operators $U_{0ij}$ and $U_{0ji}$ represent the same operator.}}
The space of single-loop configurations forms an $\mathbb{RP}^2$ structure, as illustrated in Fig.~\ref{fig: 24 step process configuration space}.
The 24-step process $\mu_{24}$ is represented on this $\mathbb{RP}^2$, explicitly manifesting the $C_3$ rotational symmetry $1 \to 2 \to 3 \to 1$.
We demonstrate that the 24-step process yields the same statistics as the 36-step process defined in Ref.~\cite{FHH21}, while being more efficient. Computational verification confirms that this 24-step sequence is the shortest way to obtain the loop statistics.

{\change
We emphasize that if all $U_f$ are Pauli operators, such that $[U_{f_3}, [U_{f_2}, U_{f_1}]] = 1$ for all faces $f_1$, $f_2$, and $f_3$, then the above statistics simplify to
\begin{equation}
    \mu_{24}^{\mathrm{Pauli}} = [U_{012}, U_{034}]^2 \, [U_{013}, U_{024}]^2 \, [U_{014}, U_{023}]^2.
\end{equation}
}

The statistics $\mu_{24}$ corresponds to a degree-5 cohomology class in 
$H^5(B^2 G, U(1))$ for the $1$-form $G$ symmetry and are also related to the third Stiefel-Whitney class $w_3$ for $G=\ZZ_2$:
\begin{equation}
    \frac{1}{2} w_3 \cup B_2 =\frac{1}{2} B_2 \cup (B_2 \cup_1 B_2) \in H^5(B^2 \ZZ_2, U(1)).
\end{equation}
{\change
To confirm the relation between $\mu_{24}$ and $w_3$, we insert the unitary operators describing the loop excitation on the (3+1)D boundary of the beyond-cohomology (4+1)D topological quantum field theory with action $S=\frac{1}{2} w_2w_3$ in Ref.~\cite{CH21} into the 24-step procedure, and obtain $\mu_{24} = -1$. This demonstrates that the statistic $\mu_{24}$ precisely detects the $w_3$ gravitational anomaly.
}
This process remains valid for the fusion group $G=\mathbb{Z}_{N}$ for even $N$, by choosing the operators $U_{ijk}$ that create a loop labeled by $\frac{N}{2} \in \mathbb{Z}_{N}$, representing an element of order 2.

\subsection{Membrane excitations}

\subsubsection{Membranes in (3+1)D}
{\change
Now, we consider membrane excitations in (3+1) spacetime dimensions.
For $G = \mathbb{Z}_N$, the fusion of membranes gives rise to the $\mathbb{Z}_N$ statistics:\footnote{\change For $N=2$, a simpler expression is $Z_5(g) := [U(g)_{D}, [U(g)_C, [U(g)_B, U(g)_A^2]]]$.}  
\begin{equation}
    \vcenter{\hbox{\includegraphics[scale=0.35]{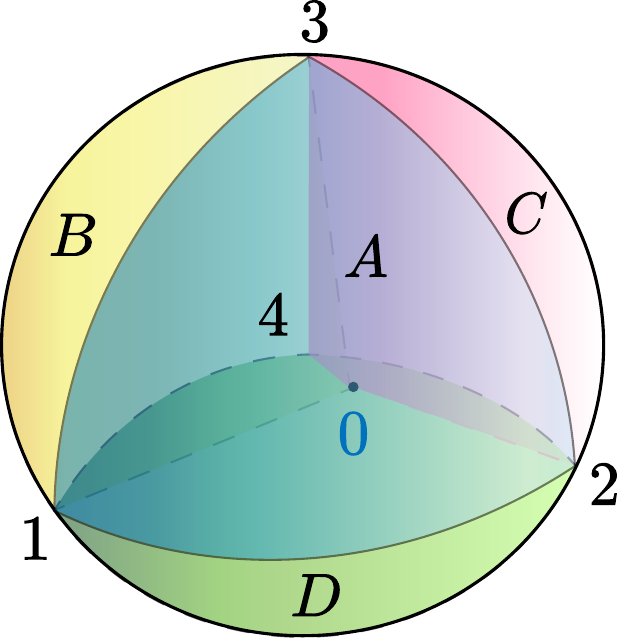}}}
    \quad \quad
    \begin{aligned}
        Z_5(g) := ( U(g)_{C+D} )^{-N}
            \left( U(g)_{C+D} [U(g)_C, [U(g)_B,U(g)_A^N ]] \right)^N~,
    \label{eq: ABCD membrane fusion process}
    \end{aligned}
\end{equation}
where $U(g)_R$ denotes the volume operator on region $R$ that creates a $g$-membrane excitation on its boundary, and $U(g)_{I+J+\cdots}$ denotes the product $U(g)_I U(g)_J \cdots$ acting on the union $I \cup J \cup \cdots$.
The formulas for $N \leq 8$ are obtained explicitly from our algorithm in Sec.~\ref{sec:Computational algorithm SNF}, and we expect that the expressions can be extended to arbitrary $N$.
Similar to the (2+1)D fusion in Eq.~\eqref{eq: ABCD loop fusion process}, the nontrivial statistics $Z_5(g)$ represents the obstruction to $U(g)_{t}^N = 1$ for all tetrahedra $t$.
The statistics $Z_5(g)$ corresponds to the degree-5 cohomology $H^5(BG, U(1))$ in Table~\ref{tab: particle loop membrane}.

To illustrate a nontrivial $Z_5(g)$, consider a cubic lattice with one qubit at each vertex.  The anomalous $\mathbb{Z}_2$ global symmetry is defined by
\vspace{-0.5em}
\begin{eqs}
    S :=& \prod_v X_v \prod_{c} \left( \vcenter{\hbox{\includegraphics[scale=.45]{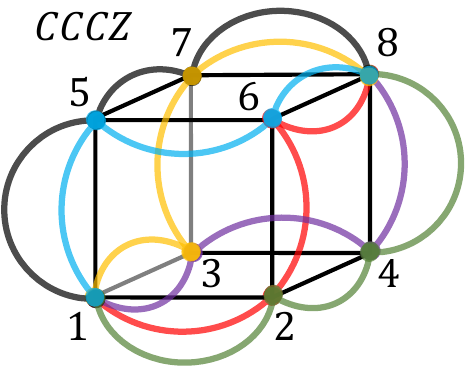}}} \right) \\
    =& \prod_v X_v
    \prod_{c = \vcenter{\hbox{\includegraphics[scale=.14]{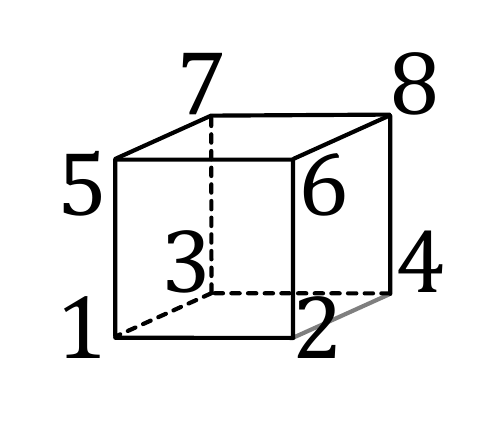}}}}
    (-1)^{a_1 a_2 a_4 a_8 + a_1 a_2 a_6 a_8 + a_1 a_3 a_4 a_8 + a_1 a_3 a_7 a_8 + a_1 a_5 a_6 a_8 + a_1 a_5 a_7 a_8},\\[-1.5em]
    \label{eq: 3d anomalous Z2 symmetry}
\end{eqs}%
where $CCCZ$ denotes the controlled–controlled–controlled–$Z$ gate. The anomalous (3+1)D theory resides on the boundary of the (4+1)D SPT phase with the cocycle
\begin{equation}
    \frac{1}{2} A_1 \cup A_1 \cup A_1 \cup A_1 \cup A_1 \in H^5(B\ZZ_2, U(1)),
\end{equation}
known as the generalized double semion model in Refs.~\cite{Freedman2016DoubleSemions, Fidkowski2020}.
In Appendix~\ref{sec: 3d anomalous Z2}, we show that, within the symmetric subspace, the domain wall excitation forms a closed membrane exhibiting nontrivial membrane fusion statistics $Z_5(g)$.
}

{\change

\subsection{Mutual statistics}

\subsubsection{Particle-loop statistics in (3+1)D}

In addition to examples in which all excitations share the same dimension, one can examine the mutual statistics between excitations of different dimensionalities. The simplest instance in (3+1)D gauge theories is the braiding of a point charge around a flux loop: as the charge encircles the loop, the wavefunction acquires a nontrivial phase.

On the lattice, this process is described by the commutator between the string operator that moves the particle and the membrane operator that creates the loop:
\vspace{-0.5em}
\begin{equation}
    \vspace{-0.5em}
    \vcenter{\hbox{\includegraphics[scale=.7]{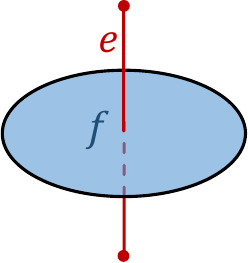}}} \quad Z_5^{\mathrm{particle-loop}} := [U_{f}, U_{e}],
\end{equation}
where $U_f$ is the membrane operator acting on a face $f$ that generates the loop excitation, $U_e$ is the string operator acting on an edge $e$ that generates the particle excitation, and the edge $e$ pierces the face $f$. This defines a $\mathbb{Z}_N$ mutual statistics, satisfying $\left(Z_5^{\mathrm{particle\text{-}loop}}\right)^N = 1$.
This braiding statistics corresponds to the 5-cocycle
\begin{equation}
    \frac{1}{N} A_2 \cup B_3 \in H^5(B^2\ZZ_N \times B^3 \ZZ_N, U(1)).
\end{equation}

\subsubsection{Loop-membrane statistics in (3+1)D}
\label{subsec: Loop membrane statistics in (3+1)D}
We now present two distinct types of mutual $\mathbb{Z}_2$ statistics between a $\mathbb{Z}_2$ loop and a $\mathbb{Z}_2$ membrane in (3+1)D. These cases are anticipated from the cohomology classification
\begin{equation}
    H^5(B \ZZ_2 \times B^2 \ZZ_2, U(1)) = \ZZ_2 \times \ZZ_2 \times \ZZ_2 \times \ZZ_2,
\end{equation}
where two $\mathbb{Z}_2$ factors correspond to the self-statistics of particles and loops, respectively, and the remaining two correspond to mutual statistics.
A $\mathbb{Z}_2$ loop along the boundary of a face $f$ is generated by the membrane operator $U_f$, which realizes the 1-form $\mathbb{Z}_2$ symmetry.
A $\mathbb{Z}_2$ membrane on the boundary surface of a tetrahedron $t$ is generated by the volume operator $U_t$, which realizes the 0-form $\mathbb{Z}_2$ symmetry.

\begin{itemize}
    \item \textbf{$\tfrac12 A_1 \cup B_2 \cup B_2$ loop-membrane statistics:}

    We first consider the following setting with three faces and one tetrahedron:
    \vspace{-0.5em}
    \begin{equation}
        \vspace{-0.5em}
        \vcenter{\hbox{\includegraphics[scale=.75]{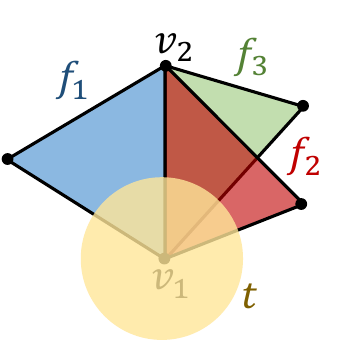}}} \quad Z_5^{\mathrm{loop\text{-}membrane\text{-}I}}
        := [U_t ,~[ U_{f_3}^{-1} U_{f_2} U_{f_1}^{-1} U_{f_3} U_{f_2}^{-1} U_{f_1}]],
        \label{eq:loop-membrane_statistics_1}
    \end{equation}
    where $f_1$, $f_2$, and $f_3$ are faces adjacent to the edge $\langle v_1 v_2 \rangle$, and $t$ is a tetrahedron (more precisely, a 3-cell) whose bulk contains the vertex $v_1$. Eq.~\eqref{eq:loop-membrane_statistics_1} can be viewed as a T-junction process of loop excitations on faces $f_1$, $f_2$, and $f_3$, with the commutator with $U_t$ isolating the contribution from the surface of tetrahedron $t$, thereby capturing the statistics of the loop–membrane intersection.
    
    This statistics is relevant for describing the boundary theory of the (4+1)D beyond-cohomology $\mathbb{Z}_2$ SPT phase with the topological action $S=\frac{1}{2} A_1 \cup w_2 \cup w_2$~\cite{Fidkowski2020beyond,CH21}, where $w_2$ denotes the second Stiefel–Whitney class.
    Ref.~\cite{CH21} shows that this beyond-cohomology SPT phase can be obtained by gauging a higher-form SPT phase with the cocycle containing
    \begin{equation}
        \frac{1}{2} w_2 \cup A_1 \cup B_2 = \frac{1}{2} A_1 \cup B_2 \cup B_2 \in H^5(B \ZZ_2 \times B^2 \ZZ_2, U(1)).
    \end{equation}
    It has been predicted that the intersection of a membrane (described by the $A_1$ field) and a loop (described by the $B_2$ field) has the fermionic particle statistics due to the presence of $w_2$.
    However, an explicit derivation of this phenomenon on a finite lattice has been lacking.
    Our novel statistics in Eq.~\eqref{eq:loop-membrane_statistics_1} resolves this gap by providing a concrete lattice definition.
    In Appendix~\ref{sec: 3d anomalous Z2^(0)xZ2^(1) 1}, we show that the (3+1)D boundary of the (4+1)D beyond-cohomology $\mathbb{Z}_2$ SPT phase exhibits the nontrivial statistics $Z_5^{\mathrm{loop\text{-}membrane\text{-}I}} = -1$.  
    More precisely, consider the anomalous 0-form $\mathbb{Z}_2$ symmetry $S^a$ and the 1-form $\mathbb{Z}_2$ symmetry $S^b_v$, where a qubit $a_v$ is placed on each vertex $v$ and a qubit $b_e$ on each edge $e$:  
    \begin{eqs}
        S^a :=& \prod_v X^a_v
        \prod_{\vcenter{\hbox{\includegraphics[scale=.14]{cube_12345678.pdf}}}}
        (-1)^{b_{12}(b_{24}+b_{48}+b_{26}+b_{68}) + b_{13}(b_{34}+b_{48}+b_{37}+b_{78}) + b_{15}(b_{56}+b_{68}+b_{57}+b_{78})}, \\[-2em]
        & \text{diagrammatically:}
        ~~\prod_v X^a_v \prod_{c} \left( \vcenter{\hbox{\includegraphics[scale=.45]{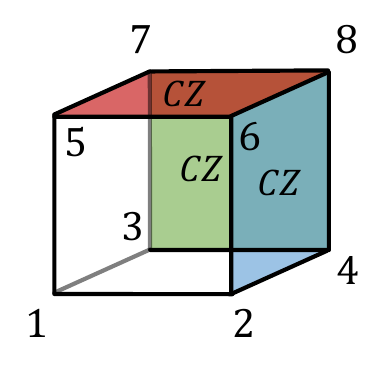}}} \right), \\
        &\text{with}
        ~~\vcenter{\hbox{\includegraphics[scale=.45]{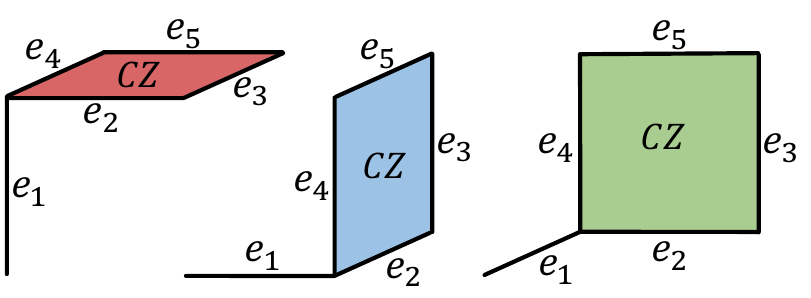}}}
        := (-1)^{b_{e_1}(b_{e_2}+b_{e_3}+b_{e_4}+b_{e_5}) }, \\
        S^b_v:=& \prod_{e \supset v} X^b_e,
        \quad \text{diagrammatically:}
        ~~ \vcenter{\hbox{\includegraphics[scale=.5]{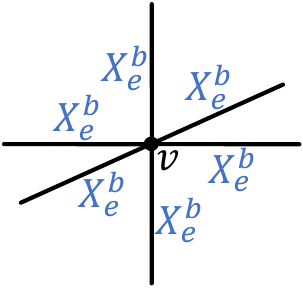}}}~.
        \label{eq: 3d anomalous loop membrane symmetry I}
    \end{eqs}
    We demonstrate that the domain wall excitations of these symmetries exhibit the nontrivial mutual loop-membrane statistics given in Eq.~\eqref{eq:loop-membrane_statistics_1}.

    \item \textbf{$\tfrac12A_1 \cup A_1 \cup A_1 \cup B_2$ loop-membrane statistics:}
    
    We next consider a different structure involving one face and two tetrahedra:
    \vspace{-0.5em}
    \begin{equation}
        \vspace{-0.5em}
        \vcenter{\hbox{\includegraphics[scale=.75]{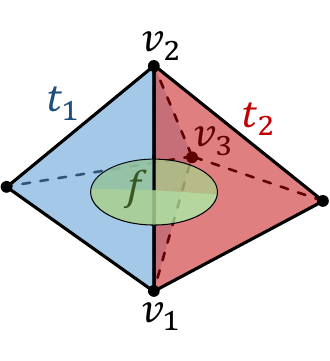}}}
        \quad \quad
        Z_5^{\mathrm{loop\text{-}membrane\text{-}II}}
        := [U_f ,[ U_{t_2}, U_{t_1}^2]],
        \label{eq:loop-membrane_statistics_2}
    \end{equation}
    where $t_1$ and $t_2$ are tetrahedra adjacent to the face $\langle v_1 v_2 v_3 \rangle$, and the edge $\langle v_1 v_2 \rangle$ pierces the face $f$.
    In Appendix~\ref{sec: 3d anomalous Z2^(0)xZ2^(1) 2}, we show that, for the anomalous 0-form $\mathbb{Z}_2$ symmetry $S^a$ and 1-form $\mathbb{Z}_2$ symmetry $S^b_v$ defined below, the corresponding domain wall excitations exhibit the nontrivial statistics $Z_5^{\mathrm{loop\text{-}membrane\text{-}II}} = -1$:  
    \begin{eqs}
        S^a :=& \prod_v X^a_v
        \prod_{\vcenter{\hbox{\includegraphics[scale=.14]{cube_12345678.pdf}}}}
        (-1)^{a_1 a_2 (b_{24}+b_{48}+b_{26}+b_{68}) + a_1 a_3 (b_{34}+b_{48}+b_{37}+b_{78}) + a_1 a_5 (b_{56}+b_{68}+b_{57}+b_{78})}~, \\[-2em]
        S^b_v:=& \prod_{e \supset v} X^b_e,
        \quad \text{diagrammatically:}
        ~~ \vcenter{\hbox{\includegraphics[scale=.5]{anomalous_loop_membrane_1b.pdf}}}~.
        \label{eq: 3d anomalous loop membrane symmetry II}
    \end{eqs}  

    These anomalous symmetries correspond to the SPT phase associated with the degree-5 cohomology class   
    \begin{equation}
        \frac{1}{2} A_1 \cup A_1 \cup A_1 \cup B_2 \in H^5(B\ZZ_2 \times B^2\ZZ_2, U(1)).
    \end{equation}
    This cocycle provides an intuitive way to interpret Eq.~\eqref{eq:loop-membrane_statistics_2}.  
    First, the term $A_1 \cup A_1 \cup A_1$ represents the 3-cocycle associated with the fusion statistics in one dimension (Eq.~\eqref{eq: (1+1)D particle fusion statistics}), motivating a structure of the form $[U_{t_2}, U_{t_1}^2]$.  
    Taking the cup product with $B_2$ corresponds to intersecting with the worldsheet of the loop excitation, which motivates the introduction of the additional commutator with $U_f$ to capture this intersection.

\end{itemize}

\subsubsection{Particle-membrane statistics in (3+1)D}

Finally, we consider the mutual statistics between a $\mathbb{Z}_2$ particle and a $\mathbb{Z}_2$ membrane in three spatial dimensions.
Let $U_e$ denote the string operator acting on an edge $e$ that creates particles at its endpoints, and let $U_t$ denote the volume operator acting on a tetrahedron $t$ that generates a membrane excitation on its boundary surface.
The mutual statistics is defined in the following structure:
\vspace{-0em}
\begin{equation}
    \vspace{-0em}
    \vcenter{\hbox{\includegraphics[scale=.75]{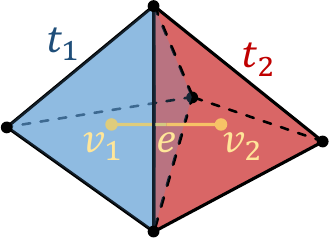}}}
    \quad \quad
    \begin{aligned}
    Z_5^{\mathrm{particle\text{-}membrane}}:=\; &[U_e^2,U_{t_1}] \,[U_{t_2},[U_{t_1},U_e]] \\
    =\; &\Big(U_{t_2}^{-1} [U_e,U_{t_1}]\, U_{t_2}\Big)\;
          \Big(U_e^{-1} [U_e,U_{t_1}]\, U_e\Big),
    \label{eq:particle_membrane_statistics}
    \end{aligned}
\end{equation}
where $t_1$ and $t_2$ are tetrahedra adjacent to opposite sides of a common face, $v_1$ is the vertex contained in $t_1$, $v_2$ is the vertex contained in $t_2$, and the edge $e = \langle v_1 v_2 \rangle$ pierces the common face.
We present two equivalent expressions for the statistics, both derived from our algorithm; either form can be used for computation, depending on convenience.

Consider the anomalous 0-form $\mathbb{Z}_2$ symmetry $S^a$ and 2-form $\mathbb{Z}_2$ symmetry $S^c_e$ defined below, with a qubit $a_v$ is placed on each vertex $v$ and a qubit $c_f$ on each face $f$: 
\begin{eqs}
    S^a :=& \prod_v X^a_v \prod_{\vcenter{\hbox{\includegraphics[scale=.14]{cube_12345678.pdf}}}}
    (-1)^{a_1 (c_{1234} + c_{1256} + c_{1357} + c_{2468} + c_{3478} + c_{5678}) }~, \\[-3em]
    S^c_e:=& \prod_{f \supset e} X^c_f,
    \quad \text{diagrammatically:}
    ~~\vcenter{\hbox{\includegraphics[scale=.45]{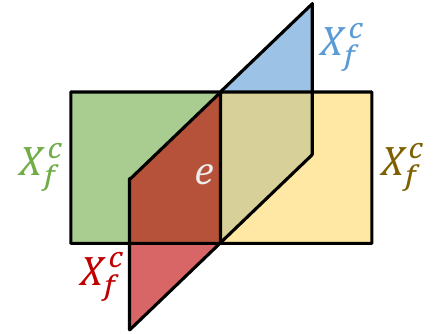}}}~,
    ~~\vcenter{\hbox{\includegraphics[scale=.45]{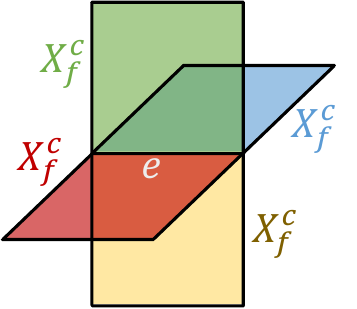}}}~,
    ~~\vcenter{\hbox{\includegraphics[scale=.45]{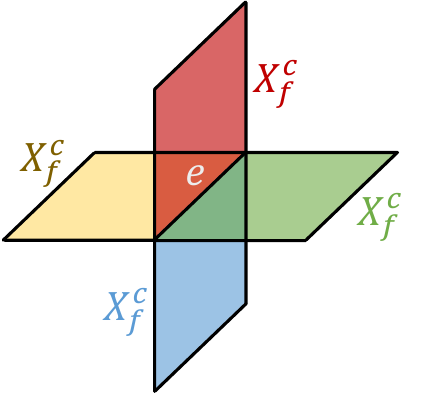}}}~.
    \label{eq: 3d anomalous particle membrane symmetry}
\end{eqs}
In Appendix~\ref{sec: 3d anomalous Z2^(0)xZ2^(2)}, we show that the domain wall excitations associated with these symmetries exhibit the nontrivial statistics $Z_5^{\mathrm{particle\text{-}membrane}} = -1$.
These symmetries arise from the (3+1)D boundary of a (4+1)D SPT phase characterized by the degree-5 cohomology class
\begin{equation}
    \frac{1}{2} A_1 \cup A_1 \cup C_3 \in H^5(B\ZZ_2 \times B^3\ZZ_2, U(1)),
\end{equation}
where $A_1$ and $C_3$ are the background gauge fields for the 0-form $\mathbb{Z}_2$ and 2-form $\mathbb{Z}_2$ symmetries, respectively.

}
\end{widetext}

\section{Framework for Generalized Statistics}\label{sec:Framework}

\subsection{Framework of invariants in generic dimensions}
\label{sec:Framework of invariants in generic dimensions}

We describe the invariants associated with extended excitations in generic dimensions.
The T-junction process, previously discussed for particle excitations, is generalized to extended excitations in any dimension through a sequence of unitaries that create these excitations.
Consider a system with a finite, invertible $(d-p-1)$-form symmetry group $G$ in a $d$-dimensional space.
The group $G$ is generated by unitary operators $U$ supported on $(p+1)$-dimensional submanifolds.
If the support has a boundary, the operator $U$ creates a $p$-dimensional extended excitation at that boundary.
For a 0-form symmetry with $p = d - 1$, $G$ can be non-Abelian; in all other cases, $G$ must be Abelian.
We first restrict attention to the Abelian fusion group $G$, and address the non-Abelian case in Sec.~\ref{subsec:nonabelian}.

\subsubsection{Excitations and unitaries on simplicial complex}
\label{subsec:complex}

Following the description of the T-junction process, the construction of the invariant starts by fixing the possible configurations of the excitations in space.  
We define the \textbf{excitation model} as follows:  
\begin{definition}
    \textnormal{An excitation model defined on a tensor product Hilbert space $\mathcal{H}$ over a $d$-dimensional spatial manifold $M$ consists of the following components:
    \begin{enumerate}    
    \item A finite Abelian group $\mathcal{A}$. Each group element $a\in \mathcal{A}$ corresponds to a configuration of the excitations in the space $M$.
    For each configuration $a \in\mathcal{A}$, there exists a state $|a\rangle \in \mathcal{H}$, such that the states that correspond to different configurations are orthogonal.\footnote{For simplicity, we typically assume that $M$ is a sphere, ensuring that the ground state has no topological degeneracy. This assumption is valid since the generalized statistics is a local property and is insensitive to the global topology.}
    \item {\change A finite set $\mathcal{S}$ and a map $\partial: \mathcal{S} \to \mathcal{A}$, such that $\text{Im}\partial$ \change{generates} $\mathcal{A}$. Each element $s \in \mathcal{S}$ represents a unitary operator that creates the excitation $\partial s$. For each $s \in \mathcal{S}$, there exists a unitary operator $U(s)$ such that $U(s) |a\rangle \propto |a + \partial s\rangle$ for all $a \in \mathcal{A}$.
    The support of the unitary $U(s)$ is the $(p+1)$-dimensional locus $\operatorname{supp}(s) \subset M$.}
    \end{enumerate}
    }
\end{definition}

We note that the above definition assumes the invertible symmetry with group-like fusion rule, and further that the fusion group is finite and Abelian.
One can extend the above definition to the case of invertible non-Abelian group symmetries, which will be studied in Sec.~\ref{subsec:nonabelian}. 

Let us take a concrete example of the excitation model based on the T-junction reviewed in Sec.~\ref{sec:Review_of_particle}. In that case, $\{ U(s) \mid s \in \mathcal{S} \}$ is a set of operators that generate all string operators with the fusion group $G$ supported on the fixed edges connecting the vertices $0,1,2,3$ of Fig.~\ref{fig: complex (b)}. There are six edges $\langle jk \rangle$ with $0 \le j < k \le 3$, so $\mathcal{S}$ is a set of generators of $G^6$. $\mathcal{A}$ is taken to be all configurations of anyons created by sequences of operators in $\{U(s)\}$ acting on the vacuum. This is isomorphic to $\mathcal{A} = G^3$, which corresponds to the configurations of anyons at four vertices fusing into the vacuum.
In the case of the T-junction, $\mathcal{S}$ can be regarded as a set of generators of the group of 1-chains $C_1(X,G)$ in the simplicial complex $X$ shown in Fig.~\ref{fig: complex (b)}, $\partial$ as the boundary map of the simplicial complex, and $\mathcal{A}$ as the boundary group $\mathcal{A} = B_0(X,G)$.  

More generally, one can construct an excitation model on any finite simplicial complex $X$ embedded in the space $M$, which can describe extended excitations in arbitrary dimensions. To illustrate these ideas, consider the excitation model on a simplicial complex. In this case, let $\mA$ be the group of $p$-dimensional simplicial boundaries  
\begin{equation}
    \mA = B_{p}(X,G)~.
\end{equation}
We take $\mS$ to be a minimal set of generators of the $(p+1)$-chains of $X$ with coefficients in $G$, i.e., generators of $C_{p+1}(X,G)$. Concretely, each element $s \in \mathcal{S}$ is given by $s = g\sigma_{p+1}$, where $g \in G$ is one of the generators of $G$, and $\sigma_{p+1}$ is a single $(p+1)$-simplex of $X$. Viewing $X$ as a topological space embedded in $M$, $\operatorname{supp}(s)$ gives the image of the simplex $\sigma_{p+1}$ in $M$ under the embedding map. The unitary $U(s)$ can then be expressed as $U(s) = U_g(\sigma_{p+1})$, namely the operator generating the $g \in G$ symmetry at the simplex $\sigma_{p+1}$ embedded in $M$. The map $\partial: \mathcal{S} \to \mathcal{A}$ is the homological boundary map of $X$, and we have $U(s)\ket{a} \propto \ket{a+\partial s}$ with $a \in \mathcal{A}, s \in \mathcal{S}$.  

In general, different choices of the simplicial complex $X$ yield different invariants. To extract the complete set of invariants observable in the spatial manifold $M$, one can choose $X$ as a simplicial decomposition of $M$. Meanwhile, the generalized statistics of excitations are typically local properties that can be extracted from a simplicial complex $X$ supported on a ball embedded in the whole space $M$. Such invariants include, for example, the T-junction of anyons, and are insensitive to the global topology of the space.  

\subsubsection{Invariant Berry phases from unitary sequences}

Now we are ready to construct the invariants out of the sequence of unitaries $\{U(s)\}$ acting on the excited states $\{\ket{a}\}$. From now we assume that the excitation model is constructed on the simplicial complex $X$ embedded in the space, according to Sec.~\ref{subsec:complex}.

For $a\in\mathcal{A}, s\in\mathcal{S}$, the unitary operator $U(s)$ transforms the state $\ket{a}$ into $\ket{a+\partial s}$ up to a phase:
\begin{align}
    U(s) \ket{a} = \exp[i\theta(s,a)]\ket{a+\partial s}~.
\label{eq: U(s) theta}
\end{align}
Accordingly, the inverse of the unitary $U(s)$ acts on the states by
\begin{align}
    U(s)^{-1} \ket{a+\partial s} = \exp[-i\theta(s,a)]\ket{a}~.
\label{eq: U(s)-1 theta}
\end{align}
The invariant is then expressed as a sequence of unitary operators starting and terminating with the same configuration of excitations
\begin{align}
     \bra{a_0}U(s_{n-1})^{\pm} \dots U(s_j)^{\pm}\dots U(s_0)^{\pm}\ket{a_0}~,
     \label{eq:sequenceU}
\end{align}
where $\pm$ is the sign which can be chosen for each unitary. This Berry phase will become the sum over the phases $\theta(s,a)$ with $s\in\mathcal{S},a\in\mathcal{A}$.

To qualify the above phase as an invariant, one needs to establish the invariance of the above quantity against possible deformations of the states as well as unitaries. 
For the formulation of the invariants, it is convenient to express the above Berry phase as the element of the formal sum of the objects $\theta(s,a)$:
\begin{align}
    E=\bigoplus_{s\in \mathcal{S},a\in \mathcal{A}} \mathbb{Z}\theta(s,a)~,
\end{align}
which we call the \textbf{expression group} associated with the excitation model. 
Each element $e\in E$ is expressed as $e=\bigoplus_{(s,a)}\epsilon(s,a)\theta(s,a)$ with the integer coefficients $\epsilon(s,a)\in\ZZ$. The condition for $e$ being the invariant will be compiled into a set of equations that the coefficients $\{\epsilon(s,a)|s\in\mathcal{S},a\in\mathcal{A}\}$ need to satisfy. 
The invariants then correspond to a specific subgroup $E_{\mathrm{inv}}\subset E$ which we now characterize.

The first condition for $E_{\mathrm{inv}}$ simply requires that the Berry phase corresponds to a sequence of unitaries initiating and terminating with the same state. This is equivalent to requiring the invariance under the redefinition of the states $\ket{a}\to e^{i\phi(a)}\ket{a}$ for $a\in\mathcal{A}$, which shifts the phases as $\theta(s,a) \to \theta(s,a) - \phi(a) + \phi(a+\partial s)$.
The element $e\in E$ is invariant under such redefinition of phases if and only if
the coefficients $\{\epsilon(s,a)\}$ satisfy
\begin{align}
    \sum_{s\in\mathcal{S}}\epsilon(s,a) - \sum_{s\in\mathcal{S}} \epsilon(s,a-\partial s) = 0, \quad \text{for any $a\in\mathcal{A}$}~.
    \label{eq:inv1}
\end{align}
This gives the first necessary condition for $e\in E_{\mathrm{inv}}$.

The second condition for $E_{\mathrm{inv}}$ is that the Berry phase is invariant under the redefinition of unitaries $U(s)$ by a phase $U(s)\to e^{i\phi(s)}U(s)$ for $s\in\mathcal{S}$, shifting the phases as $\theta(s,a)\to \theta(s,a) + \phi(s)$. The element $e\in E$ is invariant under such redefinition if and only if the coefficients $\{\epsilon(s,a)\}$ satisfy
\begin{align}
    \sum_{a\in\mathcal{A}} \epsilon(s,a) = 0, \quad\text{for any $s\in\mathcal{S}$}~.
    \label{eq:inv2}
\end{align}
This gives the second necessary condition for $e\in E_{\mathrm{inv}}$.

The rest of the conditions for $E_{\mathrm{inv}}$ is that the Berry phase is invariant under the deformations of the unitary $U(s)$ by a local operator near the boundary of the support of $U(s)$. Suppose that we add a local deformation at a point contained in the $j$-simplex $\sigma_j\in X$ embedded in the space $M$. This has the effect of locally modifying the symmetry operator $U(s)$ to the other unitary $U'(s)$, when the support $(p+1)$-simplex $\sigma_{p+1}=\text{supp}(s)$ contains the simplex $\sigma_j$. Such an inclusion of a simplex is denoted by $\sigma_j\subset \sigma_{p+1}$, meaning that a set of vertices of $\sigma_j$ is a subset of that of $\sigma_{p+1}$. 

{\change
The phase $\theta(s,a)$ gets shifted according to the redefinition of the unitary. Since $U'(s)$ differs from $U(s)$ locally, the operator $U'(s)^\dagger U(s)$ is a local operator  supported on a local region $r$ within $\sigma_j$. The excited states $\ket{a},\ket{a+\partial s}$ possibly get modified by $\ket{a}'=V\ket{a},\ket{a+\partial s}'=\tilde{V}\ket{a+\partial s}$ with some local unitary $V,\tilde V$ with the same support $r$ as $U'(s)^\dagger U(s)$. $\ket{a}$ is then an eigenstate of the operator $O(r):= V^\dagger U'(s)^\dagger\tilde VU(s)$. Let us denote the eigenvalue as the phase $\exp(i\phi(s,a))$,
\begin{align}
    e^{i\phi(s,a)}\ket{a} &= O(r)\ket{a}~.
\end{align}
Due to the locality of the perturbation, $V,\tilde V$ depend on the excitation $a$ only through its restriction to the set of $p$-simplices satisfying $\sigma_j\subset \sigma_p$. 
Furthermore, the eigenvalue $\exp(i\phi(s,a))$ of the local operator $O(r)=V^\dagger U'(s)^\dagger\tilde V U(s)$ also depends on $a$ only through the $p$-simplices with $\sigma_j\subset \sigma_p$. 
To see this, we express $\exp(i\phi(s,a))$ as
\begin{align}
\begin{split}
    e^{i\phi(s,a)} &= \bra{a} O(r)\ket{a} \\
    &= \bra{0}U^\dagger(s_a) O(r) U(s_a)\ket{0}~,
    \end{split}
\end{align}
where $s_a\in\mathcal{S}$ satisfies $a=\partial s_a$, and $0\in\mathcal{A}$ denotes the trivial element of $\mathcal{A}$.
Since $U(s_a)$ is a finite-depth circuit, the part of the circuit away from the region $r$ commutes with $O(r)$. 
This implies that $U^\dagger(s_a) O(r) U(s_a)$ is again a local operator supported on $r$, and is independent of the configuration of $a$ away from the perturbation at $r$, which lies within a simplex $\sigma_j$.
}

Now let us define $a|_{\sigma_j}$ as a $G$-valued $p$-cochain defined by $a|_{\sigma_j}=a$ at $\sigma_p$ satisfying $\sigma_j\subset\sigma_p$, otherwise zero. Due to the above argument, the action of $O(r)=V^\dagger U'(s)^\dagger \tilde VU(s)$ can be expressed as $V^\dagger U'(s)^\dagger\tilde VU(s)\ket{a}=\exp(i\phi(s,a|_{\sigma_j}))\ket{a}$, namely the dependence on $a$ is only through the configuration of $a$ nearby the simplex $\sigma_j$ where the deformation occurs. 

This implies that the action of $U'(s)$ is shifted from $U(s)$ by
    $\theta'(s,a) = \theta(s,a) + \phi(s,a|_{\sigma_j})$ when $\sigma_j\subset \text{supp}(s)$.
Requiring the invariance of $e\in E$ under this shift of $\theta$ is equivalent to requiring that the coefficients $\{\epsilon(s,a)\}$ satisfy
\begin{align}
    \sum_{\substack{a\in\mathcal{A} \\ a|_{\sigma_j}= a^{(j)}_*}} \epsilon(s, a) = 0~,
    \label{eq:inv3}
\end{align}
for any possible choices of $s\in\mathcal{S}, \sigma_j\in X$ satisfying $\sigma_j\subset \text{supp}(s)$ and any $a^{(j)}_*$, with $0\le j \le p$. 

One can further see that Eq.~\eqref{eq:inv3} labeled by $0\le j\le p$ gives a redundant set of constraints. To see this, pick a $j$-simplex $\sigma_j$ and a 0-simplex (vertex) $\sigma_0$ of $\sigma_j$, i.e., $\sigma_0\subset \sigma_j$. The chain $a^{(0)}_*$ then satisfies $(a^{(0)}_*)|_{\sigma_j} =a^{(j)}_*$. We get
\begin{align}
    \sum_{\substack{a\in\mathcal{A} \\ a|_{\sigma_j}= a^{(j)}_*}} \epsilon(s, a) = \sum_{\substack{a^{(0)}_{*} \\ (a^{(0)}_*)|_{\sigma_j} =a^{(j)}_*}}\left(\sum_{\substack{a\in\mathcal{A} \\ a|_{\sigma_0}= a^{(0)}_*}} \epsilon(s, a)\right)~,
\end{align}
which implies that the equations for $j=0$ are sufficient to fully characterize the constraints in Eq.~\eqref{eq:inv3}. Thus, a refined version of the condition equivalent to Eq.~\eqref{eq:inv3} can be written as follows:
\begin{align}
    \sum_{\substack{a\in\mathcal{A} \\ a|_{\sigma_0}= a^{(0)}_*}} \epsilon(s, a) = 0~,
    \label{eq:inv3'}
\end{align}
for any possible choices of $s\in\mathcal{S}$, a 0-simplex $\sigma_0\in X$ satisfying $\sigma_0\subset \text{supp}(s)$ and any $a^{(0)}_*$.

The above requirements complete the description of $E_{\mathrm{inv}}$; the group $E_{\mathrm{inv}}\subset E$ is defined as the group of $e\in E$ whose integral coefficients satisfy the conditions \eqref{eq:inv1}, \eqref{eq:inv2}, \eqref{eq:inv3'}. 

\subsubsection{Equivalence of Berry phases and genuine invariants}
\label{sec: genuine invariants}

Some elements of $E_{\mathrm{inv}}$ correspond to trivial invariants.
One class of trivial invariants arises from the commutators of the unitaries. For instance, if two unitaries $U(s_1), U(s_2)$ have no overlap of their support $\text{supp}(s_{1}) \cap \text{supp}(s_{2}) = \varnothing$, the commutator~\eqref{eq: Us1 Us2 commutator} $[U(s_2),U(s_1)]=1$ is trivial. This implies that the element of $E_{\mathrm{inv}}$ given by
\begin{eqs}
    &\theta(s_1,a)+\theta(s_2,a+\partial s_1)-\theta(s_1,a+\partial s_2)-\theta(s_2,a) \\
    &= 0 \pmod{2 \pi}~,
\label{eq: theta of commutator}
\end{eqs}
gives a trivial invariant. Hence, the set $\{ \theta(s,a) \}$ is not entirely independent.
More generally, let us take multiple unitaries $U(s_1),\dots,U(s_n)$ without common overlaps $\text{supp}(s_1)\cap \dots \cap \text{supp}(s_n)=\varnothing$.
Noting that each unitary $U(s_j)$ is a local finite-depth circuit, the higher commutator~\eqref{eq: Us1 Us2 ... Usn commutator}, $[U(s_n),[\cdots,[U(s_2),U(s_1)]]]$, becomes trivial.
Taking the expectation value of these higher commutators with any state $\ket{a}$ gives an element of $E_{\mathrm{inv}}$ that is a trivial invariant:\footnote{One can extend this Eq.~\eqref{eq: locality identity of n simplexes} to more generic setup:
\begin{eqs}
\begin{split}
    &\bra{a}[U(s_n),[\cdots,[U(s_2),U(s_1)]]] \ket{a} \\ =&\bra{b}[U(s_n),[\cdots,[U(s_2),U(s_1)]]]\ket{b}, 
    \end{split}
\end{eqs}
for all states $\ket{a},\ket{b}$ that are identical at the mutual support of unitaries
$\text{supp}(s_1)\cap \dots \cap \text{supp}(s_n)$. 
Note that this is equivalent to Eq.~\eqref{eq: locality identity of n simplexes} when $X \setminus \{\text{supp}(s_1)\cap \dots \cap \text{supp}(s_n) \}$ is connected, allowing us to find an additional operator that connects states $\ket{a}$ and $\ket{b}$ without overlapping with the mutual support of unitaries.
Therefore, when $X$ is a triangulation of a closed manifold, such as a sphere $S^d$, it is sufficient to consider Eq.~\eqref{eq: locality identity of n simplexes}. This paper will focus on this case.
}
\begin{eqs}
    & \bra{a}[U(s_n),[\cdots,[U(s_2),U(s_1)]]]\ket{a}=1, \\
    &\forall a\in\mathcal{A}, \forall s_1, s_2, \cdots, s_n \in \mathcal{S} | s_1 \cap s_2 \cap \dots \cap s_n = \varnothing~.
\label{eq: locality identity of n simplexes}
\end{eqs}
This equation follows from the fact that the commutator of two local finite-depth circuits are supported in the neighborhood of their intersection.

We define the subgroup $E_{\mathrm{id}} \subset E_{\mathrm{inv}}$ as the group generated by the locality identities in Eq.~\eqref{eq: locality identity of n simplexes}. It can be verified that $E_{\mathrm{id}}$ includes all invariants arising from the higher commutators of operators of the form $V = \prod_{s \in \mathcal{S}} (U(s))^{n_s}$, with $n_s \in \mathbb{Z}$.\footnote{\change This follows from the commutator decomposition $[A\,B, C] = B^{-1}[A,C]\,B\,[B,C]$, which expresses the commutator of a product as a product of commutators up to conjugation. Therefore, if all higher commutators involving each $U(s)$ are trivial, then the higher commutators of any product $V=\prod_{s\in\mathcal S} U(s)^{n_s}$ with $n_s\in\mathbb{Z}$ are also trivial, provided that the intersections of their supports vanish.}
To classify the genuine invariants, we introduce the following quotient group:
\begin{definition}
    {\bf (Generalized statistics)}
    \textnormal{The generalized statistics is defined as the quotient of $E_{\mathrm{inv}}$ by the subgroup $E_{\mathrm{id}}$,
    \begin{equation}
        T := E_{\mathrm{inv}} / E_{\mathrm{id}}~,
    \label{eq: quotient group}
    \end{equation}
    which forms an Abelian group.}
\end{definition}

We now establish the following theorem about the property of this quotient group $T$:
\begin{theorem}\label{thm: independent of initial state}
    {\bf(Initial state independence)}
    \textnormal{The generalized statistics is uniquely determined by the sequence of unitaries, independent of the initial state $\ket{a_0}$ it acts upon:
    \begin{align}
        {\bra{a_0}\prod U(s_j)^\pm\ket{a_0}} = {\bra{a'_0} \prod U(s_j)^\pm\ket{a'_0}}~,
        \label{eq: globalshiftofa}
    \end{align}
    for all $\bra{a_0}\prod U(s_j)^\pm\ket{a_0} \in E_{\mathrm{inv}}$ and $a'_0 \in \mathcal{A}$.
    }
\end{theorem}
\vspace{3mm}

To prove this, we first conjugate the entire sequence of unitaries in Eq.~\eqref{eq:sequenceU} by another unitary operator $U(s')$. Specifically, consider the equality of invariants:
\begin{align}
    \bra{a_0}\prod U(s_j)^\pm\ket{a_0} = 
    \bra{a_0+\partial s'} \prod \tilde U(s_j)^\pm \ket{a_0+\partial s'}~,
\label{eq: conjugate by U(s')}
\end{align}
with $s' \in \mathcal{S}$ and $\tilde{U}(s_j) := U(s') U(s_j) U(s')^\dagger$. One can then write $\tilde U(s_j)$ as
\begin{eqs}
    \tilde U(s_j) = U(s_j) O(\partial s_j)~,
\end{eqs}
Here, an operator $O(\partial s_j):=[U(s_j),U(s')^\dagger]$ is supported at the neighborhood of $\partial s_j$. This is because for $s' \neq s_j$, the commutator of $U(s'),U(s_j)$ has a support at $s_j \cap s' \subset \partial s_j$. For $s'=s_j$, the symmetry operators of the Abelian fusion group is commutative in the bulk, so $O(\partial s_j)$ again has a support within $\partial s_j$.
These $O(\partial s_j)$ can be treated as local perturbations acting on the boundary of each unitary $U(s)$. Then, by the definition of $E_{\mathrm{inv}}$, we have
\begin{eqs}
    &\bra{a_0+\partial s'} \prod \left(U(s_j) O(\partial s_j)\right)^\pm \ket{a_0+\partial s'} \\
    &= \bra{a_0+\partial s'} \prod \left(U(s_j) \right)^\pm \ket{a_0+\partial s'}~.
\label{eq: replace U tilde by U}
\end{eqs}
Combining Eq.~\eqref{eq: conjugate by U(s')}, and noting that $\text{Im}\partial$ generates $\mathcal{A}$, this completes the proof of Theorem~\ref{thm: independent of initial state}. 
As a result, the ratio
\begin{eqs}
    \frac{\bra{a_0}\prod U(s_j)^\pm\ket{a_0}}{\bra{a'_0} \prod U(s_j)^\pm\ket{a'_0}}
    \label{eq:ratioofglobalshift}
\end{eqs}
with $\bra{a_0}\prod U(s_j)^\pm\ket{a_0} \in E_{\mathrm{inv}}$ and $a'_0 \in \mathcal{A}$ becomes a trivial phase. In Appendix \ref{app:globalshift}, we show that the above ratio \eqref{eq:ratioofglobalshift} becomes an element of $E_{\mathrm{id}}$ by explicitly checking that it is given by a product of higher commutators.

\subsubsection{Quantization of generalized statistics}

{\change In this section, we show that the above group $T$ with any finite simplicial complex $X$ and finite Abelian group $G$ always becomes a finite Abelian group. In particular, this implies that the invariants $e^{i\Theta}$ in $T$ must take the quantized values.}

Since the number of generators of $T$ is finite upper bounded by $|\mathcal{S}|\times|\mathcal{A}|$, it suffices to show that $T$ is a torsion, i.e., a direct sum of finite Abelian groups and do not contain a free part.
Let us take an invariant $[e]\in E_{\mathrm{inv}}/E_{\mathrm{id}}$. Then take a representative $e\in E_{\mathrm{inv}}$ expressed as $e=\sum_{(s,a)}\epsilon(s,a)\theta(s,a)$. Since \eqref{eq:ratioofglobalshift} is included in $E_{\mathrm{id}}$, the equivalence class of $e$ is left invariant under the global shift of $a\to a+a_0$ with $a_0\in\mathcal{A}$ in phases $\theta(s,a)$, which leads to the expression $e'=\sum_{(s,a)}\epsilon(s,a)\theta(s,a+a_0)$ with $[e]=[e']$. We then get 
\begin{eqs}
    |\mathcal{A}|[e] &= \sum_{a_0\in\mathcal{A}} \sum_{(s,a)}\epsilon(s,a)\theta(s,a+a_0)\\
    &= \sum_{a_0\in\mathcal{A}} \sum_{(s,a)}\epsilon(s,a-a_0)\theta(s,a) \\
    &=\sum_{(s,a)}\left(\sum_{a_0\in\mathcal{A}} \epsilon(s,a_0)\right)\theta(s,a) = 0 \pmod{2\pi}~,
\end{eqs}
where we used Eq.~\eqref{eq:inv2} in the last equation. Since $\mathcal{A}=B_p(X,G)$ is finite, this implies that any element of $T$ has finite order which divides $|\mathcal{A}|$. This shows that $T$ is a finite Abelian group. 

Later in Sec.~\ref{sec:Anomaly} and Sec.~\ref{sec:SRE}, the invariants in $T$ will be identified as the microscopic definition of the 't Hooft anomalies of the global symmetry $G$. The above observation implies that the 't Hooft anomalies in the microscopic lattice systems generally become a torsion when $G$ is finite. This is consistent with a generic conjecture that the 't Hooft anomalies realized in symmetry-preserving gapped theories must be a torsion~\cite{cordova2024gappedtheoriestorsionanomalies}. We comment on the applicability of generalized statistics in generic gapless systems in Sec.~\ref{sec:discussions}.

\subsection{Generalized statistics for non-Abelian 0-form symmetries}
\label{subsec:nonabelian}
\subsubsection{Generalized statistics for non-Abelian fusion groups}
It is straightforward to extend the above formalism to the case of non-Abelian fusion group $G$, which can happen for the 0-form symmetry. Let us describe the excitation model for the non-Abelian $G$ on a simplicial complex $X$ embedded in a $d$-dimensional space. $ \mathcal{S}$ is again a set of a pair $s=(g_j,\sigma_{d})$, where $\{g_j\}$ is a minimal set of generators of $G$ and $\sigma_{d}$ is a $d$-simplex of $X$. 
Each element $s\in\mathcal{S}$ corresponds to the unitary $U(s)=U_{g_j}(\sigma_{d})$. Then $\mathcal{A}$ is identified as the set of the states $\{\ket{a}\}$ obtained by the sequence of unitaries on a fixed $G$ symmetric state $\ket{0}$. The unitary acts on the state by $U(s)\ket{a} =e^{i\theta(s,a)} \ket{a\times \partial s}$ with the group fusion $\times$, i.e., fusing the excitations at the boundary of a $d$-simplex.

The expression groups $E, E_{\mathrm{inv}},E_{\mathrm{id}}$ can be readily generalized to the non-Abelian group $G$. 

There are slight modifications to the characteristic equations for $E_{\mathrm{inv}},E_{\mathrm{id}}$ arising from the non-Abelian nature of fusion groups.
The only modification to $E_{\mathrm{inv}}$ is that one of the conditions for $E_{\mathrm{inv}}$  \eqref{eq:inv1} is expressed according to the fusion of non-Abelian groups as 
\begin{align}
    \sum_{s\in\mathcal{S}}\epsilon(s,a) - \sum_{s\in\mathcal{S}} \epsilon(s,a\times(\partial s)^{-1}) = 0, \quad \text{for any $a\in\mathcal{A}$}~.
\end{align}

For the definition of $E_{\mathrm{id}}$, the non-Abelian fusion group has the effect of modifying Eq.~\eqref{eq: globalshiftofa}. Instead of Eq.~\eqref{eq: globalshiftofa}, for non-Abelian $G$ we have the equation
\begin{align}
        {\bra{a_0}\prod U(s_j)^\pm\ket{a_0}} = {\bra{a_0\times \partial s'} \prod \rho_{s'}[U(s_j)]^\pm\ket{a_0\times \partial s'}}~,
        \label{eq: globalshiftofa_nonabelian}
    \end{align}
    for all $\bra{a_0}\prod U(s_j)^\pm\ket{a_0} \in E_{\mathrm{inv}}$
    and any choice of $s'=(g',\sigma'_{d})$. Here, we write the conjugation action of $U(s')$ on $U(s)$ with $s=(g,\sigma_{d})$ by 
\begin{align}
\rho_{s'}[U(s)]=
    \begin{cases}
        U_{g'gg'^{-1}}(\sigma_d) & \sigma_d=\sigma'_d \\
        U_{g}(\sigma_d) & \sigma_d\neq \sigma'_d 
    \end{cases}~.
\end{align}
That is, $U(s')$ acts on $U(s)$ by an automorphism if these operators have the same support. \eqref{eq: globalshiftofa_nonabelian} can be derived by a similar discussion to the proof of \eqref{eq: globalshiftofa}.

We define the group $E_{\mathrm{id}}\subset E_{\mathrm{inv}}$ as the group generated by the elements in the form of the ratio
\begin{align}
        \frac{{\bra{a_0}\prod U(s_j)^\pm\ket{a_0}}}{{\bra{a_0\times \partial s'} \prod \rho_{s'}[U(s_j)]^\pm\ket{a_0\times \partial s'}}}
        \label{eq:ratio_nonabelian}
    \end{align}
    with $\bra{a_0}\prod U(s_j)^\pm\ket{a_0} \in E_{\mathrm{inv}}$, $s'\in\mathcal{S}$,
    and the elements given by the higher commutators
    \begin{eqs}
    & \bra{a}[U(s_n),[\cdots,[U(s_2),U(s_1)]]]\ket{a}=1, \\
    &\forall a\in\mathcal{A}, \forall s_1, s_2, \cdots, s_n \in \mathcal{S} | s_1 \cap s_2 \cap \dots \cap s_n = \varnothing~.
\end{eqs}    
It is expected that the ratio \eqref{eq:ratio_nonabelian} is again given by a product of higher commutators, and the above definition gives a redundant set of generators. Verifying this expectation is left for future studies.

\subsubsection{Quantization of invariants for non-Abelian fusion groups}

One can also show the quantization of invariants for non-Abelian fusion groups. 
For simplicity, let us take $X$ to be a triangulation of a sphere $S^d$.

Consider an element $[e]\in T=E_{\mathrm{inv}}/E_{\mathrm{id}}$. Since \eqref{eq:ratio_nonabelian} is an element of $E_{\mathrm{id}}$, conjugation by a unitary $U(s')$ leaves the equivalence class $[e]$ invariant. 
Hence, conjugation by any unitary of the form $\prod_{\sigma'\in X} U_{g'(\sigma')}(\sigma')$ leaves $[e]$ invariant, where $g'(\sigma')$ is any group element $g'(\sigma')\in G$ chosen for each $d$-simplex $\sigma'$. Each $U_{g'(\sigma')}(\sigma')$ is a product of operators $U(s)$ with $s\in\mathcal{S}$ supported on the simplex $\sigma'$. 
Now let us sum over all possible conjugation actions of $[e]$ by $\prod_{\sigma'\in X} U_{g'(\sigma')}(\sigma')$. This corresponds to summing over all choices of group elements $g'(\sigma')$ for each $d$-simplex, giving $|G|^{N_{d}}$ possibilities, where $N_{d}$ is the number of $d$-simplices in $X$. 

After this summation, we obtain an expression of $|G|^{N_{d}}[e]$, and we will show that it vanishes.
To see this, pick a single $d$-simplex $\sigma_{d}\in X$, and focus on the coefficient of the phase $\theta(s,a)$ with $s$ supported on $\sigma_{d}$. Then, in the sum over conjugation actions, first sum over the $N_{d}-1$ $d$-simplices with $\sigma'\neq \sigma_{d}$,
\begin{align}
    \begin{split}
        &|G|^{N_{d}-1}[e] \\
        =& \sum_{\substack{\{g'(\sigma')\} \\ \sigma'\neq \sigma_{d}}}\sum_{a\in\mathcal{A}}\epsilon(s,a)\theta\!\left(s,a\times\prod_{\sigma'\neq\sigma_{d}} \partial(g'(\sigma'),\sigma')\right)+\cdots
        \nonumber
    \end{split}
\end{align}
where $\cdots$ denotes the other phases $\theta$. 
Note that while the excitation $a$ is shifted by the conjugation action, $s$ remains invariant under conjugation since the action avoids the simplex $\sigma_{d}$. 
Here, since $X$ is a triangulation of a sphere, $\text{Im}(\partial)$ with the domain $\sigma'\neq \sigma_{d}$ generates the entire group $\mathcal{A}$. 
Therefore, one can rewrite
\begin{align}
    \begin{split}
        |G|^{N_{d}-1}[e] &= \sum_{a'\in\mathcal{A}}\sum_{a\in\mathcal{A}}\epsilon(s,a)\theta(s,a\times a') +\cdots \\
        &= \sum_{a'\in\mathcal{A}}\sum_{a\in\mathcal{A}}\epsilon(s,a\times a'^{-1})\theta(s,a) +\cdots \\
        &= \sum_{a\in\mathcal{A}}\left(\sum_{a'\in\mathcal{A}}\epsilon(s,a')\right) \theta(s,a) +\cdots \\
        &= 0 +\cdots~,
    \end{split}
\end{align}
implying that the contribution from each $s\in\mathcal{S}$ vanishes when summing over the $N_{d}-1$ simplices avoiding $s$. This shows that $|G|^{N_{d}}[e]$ becomes zero.

Therefore, we conclude that the generalized statistics $T$ forms a finite Abelian group whose order divides $|G|^{N_{d}}$. The invariants are thus quantized into discrete values.

\section{Computation of Generalized Statistics}\label{sec:Computation}

\begin{figure*}[th]
    \centering
    \subfigure[Configuration space]{\includegraphics[scale=0.4]{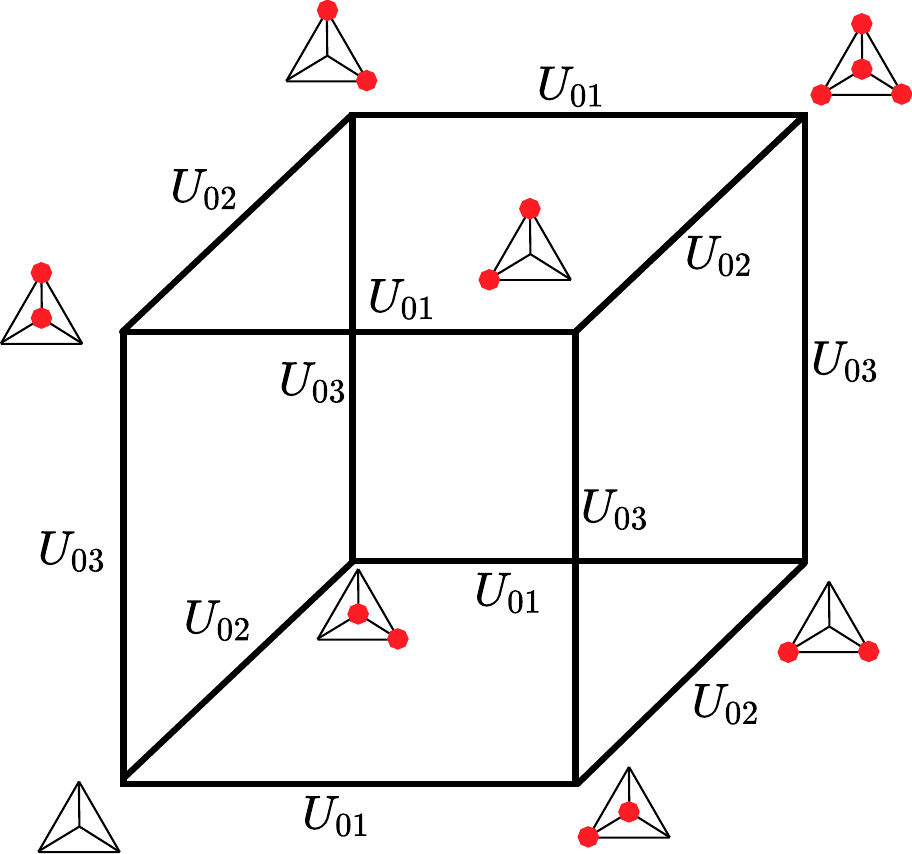}\label{fig: 2D T junction operator cube}}
    \hspace{10ex}
    \subfigure[T-junction process]{\raisebox{5ex}{\includegraphics[scale=0.4]{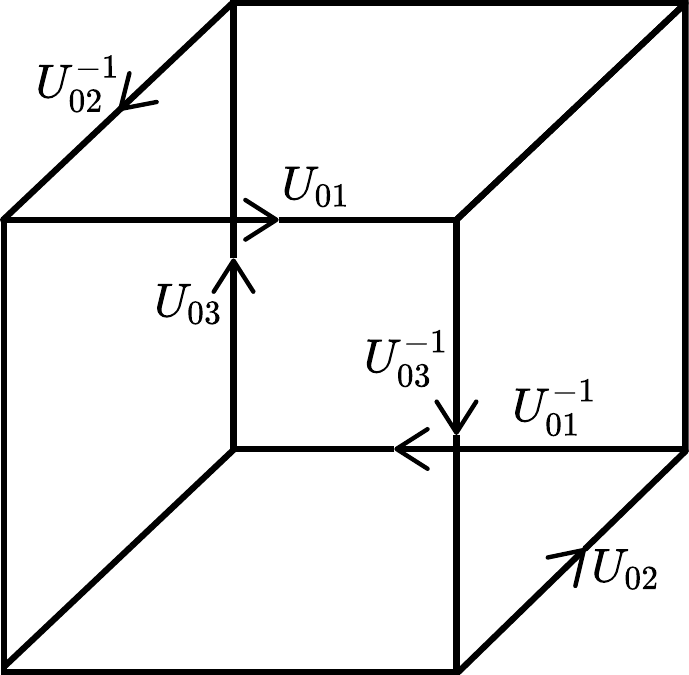}\label{fig: 2+1D T junction process}}}
    \caption{(a) Configuration space for $\ZZ_2$ particles on a triangle with a central vertex. We analyze states within the same superselection sector as the vacuum. These states are interconnected through string operators $U_{0i}$ where $i=1,2,3$, delineating the $1$-skeleton of a cube. It is important to note that transitions between states can be achieved by applying either $U_{0i}$ or $U^{-1}_{0i}$, and generally, the operator $U^2_{0i}$ does not necessarily equal $+1$. Connections between states via $U_{ij}$ are not explicitly shown in this diagram.
    (b) Visualization of the T-junction process in the configuration space. A specific initial state is selected (the outcome is independent of this choice), demonstrating how the T-junction process swaps the positions two particles.
    }
    \label{fig: T-junction configuration space}
\end{figure*}

In the previous section, we provided the conditions (Eqs.~\eqref{eq:inv1}, \eqref{eq:inv2}, and \eqref{eq:inv3'}) under which a process remains invariant under any local deformation. Solving for $ \epsilon(s, a) $ that satisfies these equations is generally challenging. For instance, Ref.~\cite{FHH21} uses computational methods to solve similar equations for loop excitations in (3+1)D, with additional constraints (specifically, restricting to processes that flip a loop). However, extending this approach to other types of excitations is not straightforward. Therefore, we adopt an alternative approach. Instead of directly solving the equations, we investigate the ``trivial solutions'' and use them to construct new solutions for genuine generalized statistics.

Specifically, for each locality identity in Eq.~\eqref{eq: locality identity of n simplexes} within $ E_{\mathrm{id}} $, a linear combination of $ \theta(s, a) $ must be zero modulo $ 2\pi $. The coefficients ${ \epsilon(s, a) }$ satisfy the conditions given in Eqs.~\eqref{eq:inv1}, \eqref{eq:inv2}, and \eqref{eq:inv3'} since the locality identities always hold, regardless of how ${U(s)}$ or ${\ket{a}}$ is deformed. Since the commutators of unitaries for non-overlapping simplexes always yield a phase factor of $+1$, the solution ${ \epsilon(s, a) }$ derived from the locality identity is considered a trivial solution.

Next, note that these equations are linear in $\epsilon(s, a)$. Specifically, if $\{ \epsilon(s, a) \}$ is a solution, then $\{ \epsilon'(s, a) := \alpha \epsilon(s, a) \}$ is also a solution, provided that all $\alpha \epsilon(s, a)$ remain integers. If we find a combination of locality identities such that all coefficients $\{ \epsilon(s, a) \}$ are multiples of an integer $k$, we can obtain a potentially nontrivial solution by dividing these coefficients by $k$. If this new solution cannot be expressed as a linear combination of identities in $E_{\mathrm{id}}$, we categorize it as a genuine invariant. Moreover, this solution imposes a constraint on the generalized statistics, which can only take the form $\exp\left( \frac{2\pi i j}{k} \right)$, where $j \in { 0, 1, 2, \dots, k-1 }$.

\subsection{Deriving generalized statistics via locality identities}
\label{subsec:T junction quantization}

{\change
In this section, we explicitly derive several statistical processes from locality identities.
These derivations can be regarded as mathematical proofs carried out by hand (without computer assistance), confirming the correctness of the results presented in the previous sections.
Specifically, the (2+1)D $\mathbb{Z}_2$ particle-exchange statistics is derived in Sec.~\ref{sec:Z2 particle in (2+1)D from locality identity},  
and the (3+1)D $\mathbb{Z}_2$ loop-flipping statistics is derived in Sec.~\ref{sec:Z2 loop in (3+1)D from locality identity}.  
In addition, the (1+1)D $\mathbb{Z}_2$ particle-fusion statistics and the (3+1)D loop–membrane mutual statistics are derived in Appendix~\ref{sec:Particle fusion process in (1+1)D in App} and Appendix~\ref{sec:Loop-membrane mutual statistics in (3+1)D in App}, respectively.
The algorithmic approach for systematically deriving these processes will be presented in the following section.
}

\subsubsection{$\mathbb{Z}_2$ particles in (2+1)D}\label{sec:Z2 particle in (2+1)D from locality identity}

As an example, we begin by considering the T-junction process in (2+1)D with the fusion group $G = \mathbb{Z}_2$, demonstrating that its anyon statistics must take the form $\exp(i \Theta) = \pm 1, \pm i$. The configuration states are shown in Fig.~\ref{fig: 2D T junction operator cube}.
We start by presenting particular locality identities for demonstration:
\begin{eqs}
    \bra{\hbox{ \raisebox{-1ex}{\includegraphics[width=.6cm]{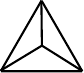}}}}
    [[U_{02},U_{03}],U_{12}]
    \ket{\hbox{ \raisebox{-1ex}{\includegraphics[width=.6cm]{T_junction_state_vaccum.pdf}}}}
    =1~.
\end{eqs}
In terms of $\theta(s,a)$, the above identity can be expanded as:
\begin{eqs}
    &\theta \left( U_{03},\hbox{ \raisebox{-1ex}{\includegraphics[width=.6cm]{T_junction_state_12.pdf}}} \right) 
    + \theta \left(U_{02},\hbox{ \raisebox{-1ex}{\includegraphics[width=.6cm]{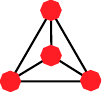}}} \right) 
    + \theta \left(U_{03}^{-1} ,\hbox{ \raisebox{-1ex}{\includegraphics[width=.6cm]{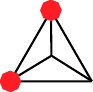}}} \right) \\ 
    &+ \theta \left(U_{02}^{-1} ,\hbox{ \raisebox{-1ex}{\includegraphics[width=.6cm]{T_junction_state_01.pdf}}} \right)
    + \theta \left(U_{02},\hbox{ \raisebox{-1ex}{\includegraphics[width=.6cm]{T_junction_state_vaccum.pdf}}} \right)
    + \theta \left(U_{03},\hbox{ \raisebox{-1ex}{\includegraphics[width=.6cm]{T_junction_state_02.pdf}}} \right) \\
    &+ \theta \left(U_{02}^{-1},\hbox{ \raisebox{-1ex}{\includegraphics[width=.6cm]{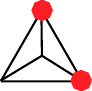}}} \right)
    + \theta \left(U_{03}^{-1},\hbox{ \raisebox{-1ex}{\includegraphics[width=.6cm]{T_junction_state_03.pdf}}} \right)
    = 0 \hspace{-1ex} \pmod{ 2 \pi}~,
\label{eq: locality identity of T-junction theta 1}
\end{eqs}
where we have adjusted the notation from $\theta(s, a)$ to $\theta(U(s), a)$. This change allows us to extend the definition of $\theta$ for $U(s)^{-1}$ as:
\begin{equation}
    \theta(U(s)^{-1}, a) := - \theta(U(s), a - \partial s)~,
\label{eq: relation between theta U-1 and theta U}
\end{equation}
which follows from Eqs.~\eqref{eq: U(s) theta} and \eqref{eq: U(s)-1 theta}.
More generally, $\theta$ can be defined for any sequence of $U(s)$ using the following property:
\begin{eqs}
    \theta(V U(s)^{\pm}, a) := \theta(V, a \pm \partial s)  + \theta(U(s)^\pm, a)~,
\end{eqs}
for all unitary operators $V$.
Using Eq.~\eqref{eq: relation between theta U-1 and theta U}, the locality identity~\eqref{eq: locality identity of T-junction theta 1} can be rewritten as:
\begin{eqs}
    &\theta \left( U_{03},\hbox{ \raisebox{-1ex}{\includegraphics[width=.6cm]{T_junction_state_12.pdf}}} \right) 
    + \theta \left(U_{02},\hbox{ \raisebox{-1ex}{\includegraphics[width=.6cm]{T_junction_state_0123.pdf}}} \right) 
    - \theta \left(U_{03},\hbox{ \raisebox{-1ex}{\includegraphics[width=.6cm]{T_junction_state_01.pdf}}} \right) \\ 
    &- \theta \left(U_{02} ,\hbox{ \raisebox{-1ex}{\includegraphics[width=.6cm]{T_junction_state_12.pdf}}} \right)
    + \theta \left(U_{02},\hbox{ \raisebox{-1ex}{\includegraphics[width=.6cm]{T_junction_state_vaccum.pdf}}} \right)
    + \theta \left(U_{03},\hbox{ \raisebox{-1ex}{\includegraphics[width=.6cm]{T_junction_state_02.pdf}}} \right) \\
    &- \theta \left(U_{02},\hbox{ \raisebox{-1ex}{\includegraphics[width=.6cm]{T_junction_state_03.pdf}}} \right)
    - \theta \left(U_{03},\hbox{ \raisebox{-1ex}{\includegraphics[width=.6cm]{T_junction_state_vaccum.pdf}}} \right)
    = 0 \pmod{ 2 \pi}~,
\label{eq: locality identity U only 1}
\end{eqs}
where we express $\theta$ solely in terms of $U$, without involving $U^{-1}$. Besides this locality identity, we can write down the following eight identities:
\begin{enumerate}
    \item $\bra{\hbox{ \raisebox{-1ex}{\includegraphics[width=.6cm]{T_junction_state_vaccum.pdf}}}}
    [[U_{01},U_{02}],U_{13}]
    \ket{\hbox{ \raisebox{-1ex}{\includegraphics[width=.6cm]{T_junction_state_vaccum.pdf}}}}
    =1~.$
    \item $\bra{\hbox{ \raisebox{-1ex}{\includegraphics[width=.6cm]{T_junction_state_vaccum.pdf}}}}
    [[U_{03},U_{01}],U_{23}]
    \ket{\hbox{ \raisebox{-1ex}{\includegraphics[width=.6cm]{T_junction_state_vaccum.pdf}}}}
    =1~.$
    \item $\bra{\hbox{ \raisebox{-1ex}{\includegraphics[width=.6cm]{T_junction_state_vaccum.pdf}}}}
    [[U_{02}^{-1},U_{03}^{-1}],U_{12}]
    \ket{\hbox{ \raisebox{-1ex}{\includegraphics[width=.6cm]{T_junction_state_vaccum.pdf}}}}
    =1~.$
    \item $\bra{\hbox{ \raisebox{-1ex}{\includegraphics[width=.6cm]{T_junction_state_vaccum.pdf}}}}
    [[U_{01}^{-1},U_{02}^{-1}],U_{13}]
    \ket{\hbox{ \raisebox{-1ex}{\includegraphics[width=.6cm]{T_junction_state_vaccum.pdf}}}}
    =1~.$
    \item $\bra{\hbox{ \raisebox{-1ex}{\includegraphics[width=.6cm]{T_junction_state_vaccum.pdf}}}}
    [[U_{03}^{-1},U_{01}^{-1}],U_{23}]
    \ket{\hbox{ \raisebox{-1ex}{\includegraphics[width=.6cm]{T_junction_state_vaccum.pdf}}}}
    =1~.$
    \item $\bra{\hbox{ \raisebox{-1ex}{\includegraphics[width=.6cm]{T_junction_state_vaccum.pdf}}}}
    \Big([[U_{02},U_{03}],U_{23}]\Big)^2
    \ket{\hbox{ \raisebox{-1ex}{\includegraphics[width=.6cm]{T_junction_state_vaccum.pdf}}}}
    =1~.$
    \item $\bra{\hbox{ \raisebox{-1ex}{\includegraphics[width=.6cm]{T_junction_state_vaccum.pdf}}}}
    \Big([[U_{01},U_{02}],U_{12}]\Big)^2
    \ket{\hbox{ \raisebox{-1ex}{\includegraphics[width=.6cm]{T_junction_state_vaccum.pdf}}}}
    =1~.$
    \item $\bra{\hbox{ \raisebox{-1ex}{\includegraphics[width=.6cm]{T_junction_state_vaccum.pdf}}}}
    \Big([[U_{03},U_{01}],U_{13}]\Big)^2
    \ket{\hbox{ \raisebox{-1ex}{\includegraphics[width=.6cm]{T_junction_state_vaccum.pdf}}}}
    =1~.$
\end{enumerate}
In Appendix~\ref{sec:T-junction process in (2+1)D in App}, we explicitly demonstrate that summing over the 9 identities above yields the resulting equation:
\begin{eqs}
    4 \Bigg(
    &\theta\Big(U_{01}^{-1},\hbox{ \raisebox{-1ex}{\includegraphics[width=.6cm]{T_junction_state_12.pdf}}}\Big)
    +\theta\Big(U_{03},\hbox{ \raisebox{-1ex}{\includegraphics[width=.6cm]{T_junction_state_02.pdf}}}\Big)
    +\theta\Big(U_{02}^{-1},\hbox{ \raisebox{-1ex}{\includegraphics[width=.6cm]{T_junction_state_23.pdf}}}\Big)\\
    &+\theta\Big(U_{01},\hbox{ \raisebox{-1ex}{\includegraphics[width=.6cm]{T_junction_state_03.pdf}}}\Big)
    +\theta\Big(U_{03}^{-1},\hbox{ \raisebox{-1ex}{\includegraphics[width=.6cm]{T_junction_state_13.pdf}}}\Big)
    +\theta\Big(U_{02},\hbox{ \raisebox{-1ex}{\includegraphics[width=.6cm]{T_junction_state_01.pdf}}}\Big)
    \Bigg) \\
    =&0 \pmod{ 2 \pi}~,
\end{eqs}
where the six $\theta$ terms can be combined into a total phase of a sequence of unitaries:
\begin{eqs}
    4 \theta\Big( U_{02} U_{03}^{-1} U_{01} U_{02}^{-1} U_{03} U_{01}^{-1},\hbox{ \raisebox{-1ex}{\includegraphics[width=.6cm]{T_junction_state_12.pdf}}}\Big)
    =0 \pmod{ 2 \pi}~.
\label{eq: T-junction process G=Z2}
\end{eqs}
This sequence of unitaries precisely corresponds to the T-junction process as defined in Eq.~\eqref{eq: T-junction process} and is illustrated within the configuration state space in Fig.~\ref{fig: 2+1D T junction process}.
Note that the coefficient $4$ in Eq.~\eqref{eq: T-junction process G=Z2} is crucial, as it constrains the phases to be
\begin{eqs}
    \theta\Big( U_{02} U_{03}^{-1} U_{01} U_{02}^{-1} U_{03} U_{01}^{-1},\hbox{ \raisebox{-1ex}{\includegraphics[width=.6cm]{T_junction_state_12.pdf}}}\Big)
    =0, \pi, \pm \frac{\pi}{2} \hspace{-1ex} \pmod{ 2 \pi}~,
\end{eqs}
indicating that the anyon with $G=\mathbb{Z}_2$ fusion can be a boson, a fermion, or an (anti-)semion.
The coefficient in Eq.~\eqref{eq: T-junction process G=Z2} depends on both the fusion group $G$ and the spatial dimensions. 

This computation can be generalized for $G = \ZZ_N$ straightforwardly.
Specifically, in (2+1)D, the coefficient is $2N$ for even $N$ and $N$ for odd $N$, while in (3+1)D, the coefficient is $2$ for even $N$ and $1$ for odd $N$, as discussed previously in Eq.~\eqref{eq: phase of T-junction for N and D}. 
This is consistent with the topological spin of $\mathbb{Z}_N$ anyons (related to the topological twist by $e^{2\pi i h}$) given by $h=\frac{p}{2N}$ mod 1 for integer $p=0,1,\cdots,2N-1$ for even $N$, and $p=0,2,4\cdots,2N-2$
for odd $N$ in a bosonic theory, i.e., no transparent local fermion \cite{Hsin:2018vcg}.

In the following sections, we present an algorithm that systematically determines these coefficients using the Smith normal form. In particular, we have reproduced the above quantization of topological spin
for $G=\ZZ_N$ with $N\leq 10$ using our personal computers, while higher $N$ demands more computational resources.  

\subsubsection{$\mathbb{Z}_2$ loops in (3+1)D}\label{sec:Z2 loop in (3+1)D from locality identity}

In addition to the well-known T-junction process for detecting the spins of particles, we now proceed to another more intricate example: the loop-flipping process for $\mathbb{Z}_2$ loops in (3+1)D. This process was first proposed in Ref.~\cite{FHH21}, where computers found a sequence of 36 unitaries to ensure invariance under any local perturbation.
In Appendix~\ref{sec:Loop-flipping process in (3+1)D in App}, we utilize our new method, which involves summing over locality identities, to derive a sequence of 24 unitaries, $\mu_{24}$, that satisfies the following equation analogous to Eq.~\eqref{eq: T-junction process G=Z2}:
\begin{eqs}
    2\theta\left(\mu_{24}, \hbox{ \raisebox{-4ex}{\includegraphics[width=1.75cm]{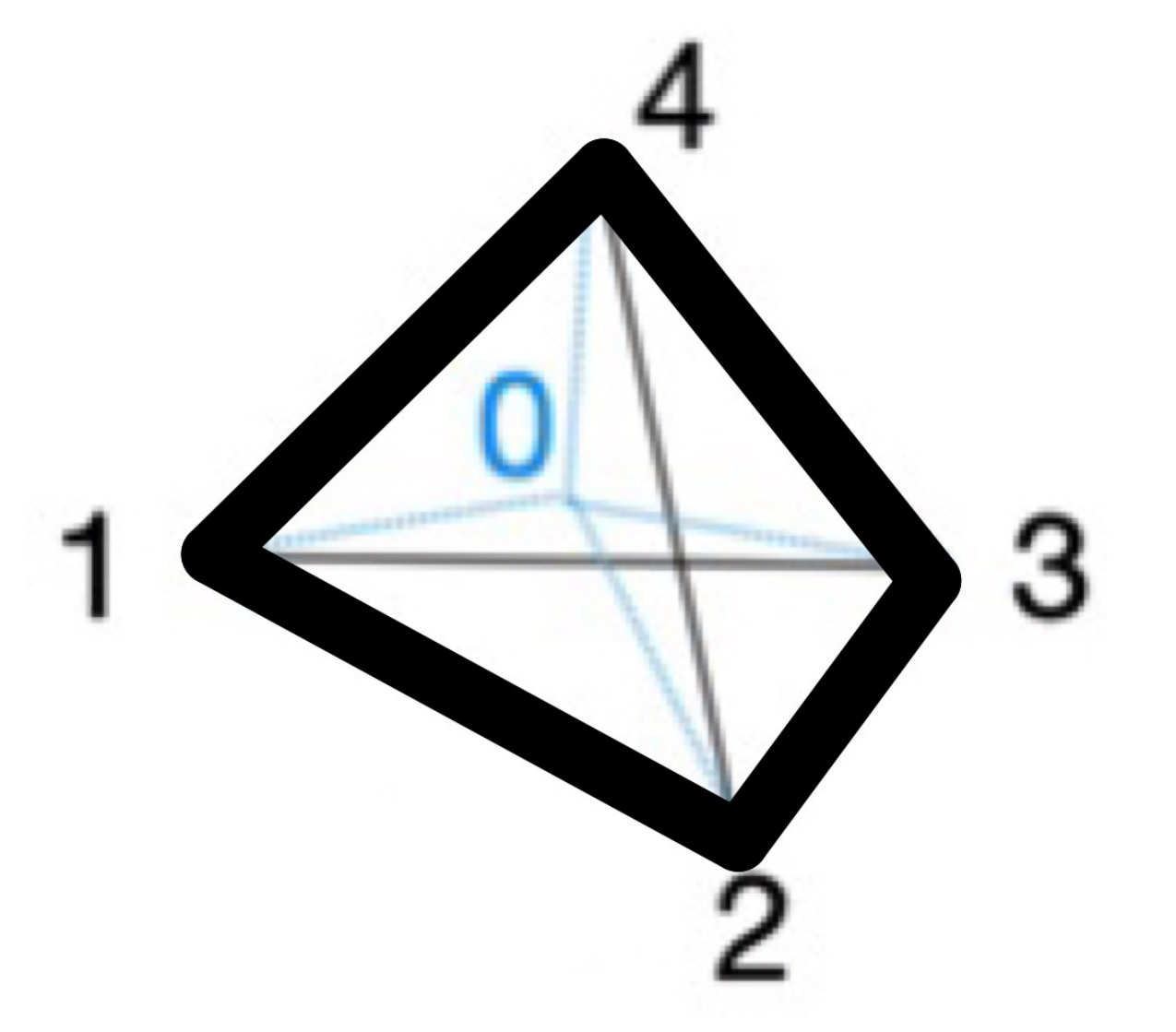}}} \right) = 0 \pmod{2\pi}~,
\end{eqs}
with $\mu_{24}$ defined in Eq.~\eqref{eq: 24-step process} and illustrated in Fig.~\ref{fig: 24 step process} and Fig.~\ref{fig: 24 step process configuration space}.
Therefore, the total phase of this sequence must be $0$ or $\pi$ mod $2\pi$, implying that a loop with $G=\ZZ_2$ fusion can be either bosonic or fermionic \cite{CH21, FHH21}. We further demonstrate that the total phase generated by the 36-unitary sequence from Ref.~\cite{FHH21} matches the phase obtained from our $\mu_{24}$ sequence, up to some locality identities. In other words, both sequences yield the same loop statistics, but our process is more efficient, requiring fewer unitaries. We have verified that the 24-step sequence is optimal, as no shorter sequence exists.

\subsection{Computational algorithm using the Smith normal form}\label{sec:Computational algorithm SNF}

In the previous section, we derived the T-junction process in (2+1)D and introduced the novel 24-step loop-flipping processes in (3+1)D. A natural question arises how to systematically determine specific linear combinations of the local identities such that the overall coefficients $\{ \epsilon(s, a) \}$ in the expression $e = \sum_{s, a} \epsilon(s, a) \theta(s, a)$ result in a greatest common divisor (gcd) greater than one.

For smaller fusion groups, such as $G = \mathbb{Z}_2$, it is often feasible to rely on human intuition to manually determine the appropriate local identities to be summed in order to cancel certain $\theta(s, a)$ terms, ultimately ensuring that the remaining coefficients are all $2$ or $4$. We demonstrated this process in detail in Appendix~\ref{app: T-junction and 24-step compute}, where we effectively exploited the structure of $\mathbb{Z}_2$ to construct explicit summations that achieve the desired gcd properties.

For larger groups or more complex settings, it becomes challenging to manually determine the appropriate summation of local identities. 
Meanwhile, it turns out that finding such a summation can be efficiently handled by computational algorithms. 
The algorithm systematically determines the appropriate summation of locality identities by computing the Smith normal form of the corresponding matrix representation.
By expressing the locality identities in matrix form, we can use the Smith normal form to identify combinations of rows that yield nontrivial gcd values for the coefficients $\epsilon(s, a)$. 
This method provides a systematic approach to finding the appropriate summations for larger fusion groups beyond $\mathbb{Z}_2$, and allows us to optimize the result by finding the shortest unitary sequence.

The algorithm proceeds as follows: we first construct the matrix $M$, whose rows represent the locality identities, with entries corresponding to the coefficients of the $\theta(s, a)$ terms in each identity. We then compute the Smith normal form (SNF) of $M$, which allows us to determine the invariant factors and, subsequently, the linear combinations of the original identities that yield the desired gcd properties for the coefficients $\epsilon(s, a)$.

To illustrate this process, let us consider a toy example with three $\theta$ terms, $\theta_1$, $\theta_2$, and $\theta_3$, and three locality identities:
\begin{eqs}
    \theta_1 + 2 \theta_2 + 3 \theta_3  &= 0 \pmod{2 \pi}~,\\
    4\theta_1 + 5 \theta_2 + 6 \theta_3  &= 0 \pmod{2 \pi}~,\\
    7 \theta_1 + 8 \theta_2 + 9 \theta_3  &= 0 \pmod{2 \pi}~.
\end{eqs}
These three locality identities are not independent, as the sum of the first and third identities equals twice the second identity. However, for the sake of demonstration, we proceed without recognizing this redundancy, as it can be challenging to identify such relationships in more extensive systems with thousands of identities in practice.

These locality identities can be represented as the \textbf{phase relation matrix}:
\begin{eqs}
    M =
    \begin{blockarray}{ccc l}
        \theta_1 & \theta_2 & \theta_3  \\
        \begin{block}{[c c c]l}
            1 & 2 & 3 \\
            4 & 5 & 6 \\
            7 & 8 & 9 \\
        \end{block}
    \end{blockarray}~~,
    \label{eq: phase relation example}
\end{eqs}
\vspace{-1.5em} \\
\noindent where each row represents a locality identity.
Row operations involving adding an integer multiple of one row to another do not change the row vector space spanned by these locality identities; they merely redefine the basis. However, column operations require more attention, as they necessitate the redefinition of phase labels for each column. Specifically, adding column $k$ to column $j$ requires a redefinition of phases as $\theta'_j = \theta_j$ and $\theta'_k = (-\theta_j + \theta_k)$:
\begin{eqs}
    M_{i,j} \theta_j + M_{i,k} \theta_k &= (M_{i,j} + M_{i,k}) \theta'_j + M_{i,k} \theta'_k \quad \forall i~.
    \label{eq: phase rearrangement for M}
\end{eqs}
Next, we calculate the Smith normal form (SNF) of $M$, obtaining integer matrices $L$, $R$, and $A$ such that 
\begin{eqs}
    M = L A R~,
    \label{eq: M=LAR}
\end{eqs}
where $A$ is a diagonal integer matrix, and $L$ and $R$ are unimodular matrices (i.e., with determinant $\pm 1$). For the matrix $M$ given in Eq.~\eqref{eq: phase relation example}, we decompose it as:
\begin{eqs}
    \begin{bmatrix}
    1 & 2 & 3 \\
    4 & 5 & 6 \\
    7 & 8 & 9
    \end{bmatrix}
    =
    \begin{bmatrix}
    1 & 0 & 0 \\
    4 & -1 & 0 \\
    7 & -2 & 1
    \end{bmatrix}
    \begin{bmatrix}
    1 & 0 & 0 \\
    0 & 3 & 0 \\
    0 & 0 & 0
    \end{bmatrix}
    \begin{bmatrix}
    1 & 2 & 3 \\
    0 & 1 & 2 \\
    0 & 0 & 1
    \end{bmatrix}~.
\end{eqs}
We claim that this system exhibits $\mathbb{Z}_3$ generalized statistics, corresponding to the second diagonal entry in the matrix $A$. This corresponds to the process $\theta_2 + 2\theta_3$, which is represented by the second row of the matrix $R$.

To justify the physical interpretation of this computation, let us proceed step by step. We begin with the matrix $M$ in Eq.~\eqref{eq: phase relation example} and perform row operations over integers to obtain its Hermite normal form (HNF). We start by eliminating the entries below the first pivot, which is $1$ in the first row and first column.
First, we subtract $4 \times \text{Row 1}$ from Row 2 and  subtract $7 \times \text{Row 1}$ from Row 3. The matrix then becomes:
\begin{eqs}
    M' =
    \begin{blockarray}{ccc l}
        \theta_1 & \theta_2 & \theta_3  \\
        \begin{block}{[c c c]l}
            1 & 2 & 3 \\
            0 & -3 & -6 \\
            0 & -6 & -12 \\
        \end{block}
    \end{blockarray}~~.
\end{eqs}
\vspace{-1.5em} \\
\noindent Next, we eliminate the entries below the second pivot. We focus on the second column, where the pivot is $-3$. We add $2 \times \text{Row 2}$ to Row 3 to eliminate the entry below the pivot. The matrix $M'$ becomes:
\begin{eqs}
    M'' =
    \begin{blockarray}{ccc l}
        \theta_1 & \theta_2 & \theta_3  \\
        \begin{block}{[c c c]l}
            1 & 2 & 3 \\
            0 & -3 & -6 \\
            0 & 0 & 0 \\
        \end{block}
    \end{blockarray}~~.
\end{eqs}
\vspace{-1.5em} \\
\noindent To make the second pivot positive, we multiply Row 2 by $(-1)$:
\begin{eqs}
    M_{\mathrm{SNF}} =
    \begin{blockarray}{ccc l}
        \theta_1 & \theta_2 & \theta_3  \\
        \begin{block}{[c c c]l}
            1 & 2 & 3 \\
            0 & 3 & 6 \\
            0 & 0 & 0 \\
        \end{block}
    \end{blockarray}~~,
\end{eqs}
\vspace{-1.5em} \\
\noindent which is the final Hermite normal form.

Subsequently, we perform column operations, using the $1$ in the first row to cancel out the entries $2$ and $3$ in the first row. Similarly, we use the $3$ in the second row to cancel out the $6$. According to the rule in Eq.~\eqref{eq: phase rearrangement for M}, we reduce $M$ to its Smith normal form:
\begin{eqs}
    M =
    \begin{blockarray}{ccc l}
        \theta'_1 & \theta'_2 & \theta'_3  \\
        \begin{block}{[c c c]l}
            1 & 0 & 0 \\
            0 & 3 & 0 \\
            0 & 0 & 0 \\
        \end{block}
    \end{blockarray}~~,
    \label{eq: example final SNF}
\end{eqs}
\vspace{-1.5em} \\
\noindent with redefined phase labels as
\begin{eqs}
    \theta'_1 &= \theta_1 + 2 \theta_2 + 3 \theta_3~, \\
    \theta'_2 &= \theta_2 + 2\theta_3~, \\
    \theta'_3 &= \theta_3~.
\end{eqs}
The physical interpretation of Eq.~\eqref{eq: example final SNF} is that all locality identities can be reduced to three equations:
\begin{eqs}
    1 \theta'_1 = 0, \quad 3 \theta'_2 = 0~, \quad 0 \theta'_3 = 0~.
\end{eqs}
This means that $\theta'_1$ is fixed, $\theta'_3$ can take any value, and $\theta'_2$ must be $\frac{2 \pi j}{3}$ for $j = 0, 1, 2$. In other words, this toy system exhibits $\mathbb{Z}_3$ statistics, corresponding to the unitary sequence labeled by $\theta'_2 = \theta_2 + 2 \theta_3$.

Inspired by this example, we can derive the following theorem:
\begin{theorem}
    {\bf (Generalized statistics)}
    \textnormal{
    Let $M$ be the phase relation matrix for a given simplicial complex and a fusion group, and consider its Smith decomposition $M = L A R$. If any diagonal entry $a_{ii}$ in $A$ satisfies $a_{ii} \neq 0, 1$, then it gives rise to generalized statistics of type $\mathbb{Z}_{a_{ii}}$, with the $i$-th row of the matrix $R$ specifying the corresponding unitary sequence. Namely, the generalized statistics $T$ forms an Abelian group characterized by:
    \begin{equation}
        T = \bigoplus_{a_{ii} \neq 0, 1} \mathbb{Z}_{a_{ii}}.
    \end{equation}
    }
\end{theorem}
\begin{remark}
    \textnormal{
    The sum of $\theta$ terms in each row of $R$ must originate from a sequence of unitary operators. This is because, in the original phase relation matrix $M$, each row corresponds to a (higher) commutator, which forms closed loops connecting the configuration states. Therefore, although each row of $R$ may be divided by the greatest common divisor of the coefficients, it still represents a superposition of multiple closed loops. By the Eulerian path theorem, we can identify a sequence that generates these loops (allowing for retracing edges to cancel their contributions).
    }
\end{remark}
\begin{remark}
    \textnormal{
    In the Smith decomposition~\eqref{eq: M=LAR}, the matrix $A$ is uniquely determined, while the matrices $L$ and $R$ are not. However, any valid choice of $R$ will yield the same generalized statistics, with the total phases differing only by linear combinations of locality identities. Typically, we choose $R$ such that its entries contain as many zeros as possible, as this simplifies the corresponding unitary sequence.
    }
\end{remark}

{\change
The phase relation matrix $M$ in Eq.~\eqref{eq: phase relation example} serves as a toy example; in practice, the computation can be much more involved. For example, when determining the Smith normal form (SNF), each diagonal element must divide the next, requiring careful manipulation. Obtaining either the Hermite normal form (HNF) or the SNF involves repeated use of the Euclidean algorithm to compute greatest common divisors between matrix entries in each row or column. For general dense matrices, this procedure can be both time- and memory-intensive for most computer implementations of SNF algorithms.  
However, in our setting the matrices are sparse and most entries are $\pm 1$, which allows programs such as SheafHom~\cite{sheafhom} to run significantly faster.  
In practice, the computation time is mainly determined by the number of columns of $M$:  
\begin{equation*}
    \operatorname{dim}E = |\text{number of simplexes}| \times |\text{generators of }G| \times |\mathcal{A}|.
\end{equation*}
On a standard personal computer, it is feasible to compute the HNF or SNF for $\operatorname{dim}E \sim 10^4$ within a reasonable time.
}

For example, in the (2+1)D T-junction process with fusion group $G = \mathbb{Z}_N$ on the simplicial complex shown in Fig.~\ref{fig: complex (b)}, there are 6 edges, $G$ has a single generator, and there are $N^3$ configuration states. Thus, the number of $\theta(s, a)$ terms is $6N^3$. In this situation, we are able to compute the cases with {\change $N \leq 12$}.

{\change
The loop fusion statistics with fusion group $G = \mathbb{Z}_N \times \mathbb{Z}_N$ is also computed on the simplicial complex shown in Fig.~\ref{fig: complex (b)}. In this case, there are 4 faces, $G$ has 2 generators, and the number of configuration states is $N^6$. Consequently, the number of $\theta(s, a)$ terms is $8N^6$. We can carry out the computation for cases with $N \leq 5$.
}

Consider another example involving loop statistics with $G = \mathbb{Z}_N$ on the (3+1)D simplicial complex shown in Fig.~\ref{fig: complex (c)}. Here, there are 10 faces, $G$ has a single generator, and there are $N^6$ configurations. The total number of $\theta(s, a)$ terms is $10N^6$, which limits our computations to $N \leq 4$.

Lastly, for membrane statistics with $G = \mathbb{Z}_N$ in the (3+1)D simplicial complex shown in Fig.~\ref{fig: complex (c)}, there are 5 tetrahedra (including the outer one), $G$ has a single generator, and there are $N^4$ configurations. Therefore, the total number of $\theta(s, a)$ terms is $5N^4$, allowing us to compute the cases with $N \leq 8$.

\subsection{Generalized statistics as anomalies: computational approach}

Using the SNF algorithm presented above, we obtain the classification of the generalized statistics in various spatial dimensions, as summarized in Sec.~\ref{sec:Examples_Statistics}. 

Below, we comment on the implications of several generalized statistics for the algebraic structure of the symmetry operators. We take the simplicial complex $X$ for the excitation model to be the minimal triangulation of $S^d$ embedded in $d$-dimensional space.\footnote{The minimal triangulation of $S^d$ is formed by $d+1$ $d$-simplices, which are identified as the boundary of a single $(d+1)$-simplex. Its vertices and edges form a complete graph $K_{d+2}$.}
\begin{itemize}
    \item In $(1+1)$D with finite 0-form symmetry $G$, we obtain the generalized statistics $T$ generated by the unitary sequence
    \begin{equation}
        Z_3(g) := [ U(g)_{02},U(g)_{01}^{|g|}]~,
    \end{equation}
    with $g\in G$. This form of the invariant immediately implies that the nontrivial invariant $Z_3(g)=e^{i\Theta}$ becomes an obstruction to having
    \begin{align}
        U(g)_{01}^{|g|} = 1~,
    \end{align}
    which means that the symmetry operator on an interval must violate the original group fusion rule. Note that an onsite symmetry satisfies $U(g)_{01}^{|g|} = 1$, since onsite operators supported on an interval again follow the $G$ fusion rule.
    This immediately leads to the physical consequence that the operator $U(g)$ with the nontrivial invariant $\Theta$ cannot be realized by an onsite symmetry on the lattice. Physically, the invariant indicates that the fusion of the operator $U(g)_{01}^{|g|}$ creates an electric charge of the $G$ symmetry localized at the end of the interval, which is a nontrivial point operator at the boundary.
    
    \item In $(2+1)$D with 0-form symmetry $G=\ZZ_N\times\ZZ_N$, the invariant is classified by $T=\ZZ_N\times\ZZ_N$. The classification is generated by a pair of invariants $Z_4(a),Z_4(b)$ shown in Eq.~\eqref{eq: ABCD loop fusion process}, with $a,b$ the generators of $\ZZ_N\times\ZZ_N$. Although not manifest from their form, it turns out that either of the invariants $Z_4(a),Z_4(b)$ becomes an obstruction to the following $\ZZ_N\times\ZZ_N$ fusion algebra of symmetry operators:
    \begin{equation}
        \hspace{3em}
        U(a)_{ijk}^N = U(b)_{ijk}^N = [U(a)_{ijk}, U(b)_{ijk} ] = 1~.
        \label{eq:ZNZNfusionrule}
    \end{equation}
    This obstruction can be explicitly demonstrated by the SNF algorithm. Specifically, we add the equations of the phases $\theta(s,a)$ originating from the three equations in \eqref{eq:ZNZNfusionrule} to the rows of the phase relation matrix $M$, and denote the new matrix by $M'$. Running the SNF algorithm on $M'$ classifies the invariants under these additional algebraic constraints. This invariant turns out to be trivial, which means that the original invariants $Z_4(a),Z_4(b)$ must be trivial under the $\ZZ_N\times\ZZ_N$ fusion rule of $U(a),U(b)$. This shows that $Z_4(a),Z_4(b)$ obstruct the fusion rule \eqref{eq:ZNZNfusionrule}. By a discussion analogous to the $(1+1)$D case, this implies that either of the invariants $Z_4(a),Z_4(b)$ becomes an obstruction to realizing the $\ZZ_N\times\ZZ_N$ symmetry by onsite operators.
\end{itemize}

In the above examples, the generalized statistics define obstructions to a certain group theoretical identity of operators $U(s)$ under its fusion. It turns out that this is a symptom of the 't Hooft anomaly of the global symmetry $G$, defined as obstructions to gauging the symmetry. 
Roughly speaking, failure of the group identities such as Eq.~\eqref{eq:ZNZNfusionrule} is directly interpreted as the failure of the Gauss law constraints in the gauge theory, where each unitary $U(s)$ is identified as a product of Gauss law operators.
This leads to the absence of a gauge invariant Hilbert space after an attempt to promoting the global symmetry to the gauge symmetry. Such a perspective of the generalized statistics as the anomalies will be discussed in details in the following sections.

In (bosonic) quantum field theory, the 't Hooft anomalies of a finite $(d-p-1)$-form symmetry $G$ in $d$ spatial dimensions are classified by the group cohomology $H^{d+2}(B^{d-p}G, U(1))$, where $B^{d-p}G$ represents the Eilenberg–MacLane space of the group $G$. In all the examples we evaluated using the SNF algorithm, the generalized statistics $T$ matches the group cohomology. This consistency leads us to the following conjecture:
\begin{conjecture}\label{conjecture: EM cohomology}
    \textnormal{
    Consider $p$-dimensional excitations with a $(d-p-1)$-form symmetry and fusion group $G$ in $d$ spatial dimensions. Let $X$ be a simplicial complex that triangulates the $d$-dimensional sphere $S^d$. Then, the generalized statistics $T$ of $p$-dimensional excitations on $X$ are classified by the cohomology of the Eilenberg–MacLane space\footnote{To be precise, $T$ behaves more similar to the homology of Eilenberg-MacLane space: $T=H_{d+2}(B^{d-p}G,\ZZ)$ (but they are isomorphic) \cite{xue2025statisticsinvertibletopologicalexcitations}.}:
    \begin{equation}
        T = H^{d+2}(B^{d-p}G, U(1))~
        \label{eq:T=EMspace}
    \end{equation}
    }
\end{conjecture}

We remark that, in principle, any finite simplicial complex $X$ on $S^d$ could be chosen to compute the generalized statistics. 
We numerically observed that different triangulations of $S^d$ yield the same result. This leads us to the above conjecture that the generalized statistics do not depend on a choice of the triangulation of $S^d$.
Hence, we have decided to use minimal triangulation for convenience.

\subsection{Stability of generalized statistics}
Beyond the evidence of Conjecture \ref{conjecture: EM cohomology} provided above by computational methods, we present additional insights to support our conjecture. Assuming the independence of $T$ under the choice of triangulations, one can see that both $T$ and the group cohomology in Eq.~\eqref{eq:T=EMspace} stabilize in the sense that for fixed $p$, both become independent of $d$ for $d\ge d_{\text{crit}} := 2p+3$.

Let us first see how the generalized statistics $T$ stabilizes.
When we define the generalized statistics of $p$-dimensional excitations, the excitations and unitaries are supported within finite $(p+1)$-skeleton $X_{p+1}$ that collects $0,\dots,(p+1)$-simplexes of $X$.
According to dimension theory in point-set topology~\cite{MunkresTopology}, every finite $(p+1)$-dimensional simplicial complex, with Lebesgue covering dimension of $p+1$, can be embedded in $S^{2p+3}$. Therefore, any complex $X_{p+1}$ can be embedded in a triangulation of $S^{2p+3}$. 
This implies that any invariants of $p$ dimensional excitations can be realized by a unitary sequence in a suitable triangulation of $S^{2p+3}$. Assuming the independence on the triangulation of $S^d$, one can conclude that the generalized statistics $T$ stabilizes for $d\ge d_{\text{crit}}=2p+3$.

Let us explicitly check this stability with several examples of small $p$.
For particle statistics ($p = 0$), all $1$-dimensional simplicial complexes can be embedded in $S^d$ for $d \geq 3$. In other words, in spatial dimensions $d \geq 3$, particle statistics stabilizes and is characterized by a single $\mathbb{Z}_2$ invariant, distinguishing bosons from fermions. This result aligns with the fact in Lorentz invariant theories that the particles with fractional statistics can exist only in (1+1)D or (2+1)D, while the statistics is restricted to bosons or fermions in higher dimensions. 
Similarly, for loop excitations ($p = 1$), the critical dimension is $d_{\text{crit}} = 2p + 3 = 5$. In spatial dimensions $d \geq 5$, loop statistics stabilizes and is characterized by a single $\mathbb{Z}_2$ invariant that distinguishes between bosonic and fermionic loops. 
An analogous result holds for membrane excitations ($p = 2$), where the generalized statistics stabilize at $d_{\text{crit}} = 7$. 

Meanwhile, the (co)homology of the Eilenberg–MacLane space also stabilizes:
\begin{theorem}
    {\bf (13.2.2 of Ref.~\cite{LodayCyclicHomology})}
    \textnormal{For any abelian group $A$, there exists a chain complex $Q_*(A)$ whose homology is isomorphic to the stable homology of the Eilenberg–MacLane space:
    \begin{align*}
        H_n(Q_*(A)) \cong H_{n+k}(K(A, k)), \quad k \geq n + 1~.
    \end{align*}
    }
\end{theorem}
\vspace{3mm}

Using the universal coefficient theorem and the long exact sequence associated with $0 \rightarrow \ZZ \rightarrow \RR \rightarrow \RR/\ZZ \rightarrow 0$, we obtain the stabilization of the cohomology of the Eilenberg–MacLane space with coefficients in $\mathbb{R}/\mathbb{Z} = U(1)$:
\begin{eqs}
    H^{n+k}(B^k G, U(1)) &= H^{2n+1}(B^{n+1} G, U(1))~,
\end{eqs}
for all $k \geq n+1$.
By choosing $n = p + 2$ and $k = d - p$, we find that $H^{d+2}(B^{d-p} G, U(1))$ stabilizes with respect to $d$ for $k \geq n + 1$, or equivalently,
\begin{equation}
    d\geq d_{\text{crit}}=2p+3~,
\end{equation}
which matches precisely with the argument from the embedding theorem. Therefore, this consistency for the pattern of stability supports Conjecture~\ref{conjecture: EM cohomology}.

The group cohomology that appears in Conjecture~\ref{conjecture: EM cohomology} classifies the topological responses in $(d+2)$ dimensions that describes the anomaly inflow.
We close this section with comments on the stable generalized statistics and the corresponding topological response:
\begin{itemize}
    \item For particle statistics $(p=0)$ with the fusion group $G=\ZZ_N$, the stable generalized statistics for $d\ge 3$ is classified by $\ZZ_{\gcd(2,N)}$; boson or fermion for even $N$. The statistics of the emergent particles are interpreted as a framing anomaly of the topological line operator, which is in turn understood as the 't Hooft anomaly of the $(d-1)$-form $\ZZ_2$ symmetry generated by this line operator. This 't Hooft anomaly is characterized by the response 
    \begin{align}
        \pi \int w_2\cup B_d=\pi\int Sq^2(B_d)~,
    \end{align}
    where $B_{d}$ is the $d$-form $\ZZ_N$ background gauge field with even $N$, and $w_2$ is the 2nd Stiefel-Whitney class.
    \item For loop statistics $(p=1)$ with the fusion group $G=\ZZ_N$, the stable generalized statistics for $d\ge 5$ is classified by $\ZZ_{\gcd(2,N)}$; bosonic or fermionic loops for even $N$. The statistics is again interpreted as a framing anomaly of the topological surface operator, which is the 't Hooft anomaly of the $(d-2)$-form $\ZZ_2$ symmetry generated by this surface operator. The 't Hooft anomaly is characterized by the response
    \begin{align}
        \pi\int w_3\cup B_{d-1}&=\pi\int w_2(dB_{d-1}/2)\cr &=\pi\int Sq^2(dB_{d-1}/2)~,
    \end{align}
    where $B_{d-1}$ is the $(d-1)$-form $\ZZ_N$ background gauge field with even $N$, and $w_3$ is the 3rd Stiefel-Whitney class. The first equality follows from $w_3=Sq^1w_2$ on orientable manifolds, and $Sq^1B_{d-1}=dB_{d-1}/2$ where the right hand side uses a lift of $\mathbb{Z}_2$ cocycle $B_{d-1}$ to $\mathbb{Z}_4$ cochain.
    
    \item 
    For membrane statistics $(p=2)$ with the fusion group $G=\ZZ_N$, numerical computations suggest that the stable generalized statistics for $d\ge 7$ is classified by $\ZZ_{\gcd(2,N)}\times\ZZ_{\gcd(3,N)}$. We expect that the $\ZZ_2$ statistics with even $N$ is again associated with the mixed gravitational 't Hooft anomaly involving the Stiefel-Whitney class,
    \begin{align}
        \hspace{3em}
        \pi\int (w_4+w_2^2)\cup B_{d-2}=\pi\int Sq^4(B_{d-2})~.
    \end{align}
    Meanwhile, for the $\ZZ_3$ statistics with $N$ multiple of 3, we expect that the anomaly is associated with the Pontryagin class $p_1$
    \begin{align}
        \frac{2\pi}{3}\int p_1\cup B_{d-2}~.
    \end{align}
    This implies that the membrane excitation is chiral--it has chiral central charge $c_-=-8$.
    See also Ref.~\cite{Yang2024gapped} for a recent discussion on this response action.
    {\change
    This $\mathbb{Z}_{\gcd(3,N)}$ invariant already appears in $d=4$, with the simplest example given in Ref.~\cite{Tomonaga1965PontryaginMod3}: 
    \begin{equation}
        \frac{2\pi}{3}\int p_1\cup B_2 \;=\; \frac{2\pi}{3}\int B_2 \cup B_2 \cup B_2~.
    \end{equation}
    It would be interesting to verify the above expectations and find the corresponding generalized statistics on lattice models.}
\end{itemize}

\section{Statistics as Obstruction to Gauging}\label{sec:Anomaly}

Let us make a direct connection between the generalized statistics and the 't Hooft anomaly of the lattice models.
Suppose that a lattice model in $d$ spatial dimensions has a finite $(d-p-1)$-form symmetry $G$.
We argue that the invariant $\Theta$ gives an obstruction to gauging the global symmetry. This implies that $\Theta$ generally gives a microscopic definition of the 't Hooft anomaly.

For simplicity, let us study the system on a $d$-dimensional hypercubic lattice. 
We assume that the generators of the symmetry $g \in G$ are expressed as a product of local unitaries: 
\begin{align}
    U_g(\Sigma_{p+1}) = \prod_{\Delta_{p+1}\in\Sigma_{p+1}} U_{\Delta_{p+1},g}~,
\end{align}
where $U_{\Delta_{p+1},g}$ is a unitary supported on a $(p+1)$-cube $\Delta_{p+1}$. We do not require the operators $\{U_g(\Delta_{p+1})\}$ to commute with each other; the above product is understood as the action of a finite-depth circuit of local unitaries $\{U_g(\Delta_{p+1})\}$.

Let us briefly recall the procedure of gauging $G$ on the lattice. 
The first step is to enlarge the Hilbert space by adding $G$ gauge fields on the $p$-cubes. 
The next step is to impose the Gauss law constraints on the Hilbert space,
\begin{align}
    G_{\Delta_{p+1},g} = 1~,  
\end{align}
with
\begin{align}
     G_{\Delta_{p+1},g} = U_{\Delta_{p+1},g}\prod_{\Delta'\in \partial \Delta_{p+1}} (A_{g,\Delta'})^{\pm}~,
\end{align}
where $A_{g,\Delta'}$ is the operator generating the $g\in G$ gauge transformation on the gauge field located on the $p$-cube $\Delta'$, and the sign $\pm$ is determined by the outgoing/ingoing orientation of the hypercubes $\Delta'$. 
One can then express the symmetry operator as a product of Gauss law operators,
\begin{align}
\tilde U_g(\Sigma_{p+1})=\prod_{\Delta_{p+1}\in \Sigma_{p+1}} G_{\Delta_{p+1},g}~,
\end{align}
which turns on the $g\in G$ gauge field at the boundary of $\Sigma_{p+1}$. 
This operator $\tilde U_g(\Sigma_{p+1})$ is regarded as the symmetry operator $U_g(\Sigma_{p+1})$ coupled to the $G$ gauge fields.

{\change
One can then construct an excitation model (see Sec.~\ref{subsec:complex}) from the Gauss law operators.
Pick a simplicial complex $X$ embedded in space. Consider a set of operators $\tilde U_{g_j}(\sigma_{p+1})$ associated with a fixed set of generators $\{g_j\}$ of $G$, where $\sigma_{p+1}$ denotes a $(p+1)$-simplex of $X$. Each unitary is labeled by a pair $s=(\sigma_{p+1},g_j)$, and we define the set $\mathcal{S}=\{s\}$ consisting of all possible choices of $g_j$ and $\sigma_{p+1}\subset X$.
Next, consider the set of states $\mathcal{A}=\{\ket{a}\}$, representing $G$ gauge field configurations $a$, which are generated by sequences of unitaries $\tilde U(s)$ acting on the fixed $G$-symmetric state with vanishing gauge field $\ket{0}$.

The pair $(\mathcal{S},\mathcal{A})$ then forms the excitation model. 
We define an invariant as the sequence of unitaries
\begin{equation}
e^{i\Theta} = \bra{0}\prod \tilde U(s)^\pm\ket{0} = \bra{0}\prod U(s)^\pm\ket{0},
\end{equation}
which is an element of $E_{\mathrm{inv}}$.
By definition, $\Theta$ is the phase obtained from the product of the Gauss law operators $G_{\Delta_{p+1},g}$. 
Thus, $\Theta\neq 0$ indicates an obstruction to the commuting Gauss law operators $\{G_{\Delta_{p+1},g}\}$.
In particular, a given symmetric state $\ket{0}$ cannot be promoted to a gauge-invariant state, since $\ket{0}$ is annihilated when projected onto the Hilbert space satisfying all Gauss law constraints. Therefore, $\Theta$ defines an obstruction to gauging the $G$ symmetry of the given $G$-symmetric state $\ket{0}$.

In our formalism, the invariant $\Theta$ is defined by evaluating the sequence of unitaries on any initial state. Consequently, 
\begin{equation}
    e^{i\Theta} = \prod U(s)^\pm
\end{equation}
holds as an operator identity, and the invariant characterizes the obstruction to gauging the symmetry in the entire Hilbert space.
}

\section{Statistics as Obstruction to Short Range Entanglement: Dynamical Consequence of Anomalies}
\label{sec:SRE}
Here let us show that the generalized statistics in $T=E_{\mathrm{inv}}/E_{\mathrm{id}}$ define the obstructions to the short-range entanglement. That is, if the bosonic state $\ket{\Psi}$ preserves the global symmetry with the nontrivial invariant $\Theta\neq 0$ in $T$, then the state cannot be short-range entangled, $\ket{\Psi}\neq V(\ket{0}^n)$ for any choice of a finite depth circuit $V$ and a product state $\ket{0}^n$. Given that the invariant $\Theta$ can be regarded as a microscopic definition of the 't Hooft anomaly, the obstruction to the SRE state is regarded as a dynamical consequence of the 't Hooft anomaly.

We generally show this statement by checking that SRE states must carry trivial invariants. Suppose that a state $\ket{\Psi}=V(\ket{0}^n)$ and the symmetry operators $U$ has an invariant $\Theta$. We can redefine the symmetry operators by $U':= V^{-1}UV$ and the input state $\ket{\Psi'}=\ket{0}^n$ without changing the invariant. 
Hence we take the input state as a product state $\ket{\Psi}=\ket{0}^n$ without loss of generality. Below we simply write $U'$ as $U$.

The first step of the proof is to notice that the state with excitations $\ket{a} = U(s)\ket{\Psi}$ with $a=\partial s$ becomes a product state away from the location of the excitations.

We first show this statement when $U(s)$ generates the $k$-form symmetry with $k\ge 1$, i.e., the excitation has the dimension smaller than $d-1$.
In that case, let us separate the system into $A \cup A^c$, where the subsystem $A$ is the locus of the excitations $a$ and $A^c$ is its complement. Let us consider the projector $\Pi_{A^c}$ onto the product state $\ket{0}^n$ within the subsystem $A^c$.
One can write $\Pi_{A^c}$ as the product of local projectors within $A^c$, $\Pi_{A^c}=\prod_{L_j\subset A^c}\Pi_{L_j}$, where the set of local regions $\{L_j\}$ satisfies $\bigcup_{j}L_j=A^c$. Since the operator $U(s)$ can be topologically deformed on $\ket{\Psi}$ as long as $a=\partial s$ is fixed, one can set $s$ to avoid each region $L_j$ by the deformation.
This implies that $U(s)$ and $\Pi_{L_j}$ are commutative within the state $\ket{\Psi}$, so $\Pi_{L_j}U(s)\ket{\Psi} =U(s)\Pi_{L_j}\ket{\Psi}=U(s)\ket{\Psi}$. This implies that $\Pi_{A^c}U(s)\ket{\Psi}=U(s)\ket{\Psi}$, so the state $U(s)\ket{\Psi}$ is expressed in the form of
\begin{align}
    U(s)\ket{\Psi} = \ket{a}_{A} \otimes \ket{0}_{A^c}~,
    \label{eq: excitation and trivial}
\end{align}
where $\ket{0}_{A^c}$ is the product state at $A^c$, and $\ket{a}_{A}$ is the state whose Hilbert space is localized along the locations of excitations $A$. 

Then, let us show Eq.~\eqref{eq: excitation and trivial} when $U(s)$ is the 0-form symmetry. In that case, take a symmetry operator $U(\overline{s})$ with $-a=\partial \overline{s}$ so that the product $U(\overline{s}) U(s)$ becomes a closed symmetry operator that preserves the state $\ket{\Psi}$, $U(\overline{s}) U(s) \ket{\Psi}\propto \ket{\Psi}$. This operator is chosen so that $\text{supp}(s)$, $\text{supp}(\overline{s})$ do not have an overlap on their bulk. 
We again separate the system into $A\cup A^c$, where $A$ has a support near the boundary of $\text{supp}(s)$. $A^c$ is naturally separated into the two subsystems by $A^c = A^c_{\text{in}}\cup A^c_{\text{out}}$, where $A^c_{\text{in}}$ is supported inside  $\text{supp}(s)$, while $A^c_{\text{out}}$ is on the outside.
Let us again write the projector onto the product state as $\Pi_{A^c} =\Pi_{A_{\text{in}}^c}\Pi_{A_{\text{out}}^c}$, and each projector $\Pi_{A_{\text{in}}^c}$, $\Pi_{A_{\text{out}}^c}$ is given by the product of projectors at local regions $\{L^{\text{in}}_j\},\{L^{\text{out}}_j\}$. The projectors $\Pi_{L^{\text{out}}_j}$ obviously commutes with $U(s)$ since their support do not overlap. For $\Pi_{L^{\text{in}}_j}$ since we have
\begin{eqs}
    U(\overline{s})U(s) \Pi_{L^{\text{in}}_j}\ket{\Psi} &= \Pi_{L^{\text{in}}_j}U(\overline{s})U(s)\ket{\Psi} \\
    &= U(\overline{s})\Pi_{L^{\text{in}}_j} U(s)\ket{\Psi}~,
\end{eqs}
so $\Pi_{L^{\text{in}}_j}$ commutes with $U(s)$ on $\ket{\Psi}$, $U(s)\Pi_{L^{\text{in}}_j}\ket{\Psi} =\Pi_{L^{\text{in}}_j} U(s)\ket{\Psi}$. This implies that $\Pi_{A^c}U(s)\ket{\Psi}=U(s)\ket{\Psi}$, so we again obtain Eq.~\eqref{eq: excitation and trivial} for 0-form symmetry. 

{\change
We remark that the discussion here assumes a tensor-product Hilbert space. There also exist fermionic SPT phases with physical fermions, which require imposing constraints of the form $\delta E = w_2$ to specify a ``spin structure'' $E$ on the underlying manifold. Such phases necessarily involve Hilbert spaces that are not tensor-product spaces and therefore lie outside the scope of the present discussion.
}

\begin{figure*}[t]
    \centering
    \includegraphics[width=0.8\textwidth]{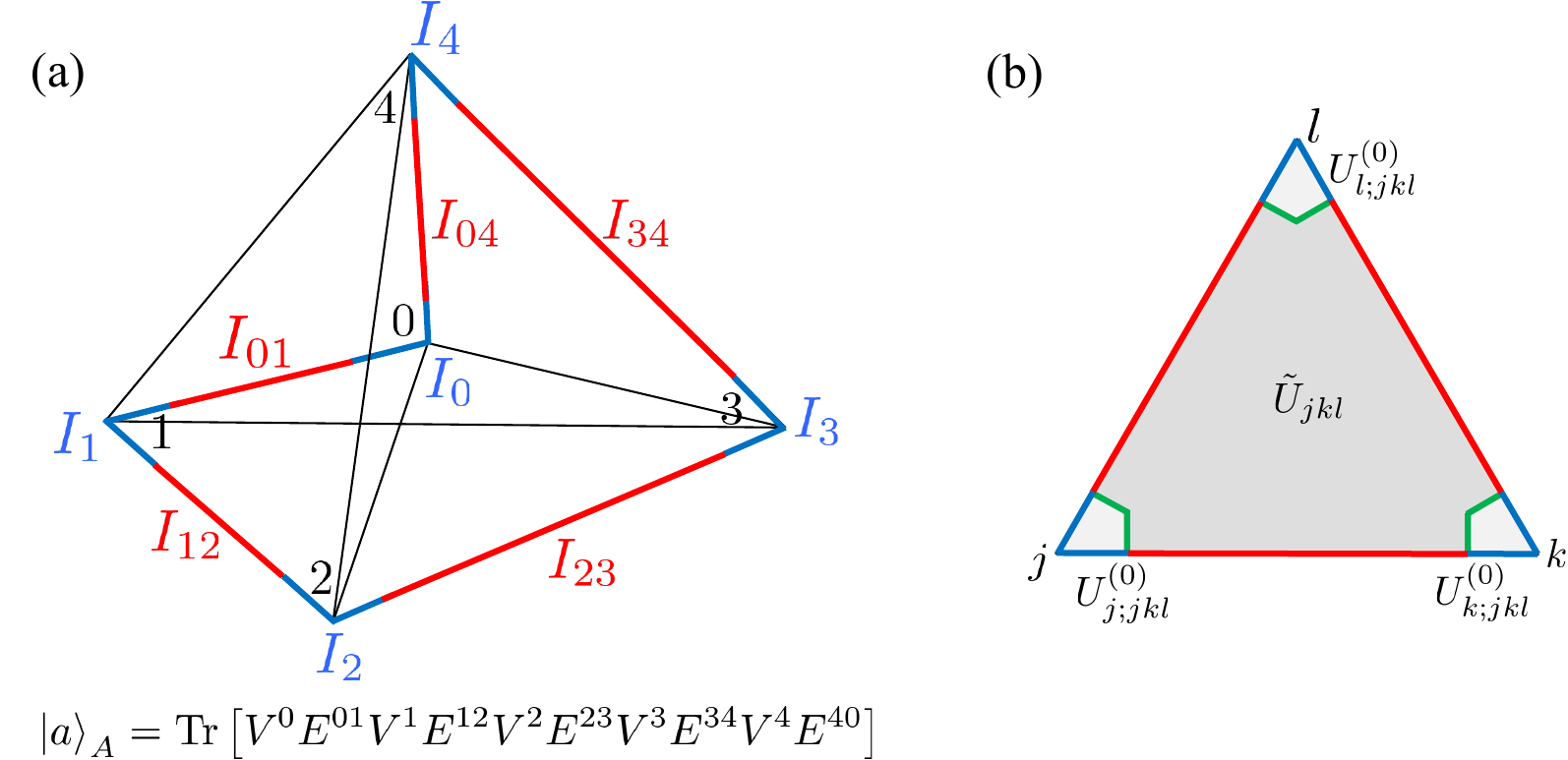}
    \caption{(a): The excited state $\ket{a}_A$ is expressed as the patchwork of MPS at the intervals $I_j, I_{jk}$. $I_j$ supports an MPS $\ket{V}^j$, $I_{jk}$ supports an MPS $\ket{E}^{jk}$.
    (b): The support of each unitary on the triangle $jkl$. $U^{(0)}_{j;jkl}$ acts nearby the vertex $j$ whose boundary contains $I_j$.}
    \label{fig:excitation_MPS}
\end{figure*}

\begin{widetext}
\subsection{Review: Anyons imply long-range entanglement}

Let us demonstrate the obstruction to SRE states when the excitations are particles with fusion rule $G=\ZZ_2$. This is a review of the result in Refs.~\cite{Bravyi2006bounds, aharonov2018quantumcircuitdepthlower}. See also Ref.~\cite{li2024entanglementneededemergentanyons} for recent discussions.
Due to the structure of excited states~\eqref{eq: excitation and trivial},\footnote{\change To be precise, Eq.~\eqref{eq: excitation and trivial} only states that the state $\ket{jk}$ has the form of $\ket{jk} = \ket{\psi}_{jk}\otimes \ket{0}_{\overline{j,k}}$, where $\ket{\psi}_{jk}$ is a possibly entangled state localized at two vertices $j,k$. Since a pair of 0-dimensional excitation is disconnected, we need an extra work to show that $\ket{\psi}_{jk}$ is factorized as $\ket{\psi}_{jk}=\ket{j}\otimes\ket{k}$. To see this, we consider a state of four Abelian anyons $\ket{jklm}$, which is expressed as either $\ket{jklm}= \ket{\psi}_{jk}\otimes \ket{\psi}_{lm}\otimes \ket{0}_{\overline{jklm}}$, or $\ket{jklm}= \ket{\psi}_{jm}\otimes \ket{\psi}_{kl}\otimes \ket{0}_{\overline{jklm}}$, depending on whether we use $U_{jk}U_{lm}$ or $U_{jm}U_{kl}$ to create them. We then have $\ket{\psi}_{jk}\otimes \ket{\psi}_{lm}=\ket{\psi}_{jm}\otimes \ket{\psi}_{kl}$, which implies that $\ket{\psi}_{jk}$ does not have entanglement between $j$ and $k$, hence $\ket{\psi}_{jk}=\ket{j}\otimes\ket{k}$.
See Ref.~\cite{li2024entanglementneededemergentanyons} for this discussion on general Abelian anyons.} each state $\ket{jk}$ with a pair of particle excitations at vertices $j$ and $k$ has the form of
\begin{align}
    \ket{jk} = \ket{j}\otimes \ket{k} \otimes \ket{0}_{\overline{j,k}}~,
\end{align}
where $\ket{j},\ket{k}$ are local states around the excitations, and $\ket{0}_{\overline{j,k}}$ is the product state on the complement.
This immediately implies that when $j,k,l,m$ are distinct positions in the space, we have $\bra{jl}U_{kj}\ket{kl} = \bra{jm}U_{kj}\ket{km}$. Hence
\begin{align}
    \theta(U_{kj},kl) = \theta(U_{kj},km)
\end{align}
for distinct $j,k,l,m$.
This immediately shows that the following invariant of the particles must be trivial,
\begin{align}
    \Theta=\;&\theta\Big( U_{02} U_{03}^{-1} U_{01} U_{02}^{-1} U_{03} U_{01}^{-1}, 12\Big)\notag\\
    =& -\theta(U_{01},02) + \theta(U_{03},02) - \theta(U_{02},03)
     +\theta(U_{01},03) - \theta(U_{03},01) + \theta(U_{02},01) = 0~.
\end{align}
Therefore, the SRE state cannot support Abelian anyons of nontrivial self-statistics.

\subsection{Example: Fermionic loops imply long-range entanglement}
Next, let us study the cases where the excitations are loops.
In a 3d space, we demonstrate that SRE states cannot support the fermionic loops with the nontrivial invariant
   \begin{eqs}
        e^{i\Theta} = ~& U_{014} U_{034} U_{023} U_{014}^{-1} U_{024}^{-1} U_{012} U_{023}^{-1} U_{013}^{-1} \\
        \times & U_{024} U_{014} U_{013} U_{024}^{-1} U_{034}^{-1} U_{023} U_{013}^{-1} U_{012}^{-1} \\
        \times & U_{034} U_{024} U_{012} U_{034}^{-1} U_{014}^{-1} U_{013} U_{012}^{-1} U_{023}^{-1}~,
    \end{eqs}
where $U_{jkl}$ is the generator of the $\ZZ_2$ 1-form symmetry is supported on the a triangle with vertices $j,k,l$.

\subsubsection{MPS representation of excitations in SRE states}

For SRE states, one can assume that the excited state $\ket{a}$ has an expression $\ket{a}=\ket{a}_A\otimes \ket{0}_{A^c}$ with the 1d state $\ket{a}_A$ localized at the position of excitations. 
One can generally express the 1d state $\ket{a}_A$ using the matrix product state (MPS) representation. 

Let us consider an MPS $\ket{a}_A$ on the 1d subsystem $A$, with the bipartition into $A= A_1\sqcup A_2$. Accordingly, the excitation $a$ allows a decomposition $a=a_1+a_2$. The state then has the expression
\begin{align}
    \ket{a}_A = \sum_{\mu,\nu} \ket{\psi}^{A_1}_{\mu\nu}\ket{\psi}_{\nu\mu}^{A_2}~,
\end{align}
where $\mu,\nu$ are bond indices.
Let us consider a symmetry operator $U(s)$ with $\partial s= -a_2+ a'_2$. That is, the unitary acts on the 2d subsystem $S$ that contains $A_2$ on its boundary, and transforms $\ket{a}$ into a state $\ket{a}'$ with excitations at $A'= A_1\sqcup A'_2$.
The unitary acts on the state $\ket{a}$ by
\begin{align}
    U(s)\ket{a} = \sum_{\mu,\nu} \ket{\psi}^{A_1}_{\mu\nu}\otimes U(s)\left[\ket{\psi}_{\nu\mu}^{A_2}\ket{0}^{S\cap A^c_2}\right]\otimes \ket{0}^{S^c}~.
\end{align}
Given that $U(s)\ket{a}$ again allows an expression $U(s)\ket{a}=\ket{a}_{{A}'}\otimes \ket{0}_{A'^c}$, the state $U(s)\left[\ket{\psi}_{\nu\mu}^{A_2}\ket{0}^{S\cap A^c_2}\right]$ has an expression
\begin{align}
U(s)\left[\ket{\psi}_{\nu\mu}^{A_2}\ket{0}^{S\cap A^c_2}\right] = \ket{\psi}_{\nu\mu}^{A'_2}\ket{0}^{S\cap {A'^{c}_2}}~.
\end{align}
That is, the operator $U(s)$ transforms the MPS $\ket{\psi}_{\nu\mu}^{A_2}$ into the new MPS $\ket{\psi}_{\nu\mu}^{A'_2}$ at the different location.

This motivates us to express the generic $\ket{a}_A$ in terms of the ``patchwork'' of MPS, see Fig.~\ref{fig:excitation_MPS}{\color{blue}(a)}. Each state $\ket{a}_A$ is expressed in terms of the following MPS of two types:
\begin{itemize}
    \item $\ket{V}^{j}$ is an MPS on a small interval $I_j\subset A$ that contains the vertex $j$. 
    \item $\ket{E}^{jk}$ is an MPS on an interval $I_{jk}\subset A$ contained in the edge $e=\langle jk \rangle$. $I_{jk}$ is adjacent to $I_j, I_k$. 
\end{itemize}
If the excitation $a$ has a support at the collection of edges $\langle jk\rangle, \langle kl\rangle,\dots$ that forms a closed loop, we have $A=I_j\sqcup I_{jk}\sqcup I_k\sqcup I_{kl}\sqcup I_l\dots$ and
the state $\ket{a}_{A}$ has an MPS representation as 
\begin{align}
    \ket{a}_A = \Tr\left[V^jE^{jk} V^k E^{kl} V^l \dots \right]~.
    \end{align}
The MPS $\{V^j, E^{jk}\}$ associated with the vertices and edges give the canonical choice of the states $\{\ket{a}\}$.

\subsubsection{Decomposition of phases}
\label{subsub:Uintopatch}
According to the expression of states $\ket{a}_A$ in terms of the patch of MPS, it is convenient to separate the symmetry generator into the operators supported near the corner vertices $j,k,l$ and the rest,
\begin{align}
    U_{jkl} = U^{(0)}_{j;jkl} U^{(0)}_{k,jkl} U^{(0)}_{l,jkl} \times \tilde{U}_{jkl}~,
\end{align}
where $U^{(0)}_{j;jkl}$ is a small defect operator acting on the region $I_j$ together with the subsystem of the surface $jkl$ near the vertex. $\tilde{U}_{jkl}$ acts on the intervals $I_{jk},I_{kl}, I_{jl}$ and the bulk of the surface $jkl$ except nearby the vertices.
See Fig.~\ref{fig:excitation_MPS} (b).

Consequently, the phase $U_{jkl}\ket{a} = \theta(U_{jkl},a)\ket{a'}$ admits an expression\footnote{To be precise, when we define the phases such as $\theta(U^{(0)}_{j;jkl},a), \theta(\tilde{U}_{jkl},a)$, we are fixing the choice of the intermediate states e.g., $\tilde U_{jkl}\ket{a}_A$ in terms of the patch of MPS. This amounts to fixing the MPS at the intermediate edges expressed as the green lines in Fig.~\ref{fig:excitation_MPS} (b).}
\begin{eqs}
    \theta(U_{jkl},a) = &~\theta(U^{(0)}_{j;jkl},a) + \theta(U^{(0)}_{k;jkl},a)
    + \theta(U^{(0)}_{l;jkl},a) + \theta(\tilde{U}_{jkl},a)~.
\end{eqs}
Given that each state is given in the form of the patch of MPS, the phases $\theta$ achieve the following properties:
\begin{itemize}
    \item The dependence of $\theta(\tilde{U}_{jkl},a)$ on $a$ is through $a$ restricted to edges $jk,kl,jl$. 
    In other words, $\theta(\tilde{U}_{jkl},a) = \theta(\tilde{U}_{jkl},a')$ as long as $a=a'$ on edges $jk,kl,jl$. 
    This is because this phase only depends on the MPS patch $E^{jk}, E^{kl}, E^{jl}$ (and the small MPS patch for the intermediate state $\tilde U_{jkl}\ket{a}$ connecting between $I_{jk},I_{kl},I_{jl}$, which is expressed as the green lines in Fig.~\ref{fig:excitation_MPS} (b)). 
    Note that this property does not hold for the original phase $\theta(U_{jkl},a)$ with the whole operator $U_{jkl}$, since this would depend on other edges that end at vertices $j,k$ or $l$ through the MPS patch $V^j,V^k,V^l$.
    \item The dependence of $\theta(U^{(0)}_{j;jkl},a)$ on $a$ is through $a$ restricted to edges ending at the vertex $j$. In other words, $\theta(U^{(0)}_{j;jkl},a) = \theta(U^{(0)}_{j;jkl},a')$ as long as $a=a'$ on edges ending at $j$. This is because this phase only depends on the MPS patch $V^j$ (and the choice of the surface $jkl$ where we act $U_{jkl}$, together with the small MPS patch for the intermediate state).
\end{itemize}
This allows us to write the above phases as $\theta(U,\{e\})$ where $\{e\}$ is the set of edges with excitations on which $\theta$ depends. We can then express the invariant as
\begin{eqs}
    \Theta =
    &- \theta(U^{(0)}_{0;023},03,02) - \theta(U^{(0)}_{3;023},03,34) - \theta(U^{(0)}_{2;023},02,12) - \theta(\tilde{U}_{023},03,02)  \\
    &- \theta(U^{(0)}_{0;012},01,03) - \theta(U^{(0)}_{1;012},01,14) - \theta(U^{(0)}_{2;012}) - \theta(\tilde{U}_{012},01) \\
    &+ \theta(U^{(0)}_{0;013},03,01) + \theta(U^{(0)}_{3;013},03,34) + \theta(U^{(0)}_{1;013},01,14) + \theta(\tilde{U}_{013},03,01)  \\
    &- \theta(U^{(0)}_{0;014},01,04) - \theta(U^{(0)}_{1;014},01,13) - \theta(U^{(0)}_{4;014},04,34) - \theta(\tilde{U}_{014},01,04) \\
    &- \theta(U^{(0)}_{0;034},01,03) - \theta(U^{(0)}_{3;034},03,13) - \theta(U^{(0)}_{4;034}) - \theta(\tilde{U}_{034},03)  \\
    &+ \theta(U^{(0)}_{0;012},01,03) + \theta(U^{(0)}_{1;012},01,13) + \theta(U^{(0)}_{2;012}) + \theta(\tilde{U}_{012},01) \\
    &+ \theta(U^{(0)}_{0;024},02,03) + \theta(U^{(0)}_{2;024},02,12) + \theta(U^{(0)}_{4;024}) + \theta(\tilde{U}_{024},02)  \\
    &+ \theta(U^{(0)}_{0;034},03,04) + \theta(U^{(0)}_{3;034},03,13) + \theta(U^{(0)}_{4;034},04,24) + \theta(\tilde{U}_{034},03,04) \\
    &- \theta(U^{(0)}_{0;012},02,01) - \theta(U^{(0)}_{2;012},02,24) - \theta(U^{(0)}_{1;012},01,31) - \theta(\tilde{U}_{012},02,01) \\ 
    &- \theta(U^{(0)}_{0;013},03,02) - \theta(U^{(0)}_{3;013},03,34) - \theta(U^{(0)}_{1;013}) - \theta(\tilde{U}_{013},03) \\
    &+ \theta(U^{(0)}_{0;023},02,03) + \theta(U^{(0)}_{2;023},02,24) + \theta(U^{(0)}_{3;023},03,34) + \theta(\tilde{U}_{023},02,03) \\
    &- \theta(U^{(0)}_{0;034},03,04) - \theta(U^{(0)}_{3;034},03,32) - \theta(U^{(0)}_{4;034},04,24) - \theta(\tilde{U}_{034},03,04) \\
    &- \theta(U^{(0)}_{0;024},03,02) - \theta(U^{(0)}_{2;024},02,32) - \theta(U^{(0)}_{4;024}) - \theta(\tilde{U}_{024},02)  \\
    &+ \theta(U^{(0)}_{0;013},03,02) + \theta(U^{(0)}_{3;013},03,32) + \theta(U^{(0)}_{1;013}) + \theta(\tilde{U}_{013},03) \\
    &+ \theta(U^{(0)}_{0;014},01,02) + \theta(U^{(0)}_{1;014},01,31) + \theta(U^{(0)}_{4;014}) + \theta(\tilde{U}_{014},01)  \\
    &+ \theta(U^{(0)}_{0;024},02,04) + \theta(U^{(0)}_{2;024},02,32) + \theta(U^{(0)}_{4;024},04,14) + \theta(\tilde{U}_{024},02,04) \\
    &- \theta(U^{(0)}_{0;013},01,03) - \theta(U^{(0)}_{1;013},01,14) - \theta(U^{(0)}_{3;013},03,23) - \theta(\tilde{U}_{013},01,03)  \\
    &- \theta(U^{(0)}_{0;023},02,01) - \theta(U^{(0)}_{2;023},02,24) - \theta(U^{(0)}_{3;023}) - \theta(\tilde{U}_{023},02) \\
    &+ \theta(U^{(0)}_{0;012},01,02) + \theta(U^{(0)}_{1;012},01,14) + \theta(U^{(0)}_{2;012},02,24) + \theta(\tilde{U}_{012},01,02)  \\
    &- \theta(U^{(0)}_{0;024},02,04) - \theta(U^{(0)}_{2;024},02,21) - \theta(U^{(0)}_{4;024},04,14) - \theta(\tilde{U}_{024},02,04) \\
    &- \theta(U^{(0)}_{0;014},02,01) - \theta(U^{(0)}_{1;014},01,21) - \theta(U^{(0)}_{4;014}) - \theta(\tilde{U}_{014},01)  \\
    &+ \theta(U^{(0)}_{0;023},02,01) + \theta(U^{(0)}_{2;023},02,21) + \theta(U^{(0)}_{3;023}) + \theta(\tilde{U}_{023},02) \\
    &+ \theta(U^{(0)}_{0;034},03,01) + \theta(U^{(0)}_{3;034},03,23) + \theta(U^{(0)}_{4;034}) + \theta(\tilde{U}_{034},03) \\
    &+ \theta(U^{(0)}_{0;014},01,04) + \theta(U^{(0)}_{1;014},01,21) + \theta(U^{(0)}_{4;014},04,34) + \theta(\tilde{U}_{014},01,04) ~.
\end{eqs}
One can explicitly check that the above phases cancel out. Therefore, we get $\Theta=0$, and the SRE state cannot support fermionic loops.
\end{widetext}

\subsection{Nontrivial statistics imply long-range entanglement}

One can extend the above discussion to the generic statistics of $p$-dimensional excitations embedded in $d$ spatial dimensions. We assume that the excitation model is defined on a certain simplicial complex $X$ embedded in space.
For SRE states, the excited state can be represented by a state $\ket{a}_A$ localized at the position $A$ of the $p$-dimensional excitations. 

\subsubsection{Patchwork of tensor network states at the excitations}

As a generalization of the MPS patchwork state for loop excitations, we assume that $\ket{a}_A$ admits an expression in terms of a tensor network state (TNS) at $A$. Note that $A$ is a locus of a simplicial complex, so it can include, for example, a junction of hypersurfaces and is generally not a manifold.

This assumption leads to the canonical choice of the state $\ket{a}_A$ using a patch of TNS, where each patch is defined in the vicinity of a simplex.
To define the support of each TNS patch, we introduce a locus for each simplex $\sigma_j$ of $X$ in space. Let $D_j(\sigma_j)$
denote the locus of the $j$-simplex $\sigma_j$ with radius $r_j$; this is given by $\sigma_j\times D^{d-j}$ in $d$-dimensional space, where $D^{d-j}$ is a $(d-j)$-ball of radius $r_j$, and its center point $\sigma_j\times \{0\}$ corresponds to $\sigma_j$. 
See Fig.~\ref{fig:TNSpatch} (a).
We also define the collection of these regions as $D_j= \bigcup_{\sigma_j\in\mathcal{T}} D_j(\sigma_j)$ for the $j$-simplices $\{\sigma_j\}$. For convenience, we set the radii $r_j$ to satisfy $r_{j-1} > \alpha r_{j}$ with a sufficiently large constant $\alpha \gg 1$. In addition, we require that the radii $\{r_j\}$ are large enough so that $A\subset \bigcup_{0\le j\le p} D_j$.

We then define a TNS patch $T^A_j(\sigma_j,a)$ with $0\le j\le p$, supported on $A\cap D_j(\sigma_j)\cap D^c_{j-1}\cap\dots\cap D^c_0$. This is the tensor network state supported on a single $j$-simplex $\sigma_j$, excluding the vicinity of its boundaries. This TNS has bond indices on the boundary that are contracted with other TNS patches at adjacent simplices. We represent the state $\ket{a}_A$ as
\begin{align}
    \ket{a}_A = \prod_{0\le j\le p}\prod_{\sigma_j\in X} T^A_j(\sigma_j,a)~,
\end{align}
where the product denotes the contraction of bond indices.
The conditions $r_{j-1}>\alpha r_j$ ensure that the supports of each pair of TNS patches $T_j^A$ do not overlap.

\subsubsection{Decomposition of phases}

Each $(p+1)$-simplex $\sigma_{p+1}\in X$ can support a unitary $U(S(\sigma_{p+1}))$ that generates the global symmetry $G$. Here $S(\sigma_{p+1})$ denotes a neighborhood of the simplex $\sigma_{p+1}$ that supports the unitary. We require $S(\sigma_{p+1})\subset D_{p+1}(\sigma_{p+1})\cup\left(\bigcup_{0\le j\le p}D_j\right)$.

\begin{figure*}[t]
    \centering
    \includegraphics[width=0.7 \textwidth]{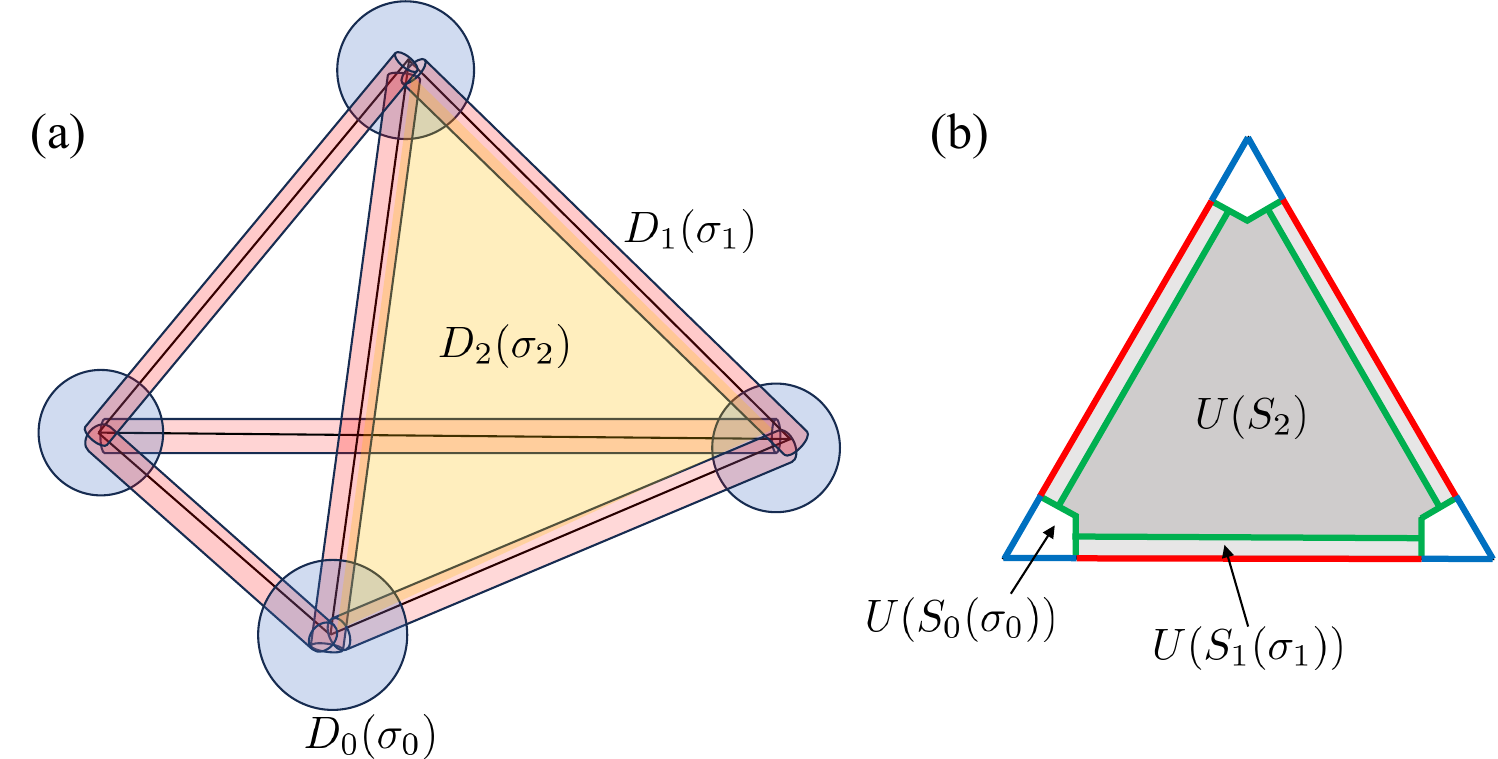}
    \caption{(a): The disk locus of each simplex is denoted by $D_j(\sigma_j)$. Each TNS patch for the state $\ket{a}_A$ corresponds to extracting the TNS state inside each locus of $\sigma_j$ with the locus of its boundary simplexes excluded. (b): The decomposition of the unitary operator in the case of $p=1$.}
    \label{fig:TNSpatch}
\end{figure*}

Following the argument in Sec.~\ref{subsub:Uintopatch}, we decompose the unitary $U(S(\sigma_{p+1}))$ into a sequence of smaller unitaries compatible with the TNS patch structure. That is, we decompose the support of the unitary as $S(\sigma_{p+1})= S_{p+1}\sqcup S_{p}\sqcup S_{p-1}\sqcup\dots\sqcup S_0$, with
\begin{align}
\begin{cases}
S_{p+1} = S \cap D_{p+1}(\sigma_{p+1}) \cap D_{p}^c\cap\dots \cap D_0^c & \\
S_{j} = S\cap D_j \cap D^c_{j-1}\cap \dots\cap D_0^c & \hspace{-3em} \text{for } 0\le j\le p~. \nonumber
\end{cases}
\end{align}
Note that $S_{j}$ with $0\le j\le p$ consists of disconnected components near each $j$-simplex of $\sigma_{p+1}$ as
\begin{align}
    S_j= \bigsqcup_{\sigma_j\subset \sigma_{p+1}} S_j(\sigma_j)~.
\end{align}
We then rewrite the unitary $U(S(\sigma_{p+1}))$ as the sequence
\begin{widetext}
    \begin{align}
        U(S(\sigma_{p+1})) = \left(\prod_{\sigma_{0}\subset \sigma_{p+1}} U(S_{0}(\sigma_{0}))\right) \times\dots\times\left(\prod_{\sigma_{p}\subset \sigma_{p+1}} U(S_{p}(\sigma_{p}))\right) \times U(S_{p+1})~.
    \end{align}
    See Fig.~\ref{fig:TNSpatch} (b). Each unitary $U(S_j(\sigma_j))$ acts on a single TNS patch on the $j$-simplex $\sigma_j$ with $0\le j\le p$.
    Accordingly, the phase $U(S(\sigma_{p+1}))\ket{a}= \theta(U(S(\sigma_{p+1})),a)\ket{a'}$ can be expressed as
    \begin{align}
        \theta(U(S(\sigma_{p+1})),a) = \left(\sum_{\sigma_0\subset \sigma_{p+1}} \theta(U(S_0(\sigma_0)),a)\right) + \dots + \left( \sum_{\sigma_{p}\subset \sigma_{p+1}} \theta(U(S_{p}(\sigma_{p})),a)\right) + \theta(U(S_{p+1}))~.
        \label{eq:decomposition_phases_TNS}
    \end{align}
\end{widetext}
Given that each small unitary $U(S_j(\sigma_j))$ acts within a single TNS patch, the phases $\theta$ satisfy the following properties:
\begin{itemize}
    \item The dependence of $\theta(U(S_j(\sigma_j)),a)$ (with $0\le j\le p$) on $a$ arises only through $a$ restricted to the set of $p$-simplices $\{\sigma_p\}$ satisfying $\sigma_{j}\subset \sigma_p$. 
    This is because the action of the unitary $U(S_j(\sigma_j))$ on $A$ is confined to the TNS patch $T_j^A(\sigma_j,a)$. 
    $\theta(U(S_j(\sigma_j)),a)$ also depends on the choice of unitary $U(\sigma_{p+1})$, namely the choice of $\sigma_{p+1}$ together with the group element $g\in G$ that corresponds to $U$.    
    \item $\theta(U(S_{p+1}))$ does not depend on $a$, but only on the choice of $U(S_{p+1})$. Note that this creates the intermediate tensor network state on the boundary sphere of $S_{p+1}$ inside the $(p+1)$-simplex. 
\end{itemize}

\subsubsection{Trivial invariants in SRE states}

Suppose that the sequence of unitaries $\bra{a}\prod U(S) \ket{a}$ 
evaluated on a given excited state becomes the sum of phases $\theta$ which gives the element of $E_{\mathrm{inv}}$,
\begin{align}
    \Theta = \sum_{U(S(\sigma_{p+1})), a} \epsilon(U(S(\sigma_{p+1})), a) \times \theta(U(S(\sigma_{p+1})),a)~,
\end{align}
with $\epsilon((U(S(\sigma_{p+1})), a))\in \ZZ$.
Recall that this phase $\Theta\in E_{\mathrm{inv}}$ satisfies the Eqs.~\eqref{eq:inv2} and \eqref{eq:inv3} given by
\begin{itemize}
    \item   The coefficients $\epsilon$ satisfy 
    \begin{align}
        \sum_{a}\epsilon(U(S(\sigma_{p+1})), a) = 0~,
    \label{eq:invariance1}
    \end{align}
    for any choice of $U(S(\sigma_{p+1}))$.

    \item The coefficients $\epsilon$ satisfy
    \begin{align}
        \sum_{\substack{a \\ a|_{\sigma_j}= a_*}} 
    \epsilon(U(S(\sigma_{p+1})), a) = 0~,
    \label{eq:invariance2}
    \end{align}
    with any choice of $U(S(\sigma_{p+1}))$ and a cochain $a_*$. 
\end{itemize}

With this in mind, one can show that such $\Theta$ satisfying the invariance condition must have $\Theta=0$ in the SRE state. 
Let us decompose the phases using Eq.~\eqref{eq:decomposition_phases_TNS} valid for the TNS patchwork state. In the sum of $\Theta$, let us extract the part involving a fixed $U(S(\sigma_{p+1}))$, and then further pick the part of $\theta$ involving $S_j(\sigma_j)$ contained in $S(\sigma_{p+1})$ with $0\le j\le p$. The extracted sum has the form of
\begin{align}
\begin{split}
    &\sum_a \epsilon(U(S(\sigma_{p+1})), a) \times \theta(U(S_j(\sigma_{j})),a) \\
    = & \sum_{a_*} \sum_{\substack{a \\ a|_{\sigma_j}= a_*}} \epsilon(U(S(\sigma_{p+1})), a) \times \theta(U(S_j(\sigma_{j})),a)~.
\end{split}
\end{align}
Here, recall that the phase $\theta(U(S_j(\sigma_{j})),a)$ depends on $a$ only through the reduced one $a|_{\sigma_j}$ which is fixed in the second sum. We then have
\begin{align}
\begin{split}
    &\sum_a \epsilon(U(S(\sigma_{p+1})), a) \times \theta(U(S_j(\sigma_{j})),a) \\
    = & \sum_{a_*} \left(\sum_{\substack{a \\ a|_{\sigma_j}= a_*}} \epsilon(U(S(\sigma_{p+1})), a) \right)  \theta(U(S_j(\sigma_{j})),a) = 0~.
\end{split}
\end{align}
So, the extracted sum with each choice of $U(S(\sigma_{p+1}))$ and $S_j(\sigma_j)$ in $\Theta$ vanishes due to Eq.~\eqref{eq:invariance2}. 

Finally, one can also verify that the phases involving $\theta(U(S_{p+1}))$ also vanishes. The part of $\Theta$ that contains  $\theta(U(S_{p+1}))$ is given by
\begin{align}
\begin{split}
    &\sum_a \epsilon(U(S(\sigma_{p+1})), a) \times \theta(U(S_{p+1})) \\
    = & \left(\sum_{a} \epsilon(U(S(\sigma_{p+1})), a)
    \right) \theta(U(S_{p+1})) = 0~,
\end{split}
\end{align}
where we used Eq.~\eqref{eq:invariance1}.
Therefore, we get $\Theta=0$ in the SRE state.

\section{Discussions}
\label{sec:discussions}

In this paper, we established a universal microscopic description for the statistics of excitations in generic spacetime dimensions. The invariants are expressed as the Berry phase associated with families of excited states, transformed by sequences of unitaries that generate finite global symmetries and move the excitations. These invariants are generally quantized into discrete values, characterizing the generalized statistics of excitations in microscopic lattice models. For instance, this framework leads to the quantization of spins for Abelian anyons, which are microscopically defined via T-junctions. These invariants can be computed using algorithms with inputs such as the symmetry group and the configurations of excitations. This approach allows us to identify new invariants in microscopic lattice models, as well as a simplified expression for the invariants of fermionic loops in (3+1)D.
These invariants can naturally be interpreted as obstructions to gauging the finite global symmetry $G$ in microscopic lattice models, providing a microscopic definition of 't Hooft anomalies. We show that these anomalies have dynamical consequences, as nontrivial invariants forbid the existence of short-range entangled states.

We close this paper with a number of future directions.
One immediate direction for future work is to extend our framework of generalized statistics to include mixtures of $(d-p-1)$-form symmetries with distinct degrees $p$, which generally form higher-group symmetries. Even the simplest gapped phases, such as the $\mathbb{Z}_2$ toric code, exhibit a rich structure of higher-group symmetries with 't Hooft anomalies~\cite{Kapustin:2017jrc, Barkeshli2023Codimension, Barkeshli2024higher, Barkeshli2024higherfermion}.
{\change
For example, in $(3+1)$D, string operators that create particle excitations may coexist with membrane operators that create loop excitations, and these excitations can exhibit nontrivial interplay; for instance, the intersection of loops may generate an additional particle. Determining the invariants associated with such higher-group structures in topologically ordered phases is an interesting direction for future work.
}

While we mainly studied the finite invertible symmetries in this paper, there are other important class of symmetries which requires further study, such as the continuous symmetries. For instance, the (2+1)D gapped phases with continuous symmetries can exhibit electric Hall conductance, which is not associated with the 't Hooft anomaly of global symmetry in a usual sense. It would be interesting to see if the Hall conductance admits a similar microscopic characterization to our current paper, e.g., through the spin of the vortex of $U(1)$ symmetry. This would help clarify if the nontrivial Hall conductance gives obstructions to gauging continuous symmetries.

The other important class of symmetries not studied in this paper is the non-invertible symmetries, which includes the non-Abelian anyons. 
The microscopic definitions of the anyon data of the non-Abelian topological order in (2+1)D has been discussed in Ref.~\cite{Kawagoe2020Microscopic}. Given that the spins of Abelian anyons are shown to be quantized using the locality identities, it would be interesting to see if the anyon data of non-Abelian anyon systems are also quantized due to the locality of topological operators. 

{\change
The ``Borromean ring'' in (2+1)D topological order is intrinsic to the non-Abelian anyons including the $D_8$ gauge theory, which is beyond the scope of our setup. 
Also, the three-loop braiding of loop excitations in (3+1)D is expected to require either non-invertible loop excitations or a nontrivial higher-group structure involving Abelian loop and particle excitations. Both scenarios lie beyond the scope of the present work, but it would be interesting to characterize the corresponding invariant for three-loop braiding.  

Recent developments on non-invertible symmetries have revealed a variety of new invariants associated with their underlying algebraic structures. For example, defects of non-invertible symmetries can carry invariants such as the Frobenius–Schur indicator and its generalizations~\cite{thorngren2019fusioncategorysymmetryi, Zhang:2023wlu, Cordova:2023bja}. It would be intriguing to investigate how such invariants can be realized through sequences of symmetry operators. Non-invertible symmetries also give rise to other invariants, including three-loop braiding of vortices in (3+1)D topological order~\cite{Wang2014braiding}, and generalizations of Hall conductance associated with continuous non-invertible symmetries~\cite{Hsin2024coset}. Exploring the characterization of these invariants remains an important direction for future work.

While the symmetries considered in this work are assumed to be implemented by finite-depth circuits, many lattice symmetries cannot be expressed in this form. Examples include crystalline symmetries such as lattice translations, and more generally symmetries generated by quantum cellular automata (QCA)~\cite{Po2016chiral, Gross2012index, Haah:2018jdf, Zhang2024Higher, Fidkowski2024Pumping, fidkowski2024qca}. It would be interesting to investigate how the corresponding invariants can be characterized for such symmetries.
}

Several open issues have arisen from our work, which deserve further study.
The genuine Berry phase invariants $T$ are conjectured to correspond to the cohomology of the Eilenberg–MacLane space $H^{d+2}(B^{d-p}G, U(1))$ classifying the anomalies of higher-form symmetries. It would be desirable to prove this correspondence using the explicit model of the Eilenberg–MacLane space. {\change A related formal mathematical treatment appears in Ref.~\cite{xue2025statisticsinvertibletopologicalexcitations}.}

Also, while the invariants in this paper were defined with an explicit reference to a specific state in the Hilbert space, but it is expected that the anomalies can be characterized solely by the algebra of symmetry operators without explicit reference to the states~\cite{Else2014Classifying}. It would be interesting to see if our invariants are promoted to operator equations independent of the choice of any states in the Hilbert space.

\section*{Acknowledgements}
Y.-A.C wants to thank Qing-Rui Wang and Meng Cheng for sharing their unpublished note~\cite{WangCheng}, and for the insightful discussions that motivated our Conjecture~\ref{conjecture: EM cohomology}.
Y.-A.C. also thanks Yitao Feng for inspiring the construction of the anomalous symmetries that demonstrate nontrivial generalized statistics.  
We are grateful to Andreas Bauer, Tyler D. Ellison, Dominic Else, Anton Kapustin, Kyle Kawagoe, Sahand Seifnashri, Wilbur Shirley, Nikita Sopenko, Nathanan Tantivasadakarn, Bowen Yang, Peng Ye, and Carolyn Zhang for valuable discussions and feedback.

Y.-A.C. is supported by the National Natural Science Foundation of China (Grant No.~12474491), and the Fundamental Research Funds for the Central Universities, Peking University.
R.K. is supported by the U.S. Department of Energy (Grant No.~DE-SC0009988) and the Sivian Fund. 
P.-S.H. is supported by the Department of Mathematics at King’s College
London. 

\appendix

\begin{widetext}

{\change
\section{Anomalous symmetries and generalized statistics}

In this appendix, we present several concrete examples exhibiting nontrivial generalized statistics.  For each (anomalous) global symmetry, we construct the operators that generate codimension-1 domain wall excitations.  The nontrivial fusion statistics of these domain walls serve as indicators of the underlying symmetry anomaly.  To our knowledge, the constructions of membrane operators for the global $\mathbb{Z}_2\times\mathbb{Z}_2$ symmetry in two dimensions and of volume operators for the global $\mathbb{Z}_2$ symmetry in three dimensions are novel.  Consequently, these generalized statistics provide an effective diagnostic for distinguishing anomalous symmetries.

\subsection{Anomalous 0-form $\mathbb{Z}_2 \times \mathbb{Z}_2$ in two spatial dimensions}\label{sec: 2d anomalous Z2xZ2}

As an explicit illustration of the loop fusion statistics in Eq.~\eqref{eq: ABCD loop fusion process}, we consider a two‐dimensional square lattice with an anomalous global $\mathbb{Z}_2\times\mathbb{Z}_2$ symmetry. 
At each vertex $v$, there are two qubits with Pauli operators $X^a_v, Z^a_v$ and $X^b_v, Z^b_v$. For convenience, we label the eigenvalues of $Z^a_v$ and $Z^b_v$ by elements of $\ZZ_2=\{0,1\}$ via
\begin{equation}
    Z^a_v := (-1)^{a_v}, \quad Z^b_v := (-1)^{b_v}.
\end{equation}
The anomalous $\ZZ_2 \times \ZZ_2$ symmetry is generated by
\begin{eqs}
    S^a &:= \prod_v X^a_v \prod_{f=\Box_{1234}} \left( \vcenter{\hbox{\includegraphics[scale=.4]{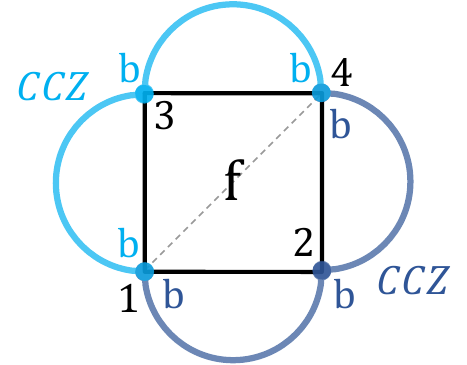}}} \right)
          = \prod_v X^a_v \prod_{f=\Box_{1234}} (-1)^{b_1 b_2 b_4 + b_1 b_3 b_4},\\
    S^b &:= \prod_v X^b_v,
\label{eq: 2d anomalous Z2 x Z2 symmetry in app}
\end{eqs}
where the $CCZ$ term represents applying a controlled–controlled–$Z$ gate on the vertices of each triangle of a chosen triangulation of the square lattice.
Focusing on the symmetric operators, we first define the domain wall operators as
\begin{equation}
    W(a)_{\langle ij\rangle} := Z^a_i Z^a_j, \quad
    W(b)_{\langle ij\rangle} := Z^b_i Z^b_j.
\end{equation}
Consider the square patch
{\vspace{-1em}
\begin{equation}
    \vspace{-1em}
    \vcenter{\hbox{\includegraphics[scale=.4]{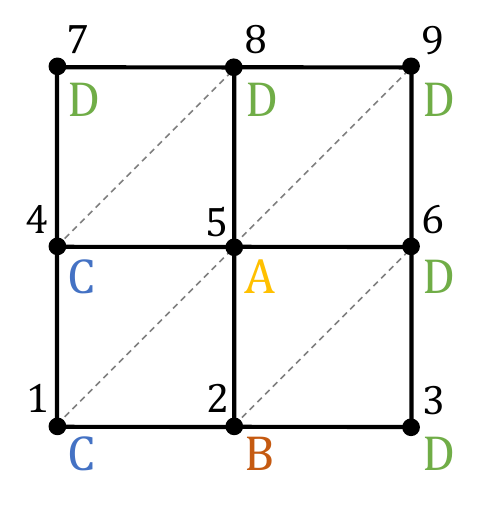}}}, 
\label{eq: 2x2_square}
\end{equation}}
the membrane operator that creates a closed loop domain wall around vertex $5$ is
\begin{eqs}
    U(a)_5 = X^a_5, \quad U(b)_5 = X^b_5\,(-1)^{\gamma_5(a,b)},
\label{eq: anomalous membrane operator in (2+1)D}
\end{eqs}
with
\begin{eqs}
    \gamma_5(a,b) :=&~ a_5 (b_1 b_2 + b_2 b_6 + b_6 b_9 + b_1 b_4 + b_4 b_8 + b_8 b_9).
\end{eqs}
Here, each product $a_5 b_i b_j$ represents a CCZ gate acting on the three qubits at vertices $\{5,i,j\}$ of each triangle adjacent to vertex 5.
Membrane operators at other vertices are defined by translation; they create loop excitations around each vertex and commute with the global symmetry in Eq.~\eqref{eq: 2d anomalous Z2 x Z2 symmetry in app}.

We compute their statistics $Z_4^{II}(a, b)$ in Eq.~\eqref{eq: ABCD loop fusion process}. Choosing region $A=\{5\}$, $B=\{2\}$, $C=\{1,4\}$, and $D$ as the set of all other vertices (including those outside the displayed patch), we start with
\begin{equation}
    U(a)_{A+B+C+D} = \prod_v X^a_v, \quad U(b)_A = X^b_5\,(-1)^{\gamma_5(a,b)},
\end{equation}
and compute their commutator
\begin{equation}
    [U(b)_A, U(a)_{A+B+C+D}]
      = (-1)^{b_1 b_2 + b_2 b_6 + b_6 b_9 + b_1 b_4 + b_4 b_8 + b_8 b_9}.
\end{equation}
Next, with $U(b)_B = X^b_2\,(-1)^{\gamma_2(a,b)}$, we find
\begin{equation}
    [U(b)_B, [U(b)_A, U(a)_{A+B+C+D}]]
      = (-1)^{b_1 + b_6} = Z^b_1 Z^b_6.
\end{equation}
Finally, since $U(b)_{B+C}$ contains $X^b_1 X^b_2 X^b_4$ up to a phase that commutes with $Z^b_1 Z^b_6$, only the anti-commutation between $X^b_1$ and $Z^b_1$ contributes to the final statistics:
\begin{equation}
    Z_4^{II}(a, b) 
      = (U(b)_{B+C})^{-2}
        \Bigl(U(b)_{B+C}\,[U(b)_B, [U(b)_A, U(a)_{A+B+C+D}]]\Bigr)^2
      = -1.
\end{equation}
This nontrivial loop fusion statistic confirms that the $\ZZ_2 \times \ZZ_2$ symmetry is anomalous.

\subsection{Anomalous 0-form $\mathbb{Z}_2$ in three spatial dimensions}\label{sec: 3d anomalous Z2}

In this section, we give an example exhibiting nontrivial membrane fusion statistics $Z_5(g)$ defined in Eq.~\eqref{eq: ABCD membrane fusion process}.
Consider a cubic lattice with one qubit at each vertex. As before, we define
\begin{equation}
Z_v := (-1)^{a_v}.
\end{equation}
The anomalous $\mathbb{Z}_2$ global symmetry on this lattice is generated by
\begin{eqs}
    S :=& \prod_v X_v \prod_{c=\Box_{12345678}} \left( \vcenter{\hbox{\includegraphics[scale=.45]{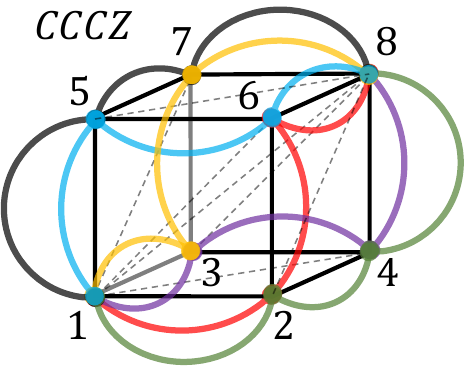}}} \right) \\
    =& \prod_v X_v \prod_{c=\Box_{12345678}} (-1)^{a_1 a_2 a_4 a_8 + a_1 a_2 a_6 a_8 + a_1 a_3 a_4 a_8 + a_1 a_3 a_7 a_8 + a_1 a_5 a_6 a_8 + a_1 a_5 a_7 a_8},
    \label{eq: 3d anomalous Z2 symmetry in app}
\end{eqs}
where $CCCZ$ denotes the controlled–controlled–controlled–$Z$ gate. It is more transparent to visualize this symmetry on a triangulated three–dimensional manifold, where the $CCCZ$ simply acts on the four vertices of each tetrahedron. To adapt this construction to the cubic lattice, we have chosen a subdivision of the cube into tetrahedra.

Similarly, we introduce the symmetric domain wall operators on each edge $\lr{ij}$
\begin{equation}
    W_{\lr{ij}} = Z_i Z_j.
\end{equation}
On the cube patch
\vspace{-0.5em}
\begin{equation}
    \vspace{-1em}
    \vcenter{\hbox{\includegraphics[scale=.4]{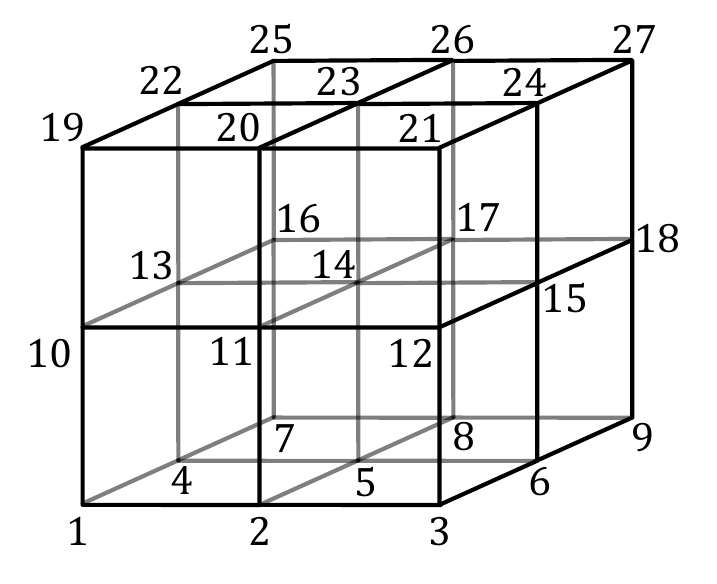}}}, 
\label{eq: 2x2_cube}
\end{equation}
the volume operator that creates a closed–membrane domain wall can be written as
\begin{eqs}
    U_{14}:=X_{14} (-1)^{\gamma_{14}},
\label{eq: Z2 volume operator}
\end{eqs}
where $\gamma_{14}$ collects the phases from all $CCCZ$ gates acting on tetrahedra containing vertex $14$:
\begin{eqs}
    \gamma_{14} =& ~(a_{1} a_{2} a_{5} a_{14} + a_{1} a_{2} a_{11} a_{14} + a_{1} a_{4} a_{5} a_{14} 
    + a_{1} a_{4} a_{13} a_{14} + a_{1} a_{10} a_{11} a_{14} + a_{1} a_{10} a_{13} a_{14} ) \\
    & + (a_{14} a_{15} a_{18} a_{27} + a_{14} a_{15} a_{24} a_{27} + a_{14} a_{17} a_{18} a_{27}
    + a_{14} a_{17} a_{26} a_{27} + a_{14} a_{23} a_{24} a_{27} + a_{14} a_{23} a_{26} a_{27}) \\
    & + (a_{2} a_{5} a_{14} a_{15} + a_{2} a_{11} a_{14} a_{15} ) + (a_{4} a_{5} a_{14} a_{17} + a_{4} a_{13} a_{14} a_{17} )
    + (a_{5} a_{14} a_{15} a_{18} + a_{5} a_{14} a_{17} a_{18}) \\
    & + (a_{10} a_{11} a_{14} a_{23} + a_{10} a_{13} a_{14} a_{23}) + (a_{11} a_{14} a_{15} a_{24} + a_{11} a_{14} a_{23} a_{24})
    + (a_{13} a_{14} a_{17} a_{26} + a_{13} a_{14} a_{23} a_{26}),
\end{eqs}
which acts as $CCCZ$ gates on the vertices of tetrahedra adjacent to vertex $14$.
It is straightforward to verify that $U_{14}$ commutes with the global symmetry $S$ in Eq.~\eqref{eq: 3d anomalous Z2 symmetry in app}.\footnote{A key fact is that $\gamma_{14}/a_{14}\pmod{2}$ remains invariant under $\prod_v X_v$, which shifts each $a_i\to a_i+1$.}

To extract the membrane fusion statistics, we select regions  
\begin{equation}
    A = \{14\},\quad B = \{11\},\quad C = \{10,13\},\quad D = \{1,2,4,5\}.
\end{equation}
First, observe  
\begin{equation}
    U_A^2 = U_{14}^2 = (-1)^{\gamma_{14}/a_{14}}.
\end{equation}
Next, the commutator gives  
\begin{eqs}
    [U_B, U_A^2] =& [X_{11}, (-1)^{a_{1} a_{2} a_{11} + a_{1} a_{10} a_{11} + a_{2} a_{11} a_{15} + a_{10} a_{11} a_{23} + a_{11} a_{15} a_{24} + a_{11} a_{23} a_{24}} ] \\
    =& (-1)^{a_{1} a_{2}  + a_{1} a_{10} + a_{2} a_{15} + a_{10} a_{23} + a_{15} a_{24} +  a_{23} a_{24}},
\end{eqs}
and hence  
\begin{equation}
    [U_C, [U_B, U_A^2]] = [X_{10} X_{13}, (-1)^{a_{1} a_{2}  + a_{1} a_{10} + a_{2} a_{15} + a_{10} a_{23} + a_{15} a_{24} +  a_{23} a_{24}}]
    = (-1)^{a_{1}+a_{23}} = Z_{1} Z_{23}.
\end{equation}
Since $Z_1$ anticommutes with the $X_1$ factor in $U_{C+D}$, the resulting fusion statistic is
\begin{equation}
    Z_5 = ( U_{C+D} )^{-2}
            \left( U_{C+D} [U_C, [U_B,U_A^2 ]] \right)^2 = -1~.
\end{equation}

\subsection{Boundary anomaly of the beyond-cohomology (4+1)D $\mathbb{Z}_2$ SPT phase}\label{sec: 3d anomalous Z2^(0)xZ2^(1) 1}

According to Ref.~\cite{CH21}, the boundary theory of the beyond-cohomology (4+1)D $\mathbb{Z}_2$ SPT phase has the anomalous symmetry described in Eq.~\eqref{eq: 3d anomalous loop membrane symmetry I} (the boundary also hosts additional particle excitations, which we omit in the following discussion).  
In the cup product notation~\cite{Chen2023Highercup}, the symmetry operators can be expressed as  
\begin{equation}
    S^a = \left( \prod_v X_v^a \right) ~(-1)^{\int \bb \cup \delta \bb} , \quad
    S^b_v = \prod_{e|\delta \bv (e) = 1} X^b_{e}.
\end{equation}
The corresponding domain wall operators are defined in the usual way:  
\begin{equation}
    W^a_{e_{ij}} := Z^a_i Z^a_j, \quad W^b_f = \prod_{e \in \partial f} Z^b_e.
\end{equation}
On the patch cube~\eqref{eq: 2x2_cube}, the associated volume operator $U_v^a$ and membrane operator $U_e^b$ are given by  
\begin{eqs}
    U^a_{5} &= X_5^a ,\\
    U^b_{\lr{14, 15}} &= X^b_{\lr{14, 15}} ~(-1)^{a_{14} (b_{\lr{15, 18}} + b_{\lr{18, 27}} + b_{\lr{15, 24}} + b_{\lr{24, 27}})
    + a_{2} (b_{\lr{2, 5}} + b_{\lr{5, 14}} + b_{\lr{2, 11}} + b_{\lr{11, 14}}) }, \\
    U^b_{\lr{14, 17}} &= X^b_{\lr{14, 17}} ~(-1)^{a_{14} (b_{\lr{17, 18}} + b_{\lr{18, 27}} + b_{\lr{17, 26}} + b_{\lr{26, 27}})
    + a_{4} (b_{\lr{4, 5}} + b_{\lr{5, 14}} + b_{\lr{4, 13}} + b_{\lr{13, 14}}) }, \\
    U^b_{\lr{14, 23}} &= X^b_{\lr{14, 23}} ~(-1)^{a_{14} (b_{\lr{23, 24}} + b_{\lr{24, 27}} + b_{\lr{23, 26}} + b_{\lr{26, 27}})
    + a_{10} (b_{\lr{10, 11}} + b_{\lr{11, 14}} + b_{\lr{10, 13}} + b_{\lr{13, 14}}) }, \\
\end{eqs}
and $U_v^a$ and $U_e^b$ on other vertices and edges are obtained by translation.
Both $U_v^a$ and $U_e^b$ commute with the symmetries.
The operator $U_v^a$ excites the domain walls $W^a_e$ on all edges $e$ adjacent to $v$, producing a membrane excitation in the dual lattice, while $U_e^b$ excites the domain walls $W^b_f$ on all faces $f$ adjacent to $e$, generating a loop excitation in the dual lattice.
These operators can be derived from the cup product formalism as  
\begin{equation}
    U^a_{v} = X_v, \quad U^b_e = X_e~(-1)^{\int \ba \cup (\be \cup \delta \bb + \delta \bb \cup \be)}.
\end{equation}
To verify the nontrivial statistics, we choose the faces $f_1$, $f_2$, and $f_3$ to be the Poincar\'e duals of the edges $\langle 14, 15\rangle$, $\langle 14, 17\rangle$, and $\langle 17, 18\rangle$, respectively:
\begin{equation}
    U_{f_3}^{-1} U_{f_2} U_{f_1}^{-1} U_{f_3} U_{f_2}^{-1} U_{f_1} = Z^a_{5} Z^a_{14}.
\end{equation}
We take $U_t$ to be the operator $U^a_5$, which can be viewed as the volume operator on the cube formed by the vertices ${5, 6, 8, 9, 14, 15, 17, 18}$, as determined by the framing implicitly chosen when defining the symmetry. This $U_t$ includes one of the endpoints of the common edge shared by $f_1$, $f_2$, and $f_3$.
With this choice, we find
\begin{equation}
    Z_5^{\mathrm{loop\text{-}membrane\text{-}I}} = [U_t, [ U_{f_3}^{-1} U_{f_2} U_{f_1}^{-1} U_{f_3} U_{f_2}^{-1} U_{f_1}] ] = -1.
\end{equation}
Therefore, the boundary theory of the beyond-cohomology (4+1)D $\mathbb{Z}_2$ SPT phase exhibits a nontrivial loop-membrane mutual statistics.

\subsection{Anomalous 0-form $\mathbb{Z}_2$ and 1-form  $\mathbb{Z}_2$ symmetries in three spatial dimensions}\label{sec: 3d anomalous Z2^(0)xZ2^(1) 2}

For the anomalous symmetries in Eq.~\eqref{eq: 3d anomalous loop membrane symmetry II}, the domain wall operators take the form  
\begin{equation}
    W^a_{e_{ij}} := Z^a_i Z^a_j, \quad W^b_f = \prod_{e \in \partial f} Z^b_e.
\end{equation}
On the patch cube~\eqref{eq: 2x2_cube}, the corresponding volume operator $U_v^a$ and membrane operator $U_e^b$ are
\begin{eqs}
    U^a_{14} &= X_{14}^a~(-1)^{\zeta_{14}},\\
    U^b_{\lr{14, 15}} &= X^b_{\lr{14, 15}} ~(-1)^{a_2 (a_5 +a_{11}) (a_{14}+1)}, \\
    U^b_{\lr{14, 17}} &= X^b_{\lr{14, 17}} ~(-1)^{a_4 (a_5+a_{13}) (a_{14}+1)}, \\
    U^b_{\lr{14, 23}} &= X^b_{\lr{14, 23}} ~(-1)^{a_{10} (a_{11}+a_{13}) (a_{14}+1)}, 
\end{eqs}
with
\begin{eqs}
    \zeta_{14} :=&~ a_{14} a_{15} (b_{\lr{15,18}}+b_{\lr{18,27}}+b_{\lr{15,24}}+b_{\lr{24,27}}) 
    +a_{13} a_{14} (b_{\lr{14,17}}+b_{\lr{17,26}}+b_{\lr{14,23}}+b_{\lr{23,26}})\\
    &+a_{14} a_{17} (b_{\lr{17,18}}+b_{\lr{18,27}}+b_{\lr{17,26}}+b_{\lr{26,27}})
    +a_{11} a_{14} (b_{\lr{14,15}}+b_{\lr{15,24}}+b_{\lr{14,23}}+b_{\lr{23,24}})\\
    &+ a_{14} a_{23} (b_{\lr{23,24}}+b_{\lr{24,27}}+b_{\lr{23,26}}+b_{\lr{26,27}})
    +a_{5} a_{14} (b_{\lr{14,15}}+b_{\lr{15,18}}+b_{\lr{14,17}}+b_{\lr{17,18}}).
\end{eqs}
The operators $U_v^a$ and $U_e^b$ at other vertices and edges are obtained by translation and commute with the anomalous symmetries~\eqref{eq: 3d anomalous loop membrane symmetry II}.

To verify the statistics in Eq.~\eqref{eq:loop-membrane_statistics_2}, we choose the cells $t_1$, $t_2$, and $f$ in the patch cube~\eqref{eq: 2x2_cube} as  
\begin{equation}
    t_1 = \{14\},~ t_2 = \{ 13 \},~f=\{\lr{14, 23}\}.
\end{equation}
More specifically, $U^a_{14}$ is the volume operator on the cube with vertices $14, 15, 17, 18, 23, 24, 26, 27$, and $U^a_{13}$ is the volume operator on the cube with vertices $13, 14, 16, 17, 22, 23, 25, 26$.  
The operator $U^b_{\langle 14, 23\rangle}$ is the membrane operator acting on the face Poincar\'e dual to the edge $\langle 14, 23\rangle$.
With this choice, we find
\begin{eqs}
    \left(U_{t_1}\right)^2 = \left(U^a_{14}\right)^2 =(-1)^{\zeta_{14}/a_{14}}.
\end{eqs}
Then,
\begin{equation}
    [U_{t_2} ,U_{t_1}^2] = [U^a_{13} ,\left(U^a_{14}\right)^2] = [X^a_{13}, (-1)^{a_{13}(b_{\lr{14,17}}+b_{\lr{17,26}}+b_{\lr{14,23}}+b_{\lr{23,26}})}] = Z^b_{\lr{14,17}} Z^b_{\lr{17,26}} Z^b_{\lr{14,23}} Z^b_{\lr{23,26}}.
\end{equation}
Finally,
\begin{equation}
    [U_f,[U_{t_2} ,U_{t_1}^2]] = [X_{\lr{14, 23}}, Z^b_{\lr{14,17}} Z^b_{\lr{17,26}} Z^b_{\lr{14,23}} Z^b_{\lr{23,26}}] = -1.
\end{equation}
Thus, the domain wall excitations realize the nontrivial loop-membrane statistics described in Eq.~\eqref{eq:loop-membrane_statistics_2}.

\subsection{Anomalous 0-form $\mathbb{Z}_2$ and 2-form  $\mathbb{Z}_2$ symmetries in three spatial dimensions}\label{sec: 3d anomalous Z2^(0)xZ2^(2)}

In the cup product notation~\cite{CH21, Chen2023Highercup}, the anomalous symmetries defined in Eq.~\eqref{eq: 3d anomalous particle membrane symmetry} can be written as  
\begin{equation}
    S^a := \left(\prod_v X^a_v \right) ~(-1)^{\int \ba \cup \delta \bc}, \quad
    S^c_e:= \prod_{f | \delta \be(f) =1} X^c_f.
\label{eq: anomalous Z2^(0) x Z2^(2) in cup product}
\end{equation}
The associated domain wall operators are  
\begin{equation}
    W^a_{e_{ij}} := Z^a_i Z^a_j, \quad W^c_t = \prod_{f \in \partial t} Z^c_f,
\end{equation}
where $W^a_{e_{ij}}$ acts on each edge $e$ and $W^c_t$ acts on each tetrahedron (3-cell) $t$.  
The corresponding volume and string operators, which commute with the symmetry~\eqref{eq: anomalous Z2^(0) x Z2^(2) in cup product}, are  
\begin{equation}
    U_v^a = X^a_v ~(-1)^{\int \bv \cup \ba \cup \delta\bc}, \quad U^c_f=X^c_f ~(-1)^{\int \ba \cup \delta\ba \cup \bface}.
\end{equation}
The volume operator $U_v^a$ excites the domain walls $W^a_e$ on all edges $e$ adjacent to $v$, producing a membrane excitation in the dual lattice, while the string operator $U_f^c$ excites the domain walls $W^c_t$ on two tetrahedra (3-cells) $t$ adjacent to $f$, generating two particle excitations in the dual lattice.

On the patch cube~\eqref{eq: 2x2_cube}, the corresponding volume operator $U_v^a$ and string operator $U_f^b$ are
\begin{eqs}
    U^a_{14} &= X_{14}^a~(-1)^{a_{14}(c_{\lr{14, 15, 17, 18}} + c_{\lr{14, 15, 23, 24}} + c_{\lr{14, 17, 23, 26}}  + c_{\lr{15, 18, 24, 27}} + c_{\lr{17, 18, 26, 27}} + c_{\lr{23, 24, 26, 27}} )},\\
    U^c_{\lr{14, 15, 17, 18}} &= X^c_{\lr{14, 15, 17, 18}} ~(-1)^{a_5 (a_{14}+1)}, \\ 
    U^c_{\lr{14, 15, 23, 24}} &= X^c_{\lr{14, 15, 23, 24}} ~(-1)^{a_{11} (a_{14}+1)}, \\ 
    U^c_{\lr{14, 17, 23, 26}} &= X^c_{\lr{14, 17, 23, 26}} ~(-1)^{a_{13} (a_{14}+1)}, 
\end{eqs}
where $\langle i,j,k,l\rangle$ denotes the face (square) formed by the four vertices $i$, $j$, $k$, and $l$.
The operators $U_v^a$ and $U_f^c$ at other vertices and faces are obtained by translation.

To verify the statistics in Eq.~\eqref{eq:particle_membrane_statistics}, we choose the cells $t_1$, $t_2$, and $e$ in the patch cube~\eqref{eq: 2x2_cube} as  
\begin{equation}
    t_1 = \{14\},~ t_2 = \{ 13 \},~e=\{\lr{14,17,23,26}\}.
\end{equation}
$U^a_{14}$ is the volume operator on the cube with vertices $14, 15, 17, 18, 23, 24, 26, 27$, and $U^a_{13}$ is the volume operator on the cube with vertices $13, 14, 16, 17, 22, 23, 25, 26$.  
The operator $U^c_{\langle 14,17,23,26\rangle}$ is the string operator acting on the edge Poincar\'e dual to the face $\langle 14,17,23,26\rangle$.

First, note that $U_e^2 = 1$ in our model, implying $[U_e^2, U_{t_1}] = 1$.
The nontrivial contribution therefore comes from the second term in Eq.~\eqref{eq:particle_membrane_statistics}.
Evaluating the first commutator, we find
\begin{equation}
    [U_{t_1}, U_e] = (-1)^{a_{13}+a_{14}} = Z^a_{13} Z^a_{14}.
\end{equation}
Then, the double commutator is
\begin{equation}
    [U_{t_2}, [U_{t_1}, U_e] ] = [X^a_{13}, Z^a_{13} Z^a_{14}] = -1.
\end{equation}
Combining these results, the statistics is
\begin{equation}
    [U_e^2, U_{t_1}]\,[U_{t_2}, [U_{t_1}, U_e] ] = -1,
\end{equation}
showing that the domain walls carry a nontrivial $\mathbb{Z}_2$ mutual statistics.

\section{Simplicial complex}

A simplicial complex $X$ is a set of finite subsets of natural numbers such that  
$\sigma \in X,\ \tau \subset \sigma \implies \tau \in X$.  
Any $\sigma \in X$ has the form $\sigma = \{a_0,a_1,\dots,a_p\}$, where $0 \le a_0 < a_1 < \cdots < a_p$, and we say $\sigma$ is a $p$-dimensional simplex of $X$.  
Subsets of $\sigma$ are called its \emph{faces}; in particular, a $(p-1)$-dimensional face of $\sigma$ is obtained by deleting the $i$th element ($0 \le i \le p$) from $\{a_0,a_1,\dots,a_p\}$, denoted by $\partial_i\sigma = \{a_0,\dots,\hat{a_i},\dots,a_p\}$.  

Geometrically, a simplex generalizes the notion of a point (0-simplex), an edge (1-simplex), a triangle (2-simplex), and a tetrahedron (3-simplex).  
Simplicial complexes are used to describe topological spaces by gluing simplexes along their faces.  
Although a topological space typically contains infinitely many points, its triangulation as a simplicial complex can be described by a finite set of combinatorial data, which greatly simplifies its study.  
For example, the simplicial complex  
$X = \{\emptyset,\{1\},\{2\},\{3\},\{1,2\},\{2,3\},\{1,3\}\}$ describes a triangle, which is a triangulation of the circle $S^1$.  
More generally, for any $d \ge 0$, we may take  
$X = \{\sigma \mid \sigma \subsetneqq \{0,1,\dots,d+1\}\}$, which is a triangulation of the $d$-sphere $S^d$.  
This triangulation is used in our study of statistics in $d$-dimensional space.  

Homology is naturally defined on simplicial complexes.  
Let $X(n)$ denote the set of $n$-simplices of $X$; the $n$th chain group is  
$C_n(X) = \sum_{\sigma \in X(n)} \mathbb{Z} \sigma$, and the boundary map $\partial: C_n(X) \to C_{n-1}(X)$ is defined by  
$\partial \sigma = \sum_{i=0}^n (-1)^i \partial_i\sigma$.  
For any discrete Abelian group $G$, we can formally multiply a group element $g \in G$ with a simplex $\sigma \in X(n)$, denoting the result by $g\sigma$.  
The $n$th chain group with coefficients in $G$ is then  
$C_n(X,G) = \bigoplus_{\sigma \in X_n} G \sigma$.  
An element of $C_n(X,G)$ is written as $\sum_{\sigma \in X_n} g_\sigma \sigma$ with $g_\sigma \in G$.  
The homological boundary map $\partial: C_n(X,G) \to C_{n-1}(X,G)$ is defined by  
$\partial (g\sigma) = \sum_{i=0}^n (-1)^i g\,\partial_i\sigma$.  
The image of $\partial$ is the $(n-1)$th boundary chain group, denoted by $B_{n-1}(X,G)$; in our excitation models, this group is the configuration group $\mathcal{A}$ for an $(n-1)$-dimensional excitation model with fusion group $G$ in the simplicial complex $X$.  

We note a related but slightly different concept: the \emph{$\Delta$-complex} (see Sec.~2 of Ref.~\cite{hatcher2002algebraic}).  
In a simplicial complex, each $p$-simplex has $(p+1)$ distinct vertices and is uniquely determined by them; in a $\Delta$-complex, vertices of a simplex may coincide, and a simplex may not be uniquely determined by its vertices.  
For example,  
\begin{equation}
    \vcenter{\hbox{\raisebox{-2.5ex}{
    \begin{tikzpicture}
			\coordinate (A) at (0,0);  
			\coordinate (C) at (2,0);   
			\draw[thick] (A) to[out=90, in=90] (C);   
			\draw[thick] (A) to[out=-90, in=-90] (C); 
			\fill[red] (A) circle (2pt) node[left] {0};
    \end{tikzpicture}}}} \;\;\;,
    \;\;\;
    \vcenter{\hbox{\raisebox{-2ex}{\begin{tikzpicture}
			\coordinate (A) at (0,0);  
			\coordinate (C) at (2,0);   
			\draw[thick] (A) to[out=60, in=120] (C);   
			\draw[thick] (A) to[out=-60, in=-120] (C); 
			\fill[red] (A) circle (2pt) node[left] {0};
			\fill[red] (C) circle (2pt) node[right] {1};
    \end{tikzpicture}}}} \;\;\;,
    \text{ and }
    \;\;\;
    \vcenter{\hbox{\raisebox{-2ex}{\begin{tikzpicture}
			\draw[thick] (0,0) -- (2,0) -- (1,1.732) -- cycle; 
            \foreach \x in {(0,0), (2,0), (1,1.732)} \fill[red] \x circle (2pt);
    \end{tikzpicture}}}} \;\;\;,
\end{equation}
are all $\Delta$-complexes, but only the last is a simplicial complex.  
The first two do not work well as excitation models for $F$-symbols because they lack locality identities.  
In general, simplicial complexes have better behavior.  

Another subtlety is that manifolds are special topological spaces.  
In a $d$-dimensional manifold, every point has a neighborhood homeomorphic to $\mathbb{R}^d$, and this property is reflected in any triangulation.  
Technically, for any $\sigma \in X$ we define $\operatorname{Lk}(\sigma)$, the \emph{link} of $\sigma$, to be the subcomplex of $X$ consisting of all $\tau$ such that $\sigma \cap \tau = \emptyset$ and the union of the vertices of $\sigma$ and $\tau$ forms the vertex set of a simplex of $X$.  
We say $X$ is a $d$-dimensional combinatorial manifold if, for any $\sigma \in X(p)$, the link $\operatorname{Lk}(\sigma)$ is homeomorphic to $S^{d-p-1}$.  

This property of local topology has important implications for statistics:  
trivial local topology implies trivial local statistics, which further implies that the statistics is independent of the choice of operator.  
This is referred to as the \emph{strong operator independence condition} in \cite{xue2025statisticsinvertibletopologicalexcitations}, where the full definition and proof are given.
}

\section{Derivations for generalized statistics with fusion group $\ZZ_2$}\label{app: T-junction and 24-step compute}

\subsection{T-junction process in (2+1)D}\label{sec:T-junction process in (2+1)D in App}

In this section, we explicitly evaluate the summation of locality identities for the (2+1)D T-junction with $G=\ZZ_2$ as outlined in Sec.~\ref{subsec:T junction quantization}. This analysis shows that a particular combination of these locality identities results in four iterations of the T-junction process, thereby verifying the quantization of the T-junction into the fourth root of unity.

For convenience, we represent a configuration of excitations by a vertex of a graph. An edge connecting two vertices represents the action of a unitary operator that changes the configuration from one vertex to the other. The direction of the edge further specifies the initial and final configurations. All possible transformations of the states using the operators $U_{0i}$ can be represented as a cube, as shown in Fig.~\ref{fig: 2D T junction operator cube}.

Since $G=\ZZ_2$, for a fixed initial configuration of excitations, the actions of the unitaries $U$ and $U^{-1}$ yield the same final state. Each direction of an edge can be associated with two distinct unitaries, $U$ or $U^{-1}$. Thus, each edge corresponds to four possible transformations, arising from the choice of edge direction and whether $U$ or $U^{-1}$ is applied. According to
\begin{align}
    \theta(U(s),a)  & = -\theta(U(s)^{-1} ,a+\partial s)~,
    \label{eq: Tjunctionreview_initial}
\end{align}
for any $a\in\mathcal{A}$ and $s\in\mathcal{S}$.
There are two relations among the four possible phases.
Taking the leftmost vertical edge in Fig.~\ref{fig: 2D T junction operator cube} with $U_{03}$ as an example, we have:
\begin{eqs}
    \theta(U_{03},\hbox{\raisebox{-1ex}{\includegraphics[width=.6cm]{T_junction_state_vaccum.pdf}}}) +\theta(U_{03}^{-1}, \hbox{\raisebox{-1ex}{\includegraphics[width=.6cm]{T_junction_state_03.pdf}}})  = 0~,
    \\
    \theta(U_{03},\hbox{\raisebox{-1ex}{\includegraphics[width=.6cm]{T_junction_state_03.pdf}}}) +\theta(U_{03}^{-1} ,\hbox{ \raisebox{-1ex}{\includegraphics[width=.6cm]{T_junction_state_vaccum.pdf}}})  = 0~.
    \label{eq: TjunctionPhase_relation}
\end{eqs}
Based on the above discussion, we describe the sum of phases using a cube with operators $\{U_{0j}\}$, where the directed edges indicate the initial and final states. We can express the phases in Eq.~\eqref{eq: TjunctionPhase_relation} diagrammatically on the edges of the cube as
\begin{eqs}
   \hbox{\raisebox{-10.5ex}{\includegraphics[scale=0.35]{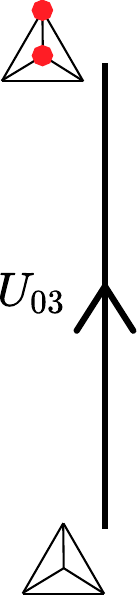}}} = -~  \hbox{\raisebox{-10.5ex}{\includegraphics[scale=0.35]{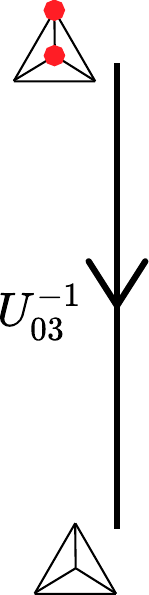}}}~,
   \quad \quad \quad \quad
   \hbox{\raisebox{-10.5ex}{\includegraphics[scale=0.35]{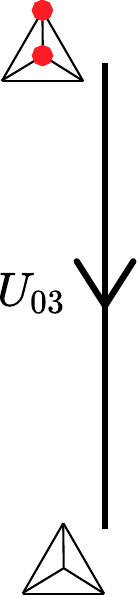}}} = -~  \hbox{\raisebox{-10.5ex}{\includegraphics[scale=0.35]{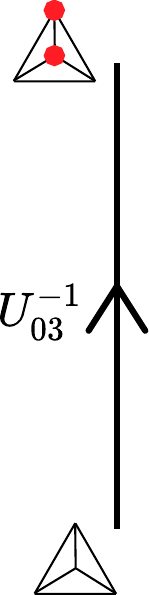}}}~.
\end{eqs}
The same relation holds for other edges in different directions.

Next, we consider the following locality identities acting on the vacuum state, represented by specific commutators of hopping operators whose supports do not overlap:
\begin{eqs}
    \text{Type 1:}& \quad
    \bra{\hbox{\raisebox{-1ex}{\includegraphics[width=.6cm]{T_junction_state_vaccum.pdf}}}}[[U_{02}, U_{03}],U_{12}] \ket{\hbox{\raisebox{-1ex}{\includegraphics[width=.6cm]{T_junction_state_vaccum.pdf}}}} = {\change 1}, \\
    \text{Type 2:}& \quad
    \bra{\hbox{\raisebox{-1ex}{\includegraphics[width=.6cm]{T_junction_state_vaccum.pdf}}}}[[U_{02}^{-1}, U_{03}^{-1}],U_{12}] \ket{\hbox{\raisebox{-1ex}{\includegraphics[width=.6cm]{T_junction_state_vaccum.pdf}}}} = {\change 1}, \\
    \text{Type 3:}& \quad
    \bra{\hbox{\raisebox{-1ex}{\includegraphics[width=.6cm]{T_junction_state_vaccum.pdf}}}}[[U_{03}, U_{02}],U_{23}] \ket{\hbox{\raisebox{-1ex}{\includegraphics[width=.6cm]{T_junction_state_vaccum.pdf}}}} = {\change 1}. \\
\end{eqs}
Each identity involves a total of eight hopping operators within the cube of configuration states, forming two distinct oriented squares.
The oriented squares associated with the Type 1 locality identity can be represented as follows:
\begin{eqs}
    &\hbox{\raisebox{-13.5ex}{\includegraphics[scale=0.3]{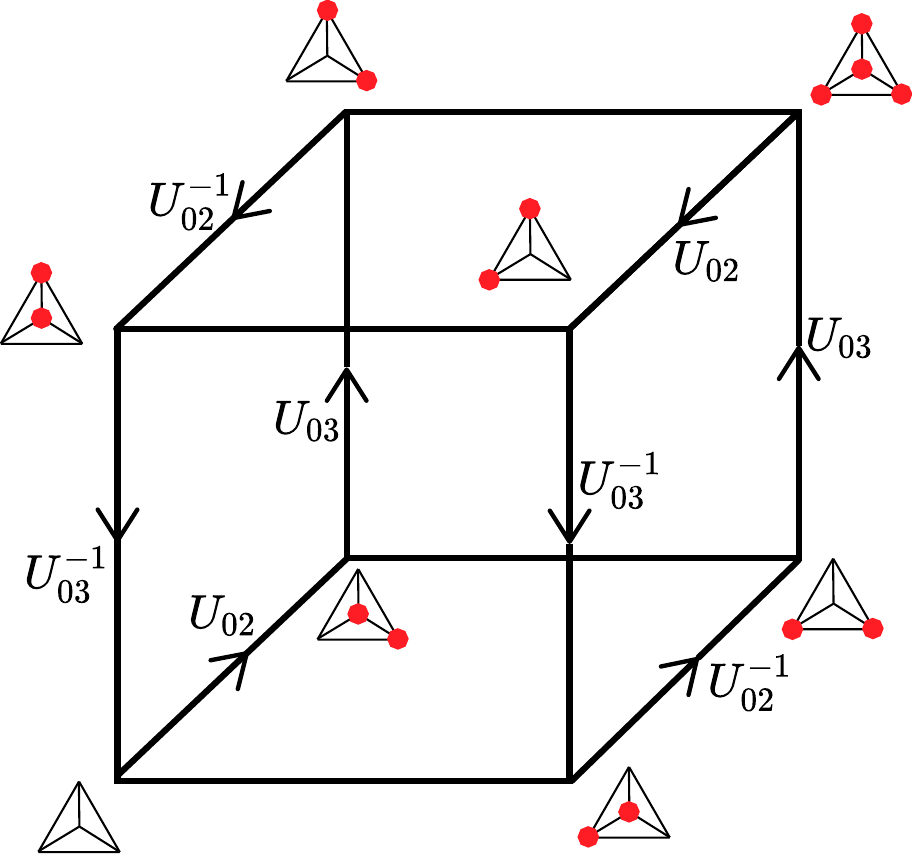}}} 
    \hspace{-1ex}= 0 \pmod{2\pi}~, 
    \label{eq: type 1 identity}
\end{eqs}
where two squares correspond to the terms $\bra{\hbox{\raisebox{-1ex}{\includegraphics[width=.6cm]{T_junction_state_12.pdf}}}}[U_{02}, U_{03}] \ket{\hbox{\raisebox{-1ex}{\includegraphics[width=.6cm]{T_junction_state_12.pdf}}}}$ and $-\bra{\hbox{\raisebox{-1ex}{\includegraphics[width=.6cm]{T_junction_state_vaccum.pdf}}}}[U_{02}, U_{03}] \ket{\hbox{\raisebox{-1ex}{\includegraphics[width=.6cm]{T_junction_state_vaccum.pdf}}}}$, respectively.
Similarly, the Type 2 locality identity can be visualized on the cube of configuration states as follows:
\begin{eqs}
    &\hbox{\raisebox{-13.5ex}{\includegraphics[scale=0.3]{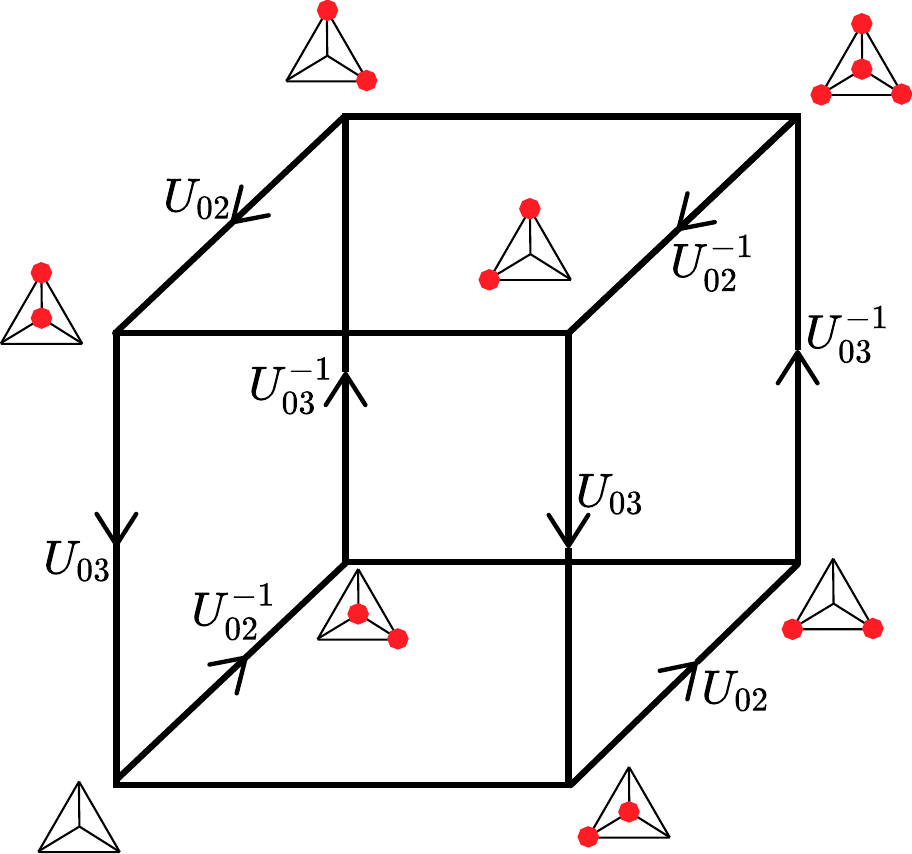}}}
    = 0 \pmod{2\pi}~.
    \label{eq: type 2 identity}
\end{eqs}
The Type 3 locality identity, on the other hand, is represented by two oriented squares located on the same face of the cube:
\begin{eqs}
    &\hbox{\raisebox{-13.5ex}{\includegraphics[scale=0.3]{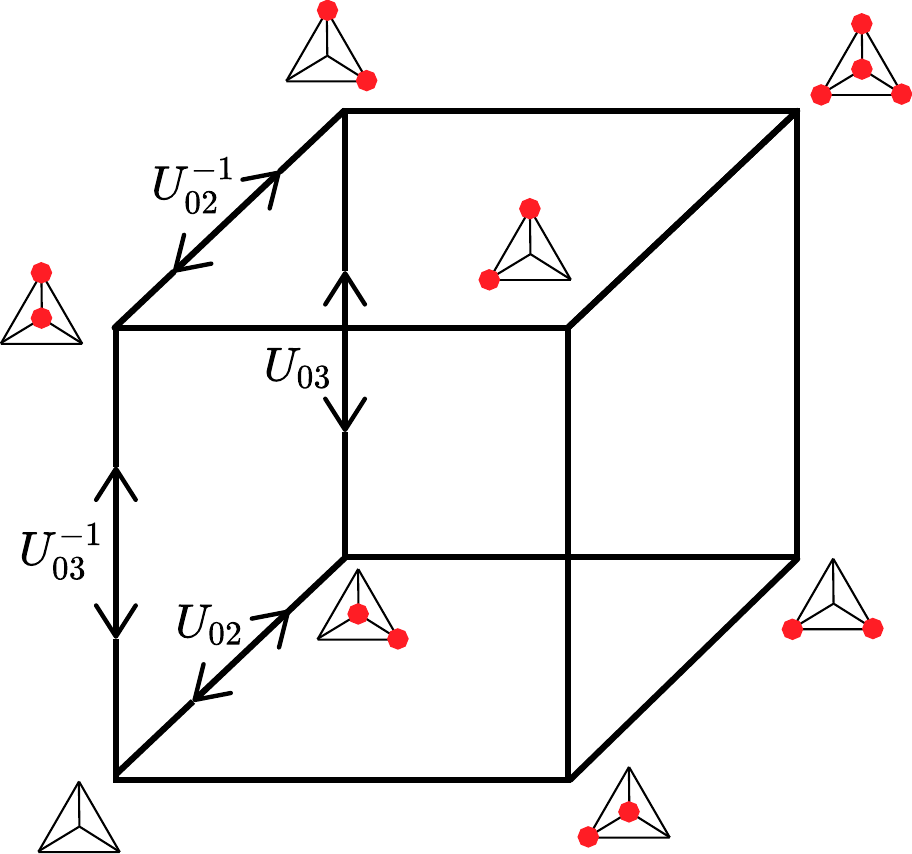}}}
    = 0 \pmod{2\pi}~.
    \label{eq: type 3 identity}
\end{eqs}
Note that each type of identity includes the one depicted above, along with two others that are related by the $C_3$ rotation along the axis extending from $\includegraphics[width=.6cm]{T_junction_state_vaccum.pdf}$ to $\includegraphics[width=.6cm]{T_junction_state_0123.pdf}$.

Below, we list all identities of the three types, which will be used in the derivation of the T-junction process.

\begin{eqs}
\text{Type 1:} \quad
    & \theta \left(U_{03},\hbox{ \raisebox{-1ex}{\includegraphics[width=.6cm]{T_junction_state_12.pdf}}} \right) 
    + \theta \left(U_{02},\hbox{ \raisebox{-1ex}{\includegraphics[width=.6cm]{T_junction_state_0123.pdf}}} \right) 
    - \theta \left(U_{03},\hbox{ \raisebox{-1ex}{\includegraphics[width=.6cm]{T_junction_state_01.pdf}}} \right)  
    - \theta \left(U_{02},\hbox{ \raisebox{-1ex}{\includegraphics[width=.6cm]{T_junction_state_12.pdf}}} \right) \\
    &-\theta \left(U_{03},\hbox{ \raisebox{-1ex}{\includegraphics[width=.6cm]{T_junction_state_vaccum.pdf}}} \right)  
    - \theta \left(U_{02},\hbox{ \raisebox{-1ex}{\includegraphics[width=.6cm]{T_junction_state_03.pdf}}} \right) 
    + \theta \left(U_{03},\hbox{ \raisebox{-1ex}{\includegraphics[width=.6cm]{T_junction_state_02.pdf}}} \right)  
    + \theta \left(U_{02},\hbox{ \raisebox{-1ex}{\includegraphics[width=.6cm]{T_junction_state_vaccum.pdf}}} \right) 
    = 0 \pmod{2\pi}~, 
    \\
    & \theta \left(U_{02},\hbox{ \raisebox{-1ex}{\includegraphics[width=.6cm]{T_junction_state_13.pdf}}} \right)  
    + \theta \left(U_{01},\hbox{ \raisebox{-1ex}{\includegraphics[width=.6cm]{T_junction_state_0123.pdf}}} \right)
    - \theta \left(U_{02},\hbox{ \raisebox{-1ex}{\includegraphics[width=.6cm]{T_junction_state_03.pdf}}} \right) 
    - \theta \left(U_{01},\hbox{ \raisebox{-1ex}{\includegraphics[width=.6cm]{T_junction_state_13.pdf}}} \right) \\
    &-\theta \left(U_{02},\hbox{ \raisebox{-1ex}{\includegraphics[width=.6cm]{T_junction_state_vaccum.pdf}}} \right)  
    - \theta \left(U_{01},\hbox{ \raisebox{-1ex}{\includegraphics[width=.6cm]{T_junction_state_02.pdf}}} \right) 
    + \theta \left(U_{02},\hbox{ \raisebox{-1ex}{\includegraphics[width=.6cm]{T_junction_state_01.pdf}}} \right)  
    + \theta \left(U_{01},\hbox{ \raisebox{-1ex}{\includegraphics[width=.6cm]{T_junction_state_vaccum.pdf}}} \right) 
    = 0 \pmod{2\pi}~,
    \\
     &\theta \left(U_{01},\hbox{ \raisebox{-1ex}{\includegraphics[width=.6cm]{T_junction_state_23.pdf}}} \right) 
    + \theta \left(U_{03},\hbox{ \raisebox{-1ex}{\includegraphics[width=.6cm]{T_junction_state_0123.pdf}}} \right) 
    - \theta \left(U_{01},\hbox{ \raisebox{-1ex}{\includegraphics[width=.6cm]{T_junction_state_02.pdf}}} \right)  
    - \theta \left(U_{03},\hbox{ \raisebox{-1ex}{\includegraphics[width=.6cm]{T_junction_state_23.pdf}}} \right) \\
    &-\theta \left(U_{01},\hbox{ \raisebox{-1ex}{\includegraphics[width=.6cm]{T_junction_state_vaccum.pdf}}} \right)  
    - \theta \left(U_{03},\hbox{ \raisebox{-1ex}{\includegraphics[width=.6cm]{T_junction_state_01.pdf}}} \right) 
    + \theta \left(U_{01},\hbox{ \raisebox{-1ex}{\includegraphics[width=.6cm]{T_junction_state_03.pdf}}} \right) 
    + \theta \left(U_{03},\hbox{ \raisebox{-1ex}{\includegraphics[width=.6cm]{T_junction_state_vaccum.pdf}}} \right)
    =0 \pmod{2\pi}~.
    \end{eqs}
\begin{eqs}
    \text{Type 2:} \quad
     &\theta \left(U_{03},\hbox{ \raisebox{-1ex}{\includegraphics[width=.6cm]{T_junction_state_13.pdf}}} \right) 
    + \theta \left(U_{02},\hbox{ \raisebox{-1ex}{\includegraphics[width=.6cm]{T_junction_state_01.pdf}}} \right) 
    - \theta \left(U_{03},\hbox{ \raisebox{-1ex}{\includegraphics[width=.6cm]{T_junction_state_0123.pdf}}} \right)  
    - \theta \left(U_{02},\hbox{ \raisebox{-1ex}{\includegraphics[width=.6cm]{T_junction_state_13.pdf}}} \right) \\
    &-\theta \left(U_{03},\hbox{ \raisebox{-1ex}{\includegraphics[width=.6cm]{T_junction_state_23.pdf}}} \right)  
    - \theta \left(U_{02},\hbox{ \raisebox{-1ex}{\includegraphics[width=.6cm]{T_junction_state_02.pdf}}} \right) 
    + \theta \left(U_{03},\hbox{ \raisebox{-1ex}{\includegraphics[width=.6cm]{T_junction_state_03.pdf}}} \right)  
    + \theta \left(U_{02},\hbox{ \raisebox{-1ex}{\includegraphics[width=.6cm]{T_junction_state_23.pdf}}} \right) 
    = 0 \pmod{2\pi}~,
    \\
     &\theta \left(U_{02},\hbox{ \raisebox{-1ex}{\includegraphics[width=.6cm]{T_junction_state_23.pdf}}} \right) 
    + \theta \left(U_{01},\hbox{ \raisebox{-1ex}{\includegraphics[width=.6cm]{T_junction_state_03.pdf}}} \right) 
    - \theta \left(U_{02},\hbox{ \raisebox{-1ex}{\includegraphics[width=.6cm]{T_junction_state_0123.pdf}}} \right)  
    - \theta \left(U_{01},\hbox{ \raisebox{-1ex}{\includegraphics[width=.6cm]{T_junction_state_23.pdf}}} \right) \\
    &-\theta \left(U_{02},\hbox{ \raisebox{-1ex}{\includegraphics[width=.6cm]{T_junction_state_12.pdf}}} \right)  
    - \theta \left(U_{01},\hbox{ \raisebox{-1ex}{\includegraphics[width=.6cm]{T_junction_state_01.pdf}}} \right) 
    + \theta \left(U_{02},\hbox{ \raisebox{-1ex}{\includegraphics[width=.6cm]{T_junction_state_02.pdf}}} \right)  
    + \theta \left(U_{01},\hbox{ \raisebox{-1ex}{\includegraphics[width=.6cm]{T_junction_state_12.pdf}}} \right) 
    = 0 \pmod{2\pi}~,
    \\
     &\theta \left(U_{01},\hbox{ \raisebox{-1ex}{\includegraphics[width=.6cm]{T_junction_state_12.pdf}}} \right) 
    + \theta \left(U_{03},\hbox{ \raisebox{-1ex}{\includegraphics[width=.6cm]{T_junction_state_02.pdf}}} \right) 
    - \theta \left(U_{01},\hbox{ \raisebox{-1ex}{\includegraphics[width=.6cm]{T_junction_state_0123.pdf}}} \right)  
    - \theta \left(U_{03},\hbox{ \raisebox{-1ex}{\includegraphics[width=.6cm]{T_junction_state_12.pdf}}} \right) \\
    &-\theta \left(U_{01},\hbox{ \raisebox{-1ex}{\includegraphics[width=.6cm]{T_junction_state_13.pdf}}} \right)  
    - \theta \left(U_{03},\hbox{ \raisebox{-1ex}{\includegraphics[width=.6cm]{T_junction_state_03.pdf}}} \right) 
    + \theta \left(U_{01},\hbox{ \raisebox{-1ex}{\includegraphics[width=.6cm]{T_junction_state_01.pdf}}} \right)  
    + \theta \left(U_{03},\hbox{ \raisebox{-1ex}{\includegraphics[width=.6cm]{T_junction_state_13.pdf}}} \right) 
    = 0 \pmod{2\pi}~.
    \end{eqs}
\begin{eqs}
    \text{Type 3:} \quad
     &\theta \left(U_{03},\hbox{ \raisebox{-1ex}{\includegraphics[width=.6cm]{T_junction_state_23.pdf}}} \right) 
    + \theta \left(U_{02},\hbox{ \raisebox{-1ex}{\includegraphics[width=.6cm]{T_junction_state_02.pdf}}} \right) 
    - \theta \left(U_{03},\hbox{ \raisebox{-1ex}{\includegraphics[width=.6cm]{T_junction_state_03.pdf}}} \right)  
    - \theta \left(U_{02},\hbox{ \raisebox{-1ex}{\includegraphics[width=.6cm]{T_junction_state_23.pdf}}} \right) \\
    &-\theta \left(U_{03},\hbox{ \raisebox{-1ex}{\includegraphics[width=.6cm]{T_junction_state_vaccum.pdf}}} \right)  
    - \theta \left(U_{02},\hbox{ \raisebox{-1ex}{\includegraphics[width=.6cm]{T_junction_state_03.pdf}}} \right) 
    + \theta \left(U_{03},\hbox{ \raisebox{-1ex}{\includegraphics[width=.6cm]{T_junction_state_02.pdf}}} \right)  
    + \theta \left(U_{02},\hbox{ \raisebox{-1ex}{\includegraphics[width=.6cm]{T_junction_state_vaccum.pdf}}} \right) 
    = 0 \pmod{2\pi}~,
    \\
     &\theta \left(U_{02},\hbox{ \raisebox{-1ex}{\includegraphics[width=.6cm]{T_junction_state_12.pdf}}} \right) 
    + \theta \left(U_{01},\hbox{ \raisebox{-1ex}{\includegraphics[width=.6cm]{T_junction_state_01.pdf}}} \right) 
    - \theta \left(U_{02},\hbox{ \raisebox{-1ex}{\includegraphics[width=.6cm]{T_junction_state_02.pdf}}} \right)  
    - \theta \left(U_{01},\hbox{ \raisebox{-1ex}{\includegraphics[width=.6cm]{T_junction_state_12.pdf}}} \right) \\
    &-\theta \left(U_{02},\hbox{ \raisebox{-1ex}{\includegraphics[width=.6cm]{T_junction_state_vaccum.pdf}}} \right)  
    - \theta \left(U_{01},\hbox{ \raisebox{-1ex}{\includegraphics[width=.6cm]{T_junction_state_02.pdf}}} \right) 
    + \theta \left(U_{02},\hbox{ \raisebox{-1ex}{\includegraphics[width=.6cm]{T_junction_state_01.pdf}}} \right)  
    + \theta \left(U_{01},\hbox{ \raisebox{-1ex}{\includegraphics[width=.6cm]{T_junction_state_vaccum.pdf}}} \right) 
    = 0 \pmod{2\pi}~,
    \\
     &\theta \left(U_{01},\hbox{ \raisebox{-1ex}{\includegraphics[width=.6cm]{T_junction_state_13.pdf}}} \right) 
    + \theta \left(U_{03},\hbox{ \raisebox{-1ex}{\includegraphics[width=.6cm]{T_junction_state_03.pdf}}} \right) 
    - \theta \left(U_{01},\hbox{ \raisebox{-1ex}{\includegraphics[width=.6cm]{T_junction_state_01.pdf}}} \right)  
    - \theta \left(U_{03},\hbox{ \raisebox{-1ex}{\includegraphics[width=.6cm]{T_junction_state_13.pdf}}} \right) \\
    &-\theta \left(U_{01},\hbox{ \raisebox{-1ex}{\includegraphics[width=.6cm]{T_junction_state_vaccum.pdf}}} \right)  
    - \theta \left(U_{03},\hbox{ \raisebox{-1ex}{\includegraphics[width=.6cm]{T_junction_state_01.pdf}}} \right) 
    + \theta \left(U_{01},\hbox{ \raisebox{-1ex}{\includegraphics[width=.6cm]{T_junction_state_03.pdf}}} \right)  
    + \theta \left(U_{03},\hbox{ \raisebox{-1ex}{\includegraphics[width=.6cm]{T_junction_state_vaccum.pdf}}} \right) 
    = 0 \pmod{2\pi}~.
\end{eqs}
By summing over all Type 1 identities, Type 2 identities, and twice the Type 3 identities, we obtain:
\begin{eqs}
     &4\theta \left(U_{03},\hbox{ \raisebox{-1ex}{\includegraphics[width=.6cm]{T_junction_state_02.pdf}}} \right) 
    - 4\theta \left(U_{02},\hbox{ \raisebox{-1ex}{\includegraphics[width=.6cm]{T_junction_state_03.pdf}}} \right) 
    + 4\theta \left(U_{01},\hbox{ \raisebox{-1ex}{\includegraphics[width=.6cm]{T_junction_state_03.pdf}}} \right) \\ 
    &-4\theta \left(U_{03},\hbox{ \raisebox{-1ex}{\includegraphics[width=.6cm]{T_junction_state_01.pdf}}} \right) 
    + 4\theta \left(U_{02},\hbox{ \raisebox{-1ex}{\includegraphics[width=.6cm]{T_junction_state_01.pdf}}} \right)  
    - 4\theta \left(U_{01},\hbox{ \raisebox{-1ex}{\includegraphics[width=.6cm]{T_junction_state_02.pdf}}} \right) 
    = 0 \pmod{ 2 \pi}~.
\end{eqs}
Using Eq.~\eqref{eq: Tjunctionreview_initial}, this expression can be simplified to
\begin{eqs}
     &4\theta \left(U_{03},\hbox{ \raisebox{-1ex}{\includegraphics[width=.6cm]{T_junction_state_02.pdf}}} \right) 
    + 4\theta \left(U_{02}^{-1},\hbox{ \raisebox{-1ex}{\includegraphics[width=.6cm]{T_junction_state_23.pdf}}} \right) 
    + 4\theta \left(U_{01},\hbox{ \raisebox{-1ex}{\includegraphics[width=.6cm]{T_junction_state_03.pdf}}} \right) \\ 
    &+4\theta \left(U_{03}^{-1} ,\hbox{ \raisebox{-1ex}{\includegraphics[width=.6cm]{T_junction_state_13.pdf}}} \right) 
    + 4\theta \left(U_{02},\hbox{ \raisebox{-1ex}{\includegraphics[width=.6cm]{T_junction_state_01.pdf}}} \right)  
    + 4\theta \left(U_{01}^{-1},\hbox{ \raisebox{-1ex}{\includegraphics[width=.6cm]{T_junction_state_12.pdf}}} \right) \\
    =&4 \theta\Big( U_{02} U_{03}^{-1} U_{01} U_{02}^{-1} U_{03} U_{01}^{-1},\hbox{ \raisebox{-1ex}{\includegraphics[width=.6cm]{T_junction_state_12.pdf}}}\Big)
    =0 \pmod{ 2 \pi}~.
\end{eqs}
This implies that four iterations of the T-junction process yield $0 \pmod{2 \pi}$. The T-junction process is illustrated in Fig.~\ref{fig: 2+1D T junction process}.

\subsection{Loop-flipping process in (3+1)D}\label{sec:Loop-flipping process in (3+1)D in App}

\begin{figure*}[t]
    \centering
    \raisebox{0ex}{\subfigure[Locality identity represented by each color or symbol]{\includegraphics[scale=0.08]{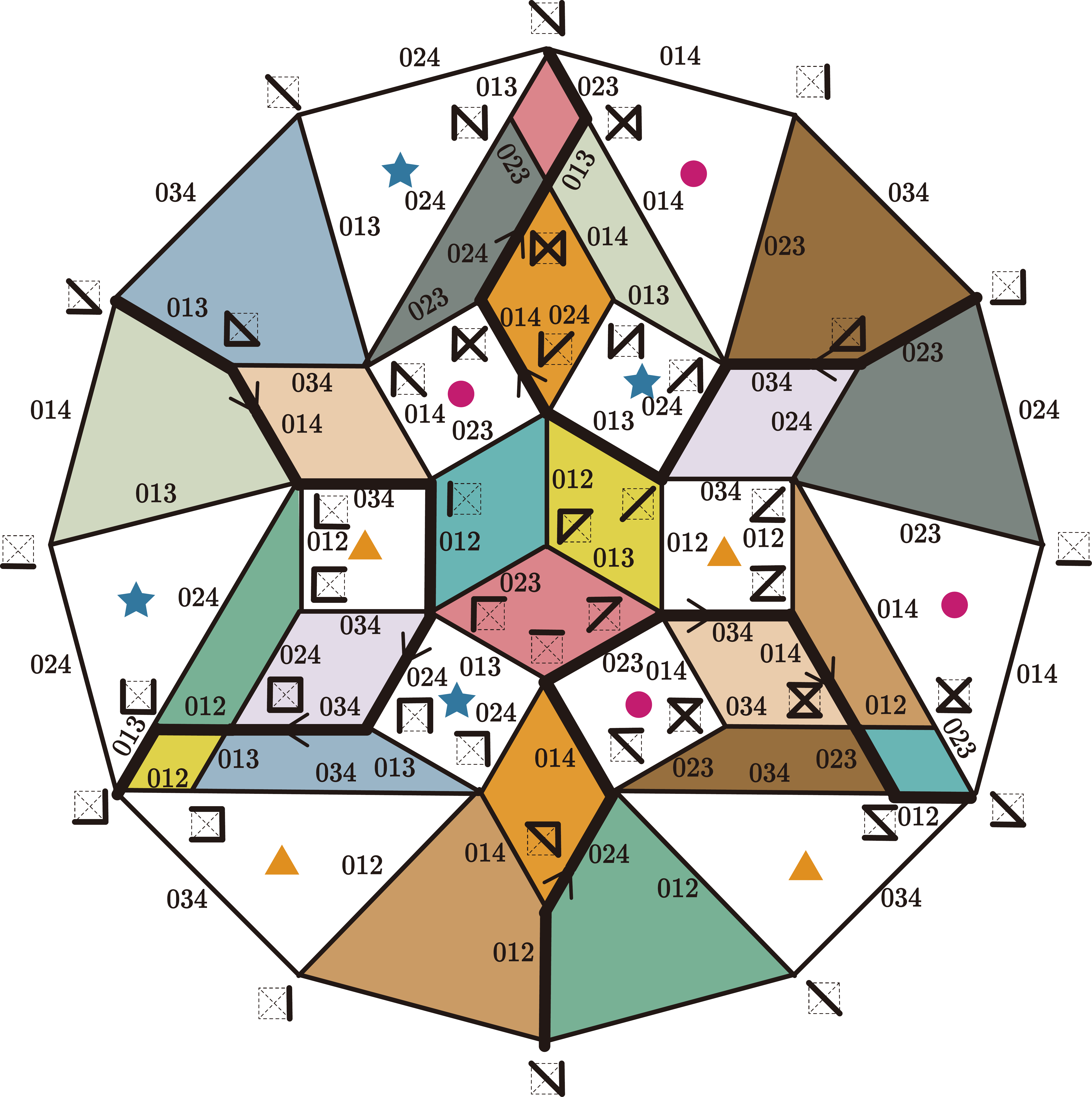}\label{fig: RP2 locality}}}
    \raisebox{0ex}{\subfigure[Orientation of plaquettes]{\raisebox{0.15ex}{\includegraphics[scale=0.08]{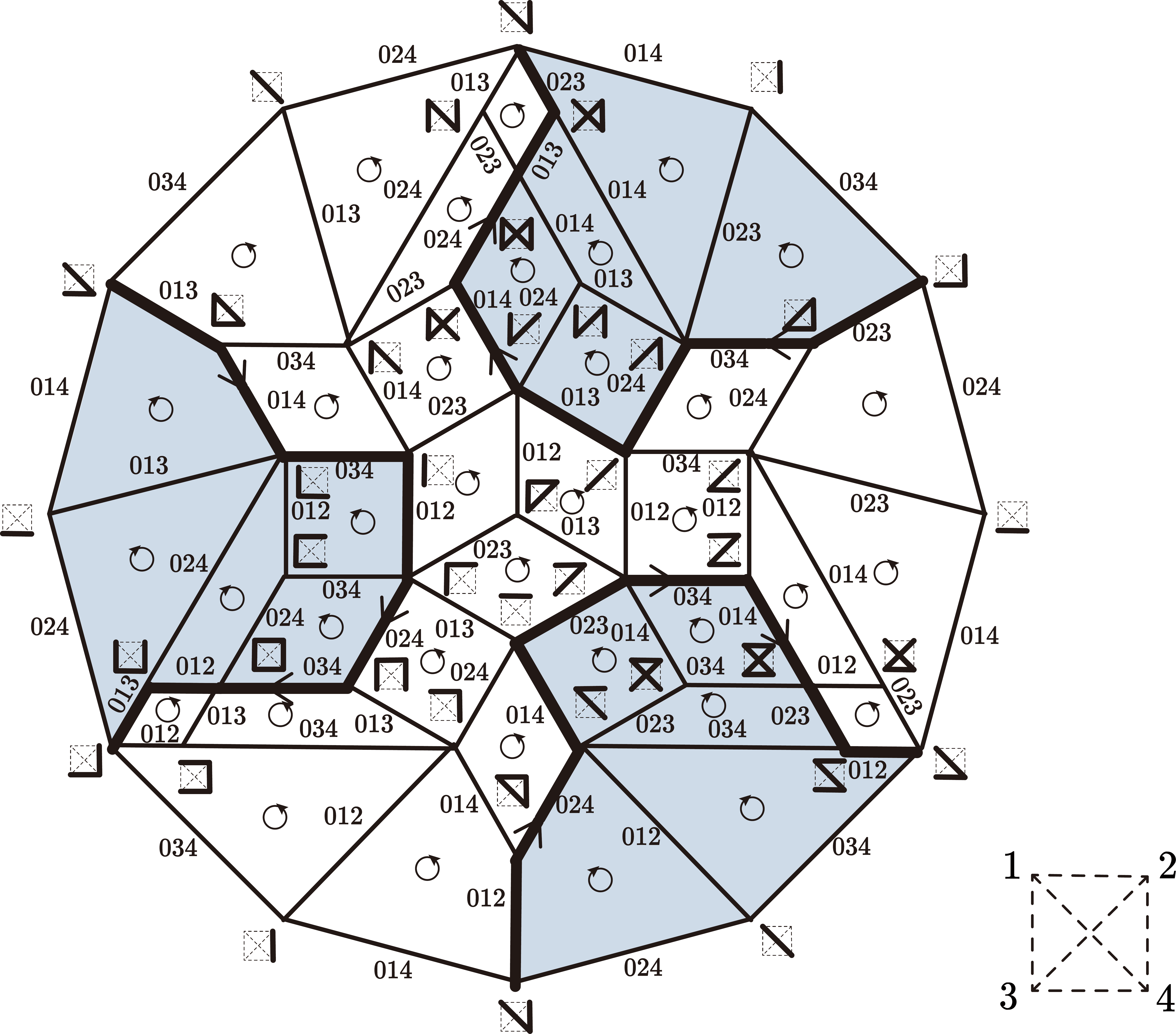}\label{fig: RP2 plaquette orientation}}}}
    \caption{Each vertex represents a single-loop configuration on the tetrahedron $\lr{1234}$ with a central vertex $0$, as shown in Fig.~\ref{fig: complex (c)}. Only the loop configuration along the edges between vertices $1$, $2$, $3$, and $4$ is depicted, as the configurations on the remaining edges can be inferred. The operator $U_{0ij}$ creates a loop excitation on edge $\lr{ij}$, while $U_{0ij}^{-1}$ annihilates it. This structure forms an $\mathbb{RP}^2$ space, as antipodal vertices represent the same configuration state \cite{FHH21}.
    (a) Each plaquette is labeled by a color or symbol, indicating that summing over plaquettes (according to the orientation in Fig.~\ref{fig: RP2 plaquette orientation}) with the same color or symbol yields a locality identity.
    (b) The orientation of each plaquette used to express the local identity. Summing over all locality identities results in twice the 24-step process (represented by the black directed line), as the $\theta(U(s), a)$ terms on all other edges are canceled out.
    }
    \label{fig: RP2 locality identities and orientations}
\end{figure*}

In this section, we derive the unitary sequence used to detect loop statistics for the fusion group $G = \ZZ_2$ in (3+1)D. This derivation is inspired by Appendix D of Ref.~\cite{FHH21}. Similar to the previous section, we first construct the graph space of configurations. In principle, there are six operators: $U_{012}$, $U_{013}$, $U_{014}$, $U_{023}$, $U_{024}$, and $U_{034}$, which create loops along the boundary of these faces, resulting in a total of $2^6 = 64$ configuration states within the same superselection sector as the vacuum state. For simplicity, we restrict our analysis to single-loop configurations, where the loop excitation is connected and does not have self-intersection. There are 37 single-loop configurations, represented as vertices in Fig.~\ref{fig: RP2 locality identities and orientations}.
Interestingly, the computation is even more straightforward than in the T-junction case. Since $G = \ZZ_2$, transitioning from one state to another allows us to apply either $U$ or $U^{-1}$, as used in the locality identities in Eqs.~\eqref{eq: type 1 identity}, \eqref{eq: type 2 identity}, and \eqref{eq: type 3 identity}. However, for loop excitations, we further constrain ourselves such that $U_{0ij}$ can only create a loop on edge $\lr{ij}$, while $U_{0ij}^{-1}$ can only annihilate it on the same edge.

The locality identities are represented by colors and symbols in Fig.~\ref{fig: RP2 locality}. For example, the two yellow plaquettes correspond to the following locality identity:
\begin{eqs}
    &\theta\left( U_{012}^{-1} U_{013}^{-1} U_{012} U_{013}, \partial \lr{023} \right) + \theta\left( U_{013}^{-1} U_{012}^{-1} U_{013} U_{012}, \partial \lr{023} + \partial \lr{234} \right) \\
    =&~\theta\left( U_{012}^{-1} U_{013}^{-1} U_{012} U_{013}, \partial \lr{023} \right) - \theta\left( U_{012}^{-1} U_{013}^{-1} U_{012} U_{013}, \partial \lr{023} + \partial \lr{234} \right) \\
    =&~\theta\left( [U_{234}, [U_{012}, U_{013}]], \partial \lr{023} \right) = 0 \pmod{2 \pi}~.
    \label{eq: locality identity on 2 yellow plaquette}
\end{eqs}
Each color corresponds to a pair of plaquettes, and summing over two plaquettes with orientations given in Fig.~\ref{fig: RP2 plaquette orientation} results in the locality identity for each color. On the other hand, each symbol (red circles, blue stars, yellow triangles) appears in four plaquettes. For instance, the red circles in Fig.~\ref{fig: RP2 locality} are present on four plaquettes spanning $\lr{014}$ and $\lr{023}$. The corresponding locality identity across these four plaquettes is more involved:
\begin{eqs}
    &~\theta\left([U_{014}, U_{023}], \partial \lr{012} \right) + \theta\left([U_{014}, U_{023}], \partial \lr{024} \right)
    +\theta\left([U_{014}, U_{023}], \partial \lr{034} \right) + \theta\left([U_{014}, U_{023}], \partial \lr{013} \right) \\
    &= 0 \pmod{2\pi}~,
    \label{eq: locality identity on 4 red circles}
\end{eqs}
which results from summing over four plaquettes following the orientations in Fig.~\ref{fig: RP2 plaquette orientation}.
To show that Eq.~\eqref{eq: locality identity on 4 red circles} is indeed a locality identity, we expand it as follows:
\begin{eqs}
    &~\Big(\theta\left(U_{023}, \partial \lr{012} \right) + \theta\left(U_{014}, \partial \lr{012} + \partial \lr{023} \right)
    - \theta\left(U_{023}, \partial \lr{012} + \partial\lr{014} \right) - \theta\left(U_{014}, \partial \lr{012} \right) \Big)\\
    &+\Big(\theta\left(U_{023}, \partial \lr{024} \right) + \theta\left(U_{014}, \partial \lr{024} + \partial \lr{023} \right)
    - \theta\left(U_{023}, \partial \lr{024} + \partial\lr{014} \right) - \theta\left(U_{014}, \partial \lr{024} \right) \Big)\\
    &+\Big(\theta\left(U_{023}, \partial \lr{034} \right) + \theta\left(U_{014}, \partial \lr{034} + \partial \lr{023} \right)
    - \theta\left(U_{023}, \partial \lr{034} + \partial\lr{014} \right) - \theta\left(U_{014}, \partial \lr{034} \right) \Big)\\
    &+\Big(\theta\left(U_{023}, \partial \lr{013} \right) + \theta\left(U_{014}, \partial \lr{013} + \partial \lr{023} \right)
    - \theta\left(U_{023}, \partial \lr{013} + \partial\lr{014} \right) - \theta\left(U_{014}, \partial \lr{013} \right) \Big)~.
    \label{eq: showing 4 plaquettes as locality identity 1}
\end{eqs}
Next, the subsequent locality identities are added to the above equation to modify the configuration states $a$ within the $\theta(U(s), a)$ terms appearing in the first three lines:
\begin{enumerate}
    \item $\theta\left([[U_{014}, U_{023}], U_{123}], \partial \lr{012} \right)=0 \pmod{ 2 \pi}$.
    \item $\theta\left([[U_{014}, U_{023}], U_{123}U_{124}], \partial \lr{024} \right) =0 \pmod{ 2 \pi}$.
    \item $\theta\left([[U_{014}, U_{023}], U_{134}], \partial \lr{034} \right) =0 \pmod{ 2 \pi}$.
\end{enumerate}
Eq.~\eqref{eq: showing 4 plaquettes as locality identity 1} is transformed to
\begin{eqs}
    &~\Big({\color{blue} \theta\left(U_{023}, \partial \lr{012} + \partial \lr{123} \right)}
    + {\color{orange} \theta\left(U_{014}, \partial \lr{012} + \partial \lr{023} + \partial \lr{123} \right)}\\
    &- {\color{red}\theta\left(U_{023}, \partial \lr{012} + \partial\lr{014} + \partial \lr{123} \right)}
    - {\color{violet}\theta\left(U_{014}, \partial \lr{012} + \partial \lr{123}\right)} \Big)\\
    &+\Big( {\color{red}\theta\left(U_{023}, \partial \lr{024} + \partial \lr{124} + \partial \lr{123}\right)}
    + {\color{cyan}\theta\left(U_{014}, \partial \lr{024} + \partial \lr{023} + \partial \lr{124} + \partial \lr{123} \right)} \\
    &- {\color{blue} \theta\left(U_{023}, \partial \lr{024} + \partial\lr{014} + \partial \lr{124} + \partial \lr{123} \right)}
    - \theta\left(U_{014}, \partial \lr{024} + \partial \lr{124} + \partial \lr{123}\right) \Big)\\
    &+\Big({\color{brown} \theta\left(U_{023}, \partial \lr{034} + \partial \lr{134} \right)}
    + \theta\left(U_{014}, \partial \lr{034} + \partial \lr{023} + \partial \lr{134} \right)\\
    &- {\color{teal} \theta\left(U_{023}, \partial \lr{034} + \partial\lr{014} + \partial \lr{134} \right)}
    - {\color{cyan}\theta\left(U_{014}, \partial \lr{034} + \partial \lr{134} \right)} \Big)\\
    &+\Big( {\color{teal}\theta\left(U_{023}, \partial \lr{013} \right)}
    + {\color{violet}\theta\left(U_{014}, \partial \lr{013} + \partial \lr{023} \right)}
    - {\color{brown} \theta\left(U_{023}, \partial \lr{013} + \partial\lr{014} \right)}
    - {\color{orange}\theta\left(U_{014}, \partial \lr{013} \right)} \Big) \\
    &=0 \pmod{ 2 \pi}~,
    \label{eq: showing 4 plaquettes as locality identity 2}
\end{eqs}
where the terms with matched colors are completely canceled out in the last equality.
Thus, we have demonstrated that Eq.~\eqref{eq: locality identity on 4 red circles} indeed represents a valid locality identity.

Note that for the locality identity of each color or symbol, we have the freedom to multiply the overall minus sign, e.g., the locality identity on the yellow plaquettes in Eq.~\eqref{eq: locality identity on 2 yellow plaquette} can be modified as:
\begin{eqs}
    0&=-\theta\left( [U_{234}, [U_{012}, U_{013}]], \partial \lr{023} \right) \\
    &=-\theta\left( U_{012}^{-1} U_{013}^{-1} U_{012} U_{013}, \partial \lr{023} \right) - \theta\left( U_{013}^{-1} U_{012}^{-1} U_{013} U_{012}, \partial \lr{023} + \partial \lr{234} \right) \\
    &= \theta\left( U_{013}^{-1} U_{012}^{-1} U_{013} U_{012}, \partial \lr{023} \right) + \theta\left( U_{012}^{-1} U_{013}^{-1} U_{012} U_{013}, \partial \lr{023} + \partial \lr{234} \right)~.
\end{eqs}
The physical interpretation of this choice of locality identity is that it reverses the orientation of the yellow plaquettes in Fig.~\ref{fig: RP2 plaquette orientation}. This same argument applies to each symbol: we can simultaneously reverse the orientations of four plaquettes with the same symbol. We can derive different unitary sequences by altering the orientation of each color or symbol. However, all these sequences are equivalent, as they differ only by locality identities. Specifically, changing the orientation corresponds to adding $(-2)$ times the locality identity, and the unitary sequence is the sum of all locality identities divided by 2. For example, by selecting a different orientation, we obtain the 36-step unitary sequence shown in Fig.~\ref{fig: 36 step unitary sequence}, which exactly matches the sequence proposed in Ref.~\cite{FHH21}.

\begin{figure}[t]
    \centering
    \hspace{5ex}
    \includegraphics[scale=0.09]{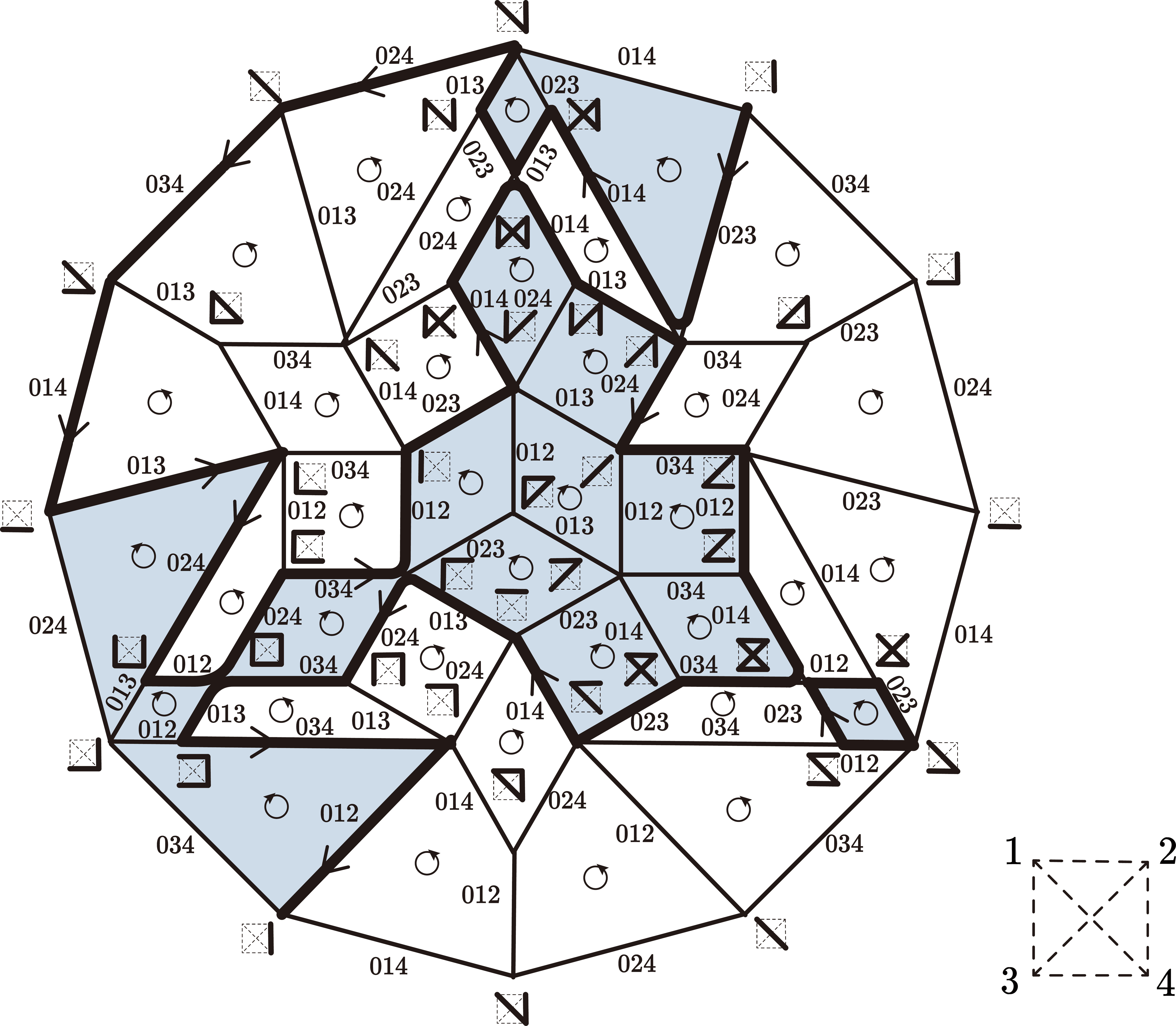}
    \caption{The 36-step unitary sequence corresponding to $\ZZ_2$ loop statistics in (3+1)D. Compared to Fig.~\ref{fig: RP2 plaquette orientation}, the orientations of certain locality identities have been reversed. Any choice of orientation for each locality identity yields a valid sequence for detecting the loop statistics. This particular orientation corresponds to the 36-step process proposed in Ref.~\cite{FHH21}, where a different computational method was used to derive the sequence.
    }
    \label{fig: 36 step unitary sequence}
\end{figure}

Finally, we used computers to enumerate all possible orientations for each locality identity. There are $2^{15} = 32768$ different configurations, corresponding to the three symbols and twelve colors. Our computations verified that the minimal number of steps required is $24$. Therefore, the unitary sequence presented in Fig.~\ref{fig: RP2 locality identities and orientations} is optimal, meaning there is no shorter sequence capable of detecting the loop statistics.

{\change 
\subsection{Particle fusion process in (1+1)D}\label{sec:Particle fusion process in (1+1)D in App}

In this section, we prove that the (1+1)D $\mathbb{Z}_2$ particle fusion statistics~\eqref{eq: (1+1)D particle fusion statistics} is an invariant in $E_{\mathrm{inv}}$. More precisely, on the complex
\begin{equation}
    \vcenter{\hbox{\includegraphics[width=0.12\linewidth]{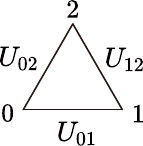}}}~~,
\end{equation}
we show that twice the phase of the (1+1)D particle fusion process can be written as a combination of locality identities.

One such identity is
\begin{eqs}
     \bra{\hbox{ \raisebox{-1ex}{\includegraphics[width=.6cm]{1d_particle_fusion_state_vaccum.pdf}}}}
    [U_{12},[U_{02},U_{01}]]
    \ket{\hbox{ \raisebox{-1ex}{\includegraphics[width=.6cm]{1d_particle_fusion_state_vaccum.pdf}}}}
    = 1,
\end{eqs}
which implies that
\begin{eqs}
    &\theta\left( [U_{02},U_{01}], \hbox{ \raisebox{-1ex}{\includegraphics[width=.6cm]{1d_particle_fusion_state_vaccum.pdf}}} \right) 
    +\theta\left( U_{12},\hbox{ \raisebox{-1ex}{\includegraphics[width=.6cm]{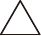}}} \right)
    +\theta\left( [U_{02},U_{01}]^{-1},\hbox{ \raisebox{-1ex}{\includegraphics[width=.6cm]{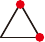}}} \right)
    +\theta\left( U_{12}^{-1},\hbox{ \raisebox{-1ex}{\includegraphics[width=.6cm]{Figures/1d_particle_fusion_state_U12.pdf}}} \right)\\
    =&~\theta\left( [U_{02},U_{01}], \hbox{ \raisebox{-1ex}{\includegraphics[width=.6cm]{1d_particle_fusion_state_vaccum.pdf}}} \right) 
    +\theta\left( [U_{02},U_{01}]^{-1},\hbox{ \raisebox{-1ex}{\includegraphics[width=.6cm]{Figures/1d_particle_fusion_state_U12.pdf}}} \right)\\
    =&~\theta\left( U_{01}, \hbox{ \raisebox{-1ex}{\includegraphics[width=.6cm]{1d_particle_fusion_state_vaccum.pdf}}} \right) 
    +\theta\left( U_{02},\hbox{ \raisebox{-1ex}{\includegraphics[width=.6cm]{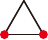}}} \right)
    +\theta\left( U_{01}^{-1}, \hbox{ \raisebox{-1ex}{\includegraphics[width=.6cm]{Figures/1d_particle_fusion_state_U12.pdf}}} \right) 
    +\theta\left( U_{02}^{-1},\hbox{ \raisebox{-1ex}{\includegraphics[width=.6cm]{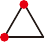}}} \right)\\
    &+\theta\left( U_{02}, \hbox{ \raisebox{-1ex}{\includegraphics[width=.6cm]{Figures/1d_particle_fusion_state_U12.pdf}}} \right) 
    +\theta\left( U_{01},\hbox{ \raisebox{-1ex}{\includegraphics[width=.6cm]{Figures/1d_particle_fusion_state_U01.pdf}}} \right)
    +\theta\left( U_{02}^{-1}, \hbox{ \raisebox{-1ex}{\includegraphics[width=.6cm]{Figures/1d_particle_fusion_state_vaccum.pdf}}} \right) 
    +\theta\left( U_{01}^{-1},\hbox{ \raisebox{-1ex}{\includegraphics[width=.6cm]{Figures/1d_particle_fusion_state_U02.pdf}}} \right)\\
    =& ~0 \pmod{2\pi}~.
\end{eqs}
In configuration space, these phases can be visualized as
\begin{equation}
    \raisebox{-7ex}{\includegraphics[scale=0.7]{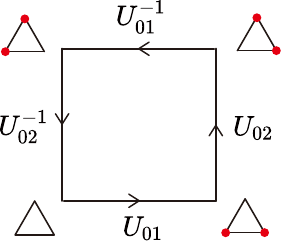}}
    \hspace{2em}+\hspace{2em}
    \raisebox{-7ex}{\includegraphics[scale=0.7]{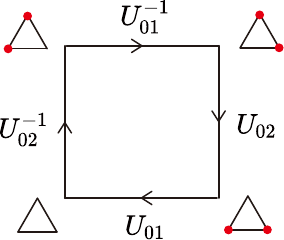}}
    \hspace{1em} = 0 \pmod{2\pi}~.
    \label{eq: 1d particle fusion locality identity 1}
\end{equation}

Another identity is 
\begin{eqs}
     \bra{\hbox{ \raisebox{-1ex}{\includegraphics[width=.6cm]{1d_particle_fusion_state_vaccum.pdf}}}}
    [U_{12},[U_{02}^{-1},U_{01}]]
    \ket{\hbox{ \raisebox{-1ex}{\includegraphics[width=.6cm]{1d_particle_fusion_state_vaccum.pdf}}}}
    = 1,
\end{eqs}
which implies that
\begin{eqs}
    &\theta\left( [U_{02}^{-1},U_{01}], \hbox{ \raisebox{-1ex}{\includegraphics[width=.6cm]{1d_particle_fusion_state_vaccum.pdf}}} \right) 
    +\theta\left( U_{12}^{-1}, \hbox{ \raisebox{-1ex}{\includegraphics[width=.6cm]{Figures/1d_particle_fusion_state_vaccum.pdf}}} \right)
    +\theta\left( [U_{02}^{-1},U_{01}]^{-1},\hbox{ \raisebox{-1ex}{\includegraphics[width=.6cm]{Figures/1d_particle_fusion_state_U12.pdf}}} \right)
    +\theta\left( U_{12},\hbox{ \raisebox{-1ex}{\includegraphics[width=.6cm]{Figures/1d_particle_fusion_state_U12.pdf}}} \right)\\
    =&~\theta\left( [U_{02}^{-1},U_{01}],\hbox{ \raisebox{-1ex}{\includegraphics[width=.6cm]{1d_particle_fusion_state_vaccum.pdf}}} \right) 
    +\theta\left( [U_{02}^{-1},U_{01}]^{-1},\hbox{ \raisebox{-1ex}{\includegraphics[width=.6cm]{Figures/1d_particle_fusion_state_U12.pdf}}} \right)\\
    =&~\theta\left( U_{01}, \hbox{ \raisebox{-1ex}{\includegraphics[width=.6cm]{1d_particle_fusion_state_vaccum.pdf}}} \right) 
    +\theta\left( U_{02}^{-1},\hbox{ \raisebox{-1ex}{\includegraphics[width=.6cm]{Figures/1d_particle_fusion_state_U01.pdf}}} \right)
    +\theta\left( U_{01}^{-1}, \hbox{ \raisebox{-1ex}{\includegraphics[width=.6cm]{Figures/1d_particle_fusion_state_U12.pdf}}} \right) 
    +\theta\left( U_{02},\hbox{ \raisebox{-1ex}{\includegraphics[width=.6cm]{Figures/1d_particle_fusion_state_U02.pdf}}} \right)\\
    &+\theta\left( U_{02}^{-1}, \hbox{ \raisebox{-1ex}{\includegraphics[width=.6cm]{Figures/1d_particle_fusion_state_U12.pdf}}} \right) 
    +\theta\left( U_{01},\hbox{ \raisebox{-1ex}{\includegraphics[width=.6cm]{Figures/1d_particle_fusion_state_U01.pdf}}} \right)
    +\theta\left( U_{02}, \hbox{ \raisebox{-1ex}{\includegraphics[width=.6cm]{Figures/1d_particle_fusion_state_vaccum.pdf}}} \right) 
    +\theta\left( U_{01}^{-1},\hbox{ \raisebox{-1ex}{\includegraphics[width=.6cm]{Figures/1d_particle_fusion_state_U02.pdf}}} \right)\\
    =& ~0 \pmod{2\pi}~.
\end{eqs}
In configuration space, these phases can be represented as
\begin{equation}
    \raisebox{-7ex}{\includegraphics[scale=0.7]{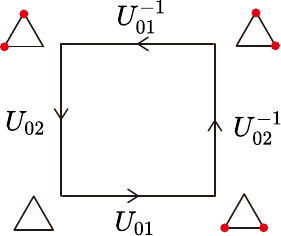}}
    \hspace{2em}+\hspace{2em}
    \raisebox{-7ex}{\includegraphics[scale=0.7]{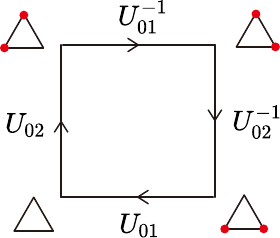}}
    \hspace{1em} = 0 \pmod{2\pi}~.
    \label{eq: 1d particle fusion locality identity 2}
\end{equation}
From the definition of $U^{-1}$ in Eq.~\eqref{eq: relation between theta U-1 and theta U}, we obtain  
\begin{eqs}
   \hbox{\raisebox{-7.5ex}{\includegraphics[scale=0.7]{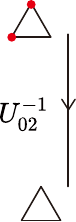}}} = -~  \hbox{\raisebox{-7.5ex}{\includegraphics[scale=0.7]{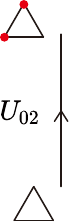}}}~,
   \quad \quad \quad \quad
   \hbox{\raisebox{-7.5ex}{\includegraphics[scale=0.7]{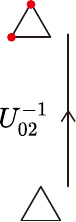}}} = -~  \hbox{\raisebox{-7.5ex}{\includegraphics[scale=0.7]{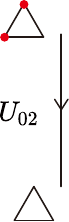}}}~.
   \label{eq: 1d particle inverse definition}
\end{eqs}
By adding Eqs.~\eqref{eq: 1d particle fusion locality identity 1} and \eqref{eq: 1d particle fusion locality identity 2} and applying Eq.~\eqref{eq: 1d particle inverse definition}, we find  
\begin{equation}
    \text{2}\hspace{1em}\times\hspace{1em}
    \raisebox{-7ex}{\includegraphics[scale=0.7]{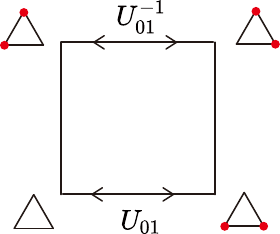}}
    \hspace{1em} = 0 \pmod{2\pi}~,
\end{equation}
which implies that $[U_{02},U_{01}^{2}]^{2} = 1$.  
This calculation produces the fusion statistics in Eq.~\eqref{eq: (1+1)D particle fusion statistics}.

\subsection{Loop-membrane mutual statistics in (3+1)D}\label{sec:Loop-membrane mutual statistics in (3+1)D in App}

In this section, we prove that the loop–membrane statistic introduced in Sec.~\ref{subsec: Loop membrane statistics in (3+1)D} is an invariant belonging to $E_{\mathrm{inv}}$.  
For the setup in Eq.~\eqref{eq:loop-membrane_statistics_2}, we introduce a new tetrahedron $t_{3}$ adjacent to $t_{1}$ and $t_{2}$ such that  
$t_{1} \cap t_{2} \cap t_{3} = v_{1}v_{2}$, the edge connecting $v_{1}$ and $v_{2}$.  
We then define $t_{4}$ as the complement of $t_{1} \cup t_{2} \cup t_{3}$, which satisfies $\partial t_{4}= \partial t_{1}+\partial t_{2}+\partial t_{3}$.  
In this configuration, we have $f \cap t_{1} \cap t_{2} \cap t_{4} = \varnothing$. Moreover, since $f$ can be replaced by its exterior without affecting the commutator, the higher commutators among $U_f$, $U_{t_1}$, $U_{t_2}$, and $U_{t_3}$ are also trivial.  
Under these conditions, the following four locality identities hold:  
\begin{enumerate}
    \item $\bra{\varnothing} [U_{f},[U_{t_{3}},[U_{t_{2}},U_{t_{1}}]]]\ket{\varnothing} =1.$
    \item$\bra{\varnothing}[U_{f},[U_{t_{3}},[U_{t_{2}}^{-1},U_{t_{1}}]]]\ket{\varnothing}=1.$
    \item $\bra{\partial U_{t_{3}}} [U_{f},[U_{t_{4}},[U_{t_{2}},U_{t_{1}}]]]\ket{\partial U_{t_{3}}} =1.$
    \item$\bra{\partial U_{t_{3}}}[U_{f},[U_{t_{4}},[U_{t_{2}}^{-1},U_{t_{1}}]]]\ket{\partial U_{t_{3}}}=1.$
\end{enumerate}
By summing these locality identities, we obtain  
\begin{eqs}
    & \theta\left( [U_{f},[U_{t_{3}},[U_{t_{2}},U_{t_{1}}]]], \varnothing \right)
    + \theta\left( [U_{f},[U_{t_{3}},[U_{t_{2}}^{-1},U_{t_{1}}]]], \varnothing \right)
    + \theta\left( [U_{f},[U_{t_{4}},[U_{t_{2}},U_{t_{1}}]]], \partial U_{t_{3}} \right)\\
    &+ \theta\left( [U_{f},[U_{t_{4}},[U_{t_{2}}^{-1},U_{t_{1}}]]], \partial U_{t_{3}} \right) \\
    =&~\theta\left( [U_{t_{3}},[U_{t_{2}},U_{t_{1}}]], \varnothing \right)
    + \theta\left( [U_{t_{3}},[U_{t_{2}},U_{t_{1}}]]^{-1}, \partial U_{f} \right)
    + \theta\left( [U_{t_{3}},[U_{t_{2}}^{-1},U_{t_{1}}]], \varnothing \right)
    + \theta\left( [U_{t_{3}},[U_{t_{2}}^{-1},U_{t_{1}}]]^{-1}, \partial U_{f} \right) \\
    &+ \theta\left( [U_{t_{4}},[U_{t_{2}},U_{t_{1}}]], \partial U_{t_{3}} \right)
    + \theta\left( [U_{t_{4}},[U_{t_{2}},U_{t_{1}}]]^{-1}, \partial U_{f}+\partial U_{t_{3}} \right)
    + \theta\left( [U_{t_{4}},[U_{t_{2}}^{-1},U_{t_{1}}]], \partial U_{t_{3}} \right)\\
    &+ \theta\left( [U_{t_{4}},[U_{t_{2}}^{-1},U_{t_{1}}]]^{-1}, \partial U_{f}+\partial U_{t_{3}} \right)\\
    =&~\theta\left( [U_{t_{3}},[U_{t_{2}},U_{t_{1}}]], \varnothing \right)+\theta\left( [U_{t_{3}},[U_{t_{2}}^{-1},U_{t_{1}}]], \varnothing \right)+ \theta\left( [U_{t_{4}},[U_{t_{2}},U_{t_{1}}]], \partial U_{t_{3}} \right)+ \theta\left( [U_{t_{4}},[U_{t_{2}}^{-1},U_{t_{1}}]], \partial U_{t_{3}} \right)\\
    &+(\theta\left( [U_{t_{3}},[U_{t_{2}},U_{t_{1}}]]^{-1}, \partial U_{f} \right)+\theta\left( [U_{t_{3}},[U_{t_{2}}^{-1},U_{t_{1}}]]^{-1}, \partial U_{f} \right)
    +\theta\left( [U_{t_{4}},[U_{t_{2}},U_{t_{1}}]]^{-1}, \partial U_{f}+\partial U_{t_{3}} \right)\\
    &+\theta\left( [U_{t_{4}},[U_{t_{2}}^{-1},U_{t_{1}}]]^{-1}, \partial U_{f}+\partial U_{t_{3}} \right))\\
    =&~2\,\theta\left( [U_{t_{2}},U_{t_{1}}^{2}], \varnothing \right)+ 2\,\theta\left( [U_{t_{2}},U_{t_{1}}^{2}]^{-1}, \partial U_{f} \right)\\
    =&~2\,\theta\left( [U_{f},[U_{t_{2}},U_{t_{1}}^{2}]], \varnothing \right)\\
    =&~\theta\left( [U_{f},[U_{t_{2}},U_{t_{1}}^{2}]]^{2}, \varnothing \right)\\
    =&~0 \pmod{2\pi}~.
\end{eqs}
Therefore, the loop–membrane statistical process $[U_{f},[U_{t_{2}},U_{t_{1}}^{2}]]$ follows directly from the locality identities.
}

\section{Detailed analysis on the structure of $E_{\mathrm{id}}$}
\label{app:globalshift}

In this Appendix, we show the following statement. Suppose we have an invariant $\Theta$ in $E_{\mathrm{inv}}$ for an Abelian symmetry group $G$, expressed as
\begin{align}
    e^{i\Theta}= \bra{a}U(s_n)^\pm\dots  U(s_1)^\pm\ket{a}~.
\end{align}
Let us pick any element $s_0\in\mathcal{S}$. Then the following ratio is an element of $E_{\mathrm{id}}$:
\begin{align}
    e^{i\Phi} = \frac{\bra{a}U(s_n)^\pm\dots  U(s_1)^\pm\ket{a}}{\bra{a + \partial s_0}U(s_n)^\pm\dots U(s_1)^\pm\ket{a+\partial s_0}}~.
\end{align}
In other words, we show that the phase $e^{i\Phi}$ can be expressed as the product of higher commutators of the form
\begin{eqs}
    & \bra{a'}[U(s'_m),[\cdots,[U(s'_2),U(s'_1)]]] \ket{a'}=1, \\
    &a'\in\mathcal{A}, s'_1, s'_2, \cdots, s'_m \in \mathcal{S} | s'_1 \cap s'_2 \cap \dots \cap s'_m = \varnothing~,
\end{eqs}
which we refer to as the locality identity in the main text. {\change This property holds when $X$ is a combinatorial manifold (Theorems VI.3 and VI.10 of Ref.~\cite{xue2025statisticsinvertibletopologicalexcitations}), but it fails for a generic simplicial complex $X$. A crucial step in the proof is the introduction of localization for excitation models. For completeness, we present a less abstract argument here.}

We first rewrite the phase $\Phi$ in the following fashion:
\begin{align}
    \begin{split}
        \Phi 
        =& \sum_{j=0}^{n-1} \theta(U(s_{j+1})^\pm, a+t_j) -\theta(U(s_{j+1})^\pm, a+\partial s_0 + t_j) \\
        =& \sum_{j=0}^{n-1} \theta(U(s_{j+1})^\pm, a+t_j) -\theta(U(s_{j+1})^\pm, a+\partial s_0 + t_j) 
        + \theta(U(s_0),a+t_{j+1}) - \theta(U(s_0),a+t_j)~,
    \end{split}
\end{align}
where $t_j = \sum_{k=1}^j \pm \partial s_j$. In the second equation, we used $t_n=t_0=0$.
Noting that
\begin{align}
\begin{split}
    &\arg(\bra{a} [U(s_2),U(s_1)]\ket{a} ) \\ &=\theta(s_1,a)+\theta(s_2,a+\partial s_1)-\theta(s_1,a+\partial s_2)-\theta(s_2,a) ~,
    \end{split}
\end{align}
and $\theta(U(s),a) = -\theta(U(s)^{-1} ,a+\partial s),$ we get
\begin{align}
    \begin{split}
        \Phi &= \sum^{n-1}_{\substack{j=0 \\ \text{$+$ sign}}} \bra{a + t_j} [U(s_0),U(s_{j+1})]\ket{a + t_j} 
        + \sum^{n-1}_{\substack{j=0 \\ \text{$-$ sign}}} \bra{a + t_{j+1}} [U(s_0),U(s_{j+1})]^{-1}\ket{a + t_{j+1}}~,
    \end{split}
    \label{eq:Thetacommutators}
\end{align}
where the sum breaks into the two parts depending on the sign on $U(s_j)^\pm$.
Let us fix an element $s\in\mathcal{S}$. In the expression \eqref{eq:Thetacommutators}, let us collect the commutators involving $U(s_j)^\pm$ with $s_j=s$.
Focusing on these commutators, the sum is schematically written as
\begin{align}
    \Phi = \sum_{a'} \bra{a'}[U(s_0),U(s)]^{\pm}\ket{a'} +\dots~,
    \label{eq:Phi sum commutator s}
\end{align}
where $\dots$ denotes the commutators involving $s_j\neq s$.
If $\text{supp}(s_0)\cap \text{supp}(s)=\varnothing$, the commutator can be regarded as the locality identity. Hence we assume $\text{supp}(s_0)\cap \text{supp}(s)$ is nontrivial.

Let us first study the case with $\text{supp}(s_0)\neq \text{supp}(s)$.
For $a',b'\in\mathcal{A}$, if $a'=b'$ at $\text{supp}(s_0)\cap \text{supp}(s')$, there exists $s_{ab}\in\mathcal{S}$ satisfying $ U(s_{ab})\ket{a'} = \ket{b'}$ and $\text{supp}(s_0)\cap \text{supp}(s')\cap \text{supp}(s_{ab})=\varnothing$. In that case one can convert $a'$ into $b'$ by the locality identity:
\begin{align}
    \frac{ \bra{a'}[U(s_0),U(s)]^{\pm}\ket{a'}}{\bra{b'}[U(s_0),U(s)]^{\pm}\ket{b'}} = \langle [U(s_{ab}),[U(s_0),U(s)]^\pm]\rangle = 1~.
\end{align}
This implies that the phase $\Phi$ in Eq.~\eqref{eq:Phi sum commutator s} depends on $a'$ only through the configuration of $a'$ in the vicinity of the mutual support $\text{supp}(s_0)\cap \text{supp}(s)$. Concretely, the configuration of $a'$ near the mutual support of the unitaries can be specified as follows. 
The mutual support $\text{supp}(s_0)\cap \text{supp}(s)$ generally becomes a $p$-simplex $\sigma_p$. Pick $(p+1)$ vertices $v_0,\dots, v_{p}$ of this $p$-simplex $\sigma_p$. Then, $a'$ in the expression~\eqref{eq:Phi sum commutator s} can be represented by a $(p+1)$-tuple of the restricted configurations $(a'|_{v_0},\dots, a'|_{v_p})$. In other words, we can replace $\ket{a'}$ with a canonical representative of the state $\ket{(a'|_{v_0},\dots, a'|_{v_p})}$ in $\mathcal{A}$ whose restriction to vertices become the fixed ones. This can be done by using the locality identity, and leaves the value of $\Phi$ invariant.

Let us write the set of restricted configurations $a|_{v_j}$ as $\mathcal{A}_{v_j}$. 
We also introduce a shorthand notation $a|_{v_j}=\alpha_j$.
We can express $\Phi$ as
\begin{align}
\begin{split}
    \Phi =& \sum_{\bigoplus_j\mathcal{A}_{v_j}}\epsilon_+(s,\{\alpha_j \}) \bra{\{\alpha_j\}} [U(s_0),U(s)] \ket{\{\alpha_j\}}
    +  \sum_{\bigoplus_j\mathcal{A}_{v_j}}\epsilon_-(s,\{\alpha_j\}) \bra{\{\alpha_j\}} [U(s_0),U(s)]^{-1} \ket{\{\alpha_j\}}~,
\end{split}
\end{align}
where $\epsilon_{\pm}$ is the positive integer coefficient counting the number of the commutator $\bra{\{\alpha_j\}} [U(s_0),U(s)]^{\pm} \ket{\{\alpha_j\}}$ appears in the sum of $\Phi$. Let us define $\epsilon$ as
\begin{align}
    \epsilon(s,\{\alpha_j \}) = \epsilon_+(s,\{\alpha_j \}) - \epsilon_-(s,\{\alpha_j \})~.
\end{align}
Then, due to the condition of $E_{\mathrm{inv}}$ shown in Eq.~\eqref{eq:inv3'} satisfied by the invariant $\Theta$, $\epsilon$ satisfies
\begin{align}
    \sum_{\bigoplus_{j\neq k} \mathcal{A}_{v_j}} \epsilon(s,(\alpha_0,\dots,\alpha_p)) = 0~,
    \label{eq:vertexconstraint0}
\end{align}
for any $\alpha_k\in\mathcal{A}_{v_k}$, and any $0\le k\le p$. The sum is over $\alpha_j\in\mathcal{A}_{v_j}$ with $j\neq k$ while $\alpha_k$ is fixed. Further, once we fix a choice of $\alpha_{v_k}$, it fixes the configuration of the excitation at the $p$-simplex $\sigma_p$ which we write $a_{\sigma_p}$. Then $\epsilon$ is nonzero iff $\alpha_0,\dots,\alpha_p$ all specify the same $a_{\sigma_p}$ at $\sigma_p$. This allows us to write a refined version of Eq.~\eqref{eq:vertexconstraint0} as
\begin{align}
    \sum_{\bigoplus_{j\neq k} \mathcal{A}_{v_j}(a_{\sigma_p})} \epsilon(s,(\alpha_0,\dots,\alpha_p)) = 0~,
    \label{eq:vertexconstraint}
\end{align}
where we defined $\mathcal{A}_{v_j}(a_{\sigma_p})\subset\mathcal{A}_{v_j}$ as a set of $\alpha_j$ which has the fixed configuration $a_{\sigma_p}$ at the simplex $\sigma_p$.

Eq.~\eqref{eq:vertexconstraint} gives a number of constraints on $\epsilon(s,\{\alpha_j\})$.
One can show that the solution to Eq.~\eqref{eq:vertexconstraint} is generated by the following (overcomplete) basis; let us pick a pair of elements $\alpha^{(0)}_j,\alpha^{(1)}_j$ from each $\mathcal{A}_{v_j}(a_{\sigma_p})$. Then, the basis is given by
\begin{align}
    \epsilon(s,\{\alpha^{(n_j)}_j\}) = (-1)^{\sum_j n_j}~,
    \label{eq:epsbasis}
\end{align}
with $n_j=0,1$, with the other $\epsilon$ zero.

Let us show that the basis \eqref{eq:epsbasis} corresponds to the trivial phase due to the locality identity. To see this, take a unitary $U(s_{\alpha_k})$ that shifts $\alpha^{(0)}_k$ to $\alpha^{(1)}_k$, but leaves the other $\alpha_j$ with $j\neq k$ invariant.
Note that the mutual support of the unitaries $\text{supp}(s_0)\cap \text{supp}(s)\cap \text{supp}(s_{\alpha_k})$ becomes a 0-simplex $v_k$. In particular, this implies that
\begin{align}
    \left(\bigcap_{j=0}^p \text{supp}(s_{\alpha_j})\right)\cap  \text{supp}(s_0)\cap \text{supp}(s) = \varnothing~.
\end{align}
Now, the basis \eqref{eq:epsbasis} corresponds to the invariant given by the following higher commutator:
\begin{align}
    \bra{\{\alpha^{(0)}_j\}}[U(s_{\alpha_0}),[\dots,[U(s_{\alpha_p}), [U(s_0),U(s)]]]]\ket{\{\alpha^{(0)}_j\}}~,
\end{align}
so each basis corresponds to the locality identity. Noticing this for each choice of $a_{\sigma_p}$ at $\sigma_p$ completes the proof for $\text{supp}(s_0)\neq \text{supp}(s)$.

When $\text{supp}(s_0)=\text{supp}(s)$, the proof is done in a similar logic: the difference is that the support of $[U(s_0),U(s)]$ is given by $\partial(\text{supp}(s))$, so we need to take $p+2$ vertices of a $(p+1)$-simplex $s$ to fix the configuration of excitations in the vicinity of the operator support.
Let us write these vertices as $v_0,\dots, v_{p+1}$. The boundary $p+1$-simplexes of $s$ are written as $\sigma_{\hat 0}, \dots \sigma_{\widehat{p+1}}$, where $\sigma_{\hat j}$ does not contain the vertex $v_j$. 
We again write the set of restricted configurations $a|_{v_j}=\alpha_j$ as $\mathcal{A}_{v_j}$, with $0\le j\le p+1$. One can again express $\Phi$ as
\begin{align}
\begin{split}
    \Phi =& \sum_{\bigoplus_j\mathcal{A}_{v_j}}\epsilon_+(s,\{\alpha_j \}) \bra{\{\alpha_j\}} [U(s_0),U(s)] \ket{\{\alpha_j\}}
    +  \sum_{\bigoplus_j\mathcal{A}_{v_j}}\epsilon_-(s,\{\alpha_j\}) \bra{\{\alpha_j\}} [U(s_0),U(s)]^{-1} \ket{\{\alpha_j\}}~,
\end{split}
\end{align}
where $\epsilon_{\pm}$ is the positive integer coefficient counting the number of the given commutator appears in the sum of $\Phi$. We again define $\epsilon$ as
\begin{align}
    \epsilon(\{\alpha_j \}) = \epsilon_+(s,\{\alpha_j \}) - \epsilon_-(s,\{\alpha_j \})~.
\end{align}
where we suppressed the dependence on $s$.
Then, due to the condition of $E_{\mathrm{inv}}$ in Eq.~\eqref{eq:inv3'} satisfied by the invariant $\Theta$, $\epsilon$ satisfies
\begin{align}
    \sum_{\bigoplus_{j\neq k} \mathcal{A}_{v_j}} \epsilon(\alpha_0,\dots,\alpha_{p+1}) = 0~,
    \label{eq:vertexconstraint0'}
\end{align}
for any $\alpha_k\in\mathcal{A}_{v_k}$, and any $0\le k\le p+1$. The sum is over $\alpha_j\in\mathcal{A}_{v_j}$ with $j\neq k$ while $\alpha_k$ is fixed. 
Once we fix the restricted configuration $\alpha_k$, it fixes the configuration of the excitation $a$ at $p$-simplexes $\{\sigma_{\hat{j}}\}$ except for $\sigma_{\hat{k}}$.

Let us define the set 
$\mathcal{A}_{v_k}(\{a_{\sigma_{\hat j}}\})\subset\mathcal{A}_{v_k}$ as a set of $\alpha_k$ which has the fixed configurations of excitations $\{a_{\sigma_{\hat j}}\}$ at the $p$-simplexes $\sigma_{\hat{0}},\dots\sigma_{\widehat{p+1}}$ of $s$. Note that this set does not depend on $a_{\sigma_{\hat k}}$.

One can show that the solution to Eq.~\eqref{eq:vertexconstraint0'} is generated by the following basis:
\begin{enumerate}
    \item Let us fix a configuration of excitations $\{a_{\sigma_{\hat j}}\}$ at the $p$-simplexes of $s$, and consider $\mathcal{A}_{v_k}(\{a_{\sigma_{\hat j}}\})$. Let us pick two elements $\alpha_k^{(0)},\alpha_k^{(1)}\in\mathcal{A}_{v_k}(\{a_{\sigma_{\hat j}}\})$ for each $0\le k\le p+1$.
    Then, the basis is given by
\begin{align}
    \epsilon(\{\alpha^{(n_j)}_j\}) = (-1)^{\sum_j n_j}~,
    \label{eq:epsbasis1}
\end{align}
with $n_j=0,1$, with the other $\epsilon$ zero. 

\item Let us fix a choice of the vertex $v_k$. Take two configurations of the excitations $\{a_{\sigma_{\hat j}}\}$, $\{a'_{\sigma_{\hat j}}\}$, where $a_{\sigma_{\hat j}}=a'_{\sigma_{\hat j}}$ for $j\neq k$.
Pick two elements $\alpha_k^{(0)}, \alpha_k^{(1)}\in\mathcal{A}_{v_k}(\{a_{\sigma_{\hat j}}\})$. Note that $\alpha_k^{(0)}, \alpha_k^{(1)}\in\mathcal{A}_{v_k}(\{a'_{\sigma_{\hat j}}\})$. For $l\neq k$, let us pick a set of elements $\{\alpha_{l}\}_{l\neq k},\{\alpha'_{l}\}_{l\neq k}$ from $\{\mathcal{A}_{v_l}(\{a_{\sigma_{\hat j}}\})\}_{l\neq k}$, $\{\mathcal{A}_{v_l}(\{a'_{\sigma_{\hat j}}\})\}_{l\neq k}$. Then, the basis is given by
\begin{align}
    \begin{split}
        \epsilon(\alpha_k^{(0)},\{\alpha_{l}\}_{l\neq k}) &= 1~, \\
        \epsilon(\alpha_k^{(1)},\{\alpha_{l}\}_{l\neq k}) &= -1~, \\
        \epsilon(\alpha_k^{(0)},\{\alpha'_{l}\}_{l\neq k}) &= -1~, \\
        \epsilon(\alpha_k^{(1)},\{\alpha'_{l}\}_{l\neq k}) &= 1~, \\
    \end{split}
    \label{eq:epsbasis2}
\end{align}
with the other $\epsilon$ zero. 

\item For each $p$-simplex $\sigma_{\hat{j}}$, let us take a pair of  configurations of excitation $a^{(0)}_{\sigma_{\hat{j}}},a^{(1)}_{\sigma_{\hat{j}}}$. For each configuration $\{a^{(n_j)}_{\sigma_{\hat{j}}}\}$ at $p$-simplexes, we choose a single excitation configuration $\{\alpha_j\}_{\{n_j\}}\in \mathcal{A}(\{a^{(n_j)}_{\sigma_{\hat{j}}}\})$. 
These excitation configurations are chosen so that when $n_j=n'_j$ for $j\neq k$, we have $(\alpha_k)_{\{n_j\}} =(\alpha_k)_{\{n'_j\}}$.
Then, the basis is given by
\begin{align}
    \epsilon(\{\alpha_j\}_{\{n_j\}}) = (-1)^{\sum_j n_j}~.
    \label{eq:epsbasis3}
\end{align}
\end{enumerate}
 Each type of the basis \eqref{eq:epsbasis1} corresponds to the higher commutators:
 \begin{enumerate}
     \item For the first type of the basis, let us take a unitary $U(s_{\alpha_k})$ that transforms $\alpha_k^{(0)}$ into $\alpha_k^{(1)}$, leaving the other $\{\alpha_j\}$ invariant. Note that the mutual support of the unitaries $\text{supp}(s)\cap \text{supp}(s_{\alpha_k})$ becomes a 0-simplex $v_k$. In particular, this implies that
\begin{align}
    \left(\bigcap_{j=0}^{p+1} \text{supp}(s_{\alpha_j})\right)\cap   \text{supp}(s) = \varnothing~.
\end{align}
Now, the basis \eqref{eq:epsbasis1} corresponds to the invariant given by the following higher commutator:
\begin{align}
    \bra{\{\alpha^{(0)}_j\}}[U(s_{\alpha_0}),[\dots,[U(s_{\alpha_{p+1}}), [U(s_0),U(s)]]]]\ket{\{\alpha^{(0)}_j\}}~.
\end{align}
\item For the second type of the basis \eqref{eq:epsbasis2}, let us take a unitary $U(s_{\alpha_k})$ that transforms $\alpha_k^{(0)}$ into $\alpha_k^{(1)}$, leaving the other $\{\alpha_j\}$ invariant.
Let us also take a unitary $U(s_{\hat k})$ that transforms $\{\alpha_{l}\}_{l\neq k}$ into $\{\alpha'_{l}\}_{l\neq k}$, leaving $\alpha_k$ invariant. The mutual support of the unitaries becomes $\text{supp}(s_{\alpha_{k}})\cap\text{supp}(s) = v_k$, $\text{supp}(s_{\hat{k}})\cap\text{supp}(s) = \sigma_{\hat k}$.
We then have
\begin{align}
    \text{supp}(s_{\alpha_{k}})\cap\text{supp}(s_{\hat{k}})\cap\text{supp}(s)=\varnothing~.
\end{align}
The basis \eqref{eq:epsbasis2} corresponds to the invariant given by the following higher commutator:
\begin{align}
    \bra{\alpha^{(0)}_k,\{\alpha_{l}\}_{l\neq k}}[U(s_{\alpha_k}),[U(s_{\hat{k}}),[U(s_0),U(s)]]]\ket{\alpha^{(0)}_k,\{\alpha_{l}\}_{l\neq k}}~.
\end{align}

\item For the third type of the basis \eqref{eq:epsbasis3}, let us take a unitary $U(s_{\hat{k}})$ that transforms $\{\alpha_j\}_{\{n_j\}}$ into $\{\alpha_j\}_{\{n'_j\}}$, with $n'_k= n_k+1$ mod 2, and $n_j=n'_j$ with $j\neq k$. Note that the mutual support $\text{supp}(s)\cap \text{supp}(s_{\hat{k}})$ becomes a $p$-simplex $\sigma_{\hat{k}}$. This implies that
\begin{align}
    \left(\bigcap_{j=0}^{p+1} \text{supp}(s_{\hat{j}})\right)\cap   \text{supp}(s) = \varnothing~.
\end{align}
The basis \eqref{eq:epsbasis3} corresponds to the invariant given by the following higher commutator:
\begin{align}
    \bra{\{\alpha_j\}_{\{n_j=0\}}}[U(s_{\hat{0}}),[\dots,[U(s_{\widehat{p+1}}), [U(s_0),U(s)]]]]\ket{\{\alpha_j\}_{\{n_j=0\}}}~.
\end{align}

\end{enumerate}
This completes the proof that $\Phi$ is given by linear combinations of the locality identities.
\end{widetext}

\bibliography{bibliography}

\end{document}